\documentclass[aps, prd, reprint, onecolumn, 10pt, superscriptaddress, nofootinbib, showkeys]{revtex4-2}

\usepackage{amsmath}

\usepackage{amsfonts}
\usepackage{amssymb}

\usepackage{graphicx}
\usepackage{subfig}
\usepackage[colorlinks=true, allcolors=blue]{hyperref}

\usepackage{ragged2e}
\usepackage[compatibility=false]{caption}
\usepackage{subcaption}

\DeclareCaptionJustification{fulljustified}{\justifying}

\usepackage[normalem]{ulem}

\usepackage[dvipsnames]{xcolor}

\begin{document}

	\title{Seven- and eight-loop critical exponents of the three-dimensional Ising model}
	
	\author{D. Shapoval}
        \affiliation{Yukhnovskii Institute for Condensed Matter Physics of the National Academy of Sciences of Ukraine, Lviv 79011, Ukraine}
        \affiliation{${\mathbb L}^4$ Collaboration \& Doctoral College for the Statistical Physics of Complex Systems, Leipzig-Lorraine-Lviv-Coventry, Europe}
        
    \author{Yu. Honchar}
        \affiliation{Yukhnovskii Institute for Condensed Matter Physics of the National Academy of Sciences of Ukraine, Lviv 79011, Ukraine}
        \affiliation{${\mathbb L}^4$ Collaboration \& Doctoral College for the Statistical Physics of Complex Systems, Leipzig-Lorraine-Lviv-Coventry, Europe}
        
    \author{B. Delamotte}
        \affiliation{Sorbonne Universit{\'e}, CNRS, Laboratoire de Physique Th{\'e}orique de la Mati{\`e}re Condens{\'e}e, LPTMC, Paris F-75005, France}
        
    \author{M. Dudka}
        \affiliation{Yukhnovskii Institute for Condensed Matter Physics of the National Academy of Sciences of Ukraine, Lviv 79011, Ukraine}
        \affiliation{${\mathbb L}^4$ Collaboration \& Doctoral College for the Statistical Physics of Complex Systems, Leipzig-Lorraine-Lviv-Coventry, Europe}
        \affiliation{Lviv Polytechnic National University, Lviv 79013, Ukraine}
        
    \author{Yu. Holovatch}
        \affiliation{Yukhnovskii Institute for Condensed Matter Physics of the National Academy of Sciences of Ukraine, Lviv 79011, Ukraine}
        \affiliation{${\mathbb L}^4$ Collaboration \& Doctoral College for the Statistical Physics of Complex Systems, Leipzig-Lorraine-Lviv-Coventry, Europe}
        \affiliation{Centre for Fluid and Complex Systems, Coventry University, Coventry CV1 5FB, United Kingdom}
        \affiliation{Complexity Science Hub, Vienna 1030, Austria}
	
	\begin{abstract}
        We determine the critical exponents $\eta$, $\nu$, and the correction-to-scaling exponent $\omega$ of the three-dimensional Ising universality class by resumming the recently computed seven- and eight-loop renormalization-group series in the $\epsilon=4-d$ expansion [O.~Schnetz, \textit{Phys. Rev. D} \textbf{97}, 085018 (2018); O.~Schnetz, \textit{Phys. Rev. D} \textbf{107}, 036002 (2023)]. The resummation combines conformal mapping with a homographic transformation, while the resummation parameters are optimized according to two complementary criteria. This approach yields precise estimates of the critical exponents together with quantitative uncertainty estimates. We find that the error bar on $\eta$ decreases rapidly with increasing loop order, whereas this is the case neither for $\nu$ nor for $\omega$. Unexpectedly, although the estimated values are  accurate in absolute terms, their slow convergence with the loop order leads to a slight but systematic tension with the conformal bootstrap estimates that are currently considered as the benchmark. We discuss several possible origins of this behavior and its implications for high-order resummations of perturbative renormalization-group series.
    \end{abstract}
    
    
    \maketitle
    
	\section{Introduction}\label{intro}

Over the past decades, field-theoretic methods have become indispensable in statistical physics, providing a unified framework for the description of strongly correlated systems, whether classical or quantum, at or out of thermal equilibrium. This development stems from the profound and highly nontrivial analogies between classical and quantum fluctuations, which have fostered a fruitful exchange of concepts and techniques between statistical mechanics and quantum field theory. As a result, methods originally devised in the latter have become standard tools in the former, while statistical-physics approaches have significantly influenced the modern formulation of quantum field theories.

Perturbative renormalization-group (RG) methods have played a central role in this cross-fertilization, as they constitute one of the most successful theoretical frameworks for the study of strongly correlated systems, for instance for systems near a continuous phase transition. While alternative nonperturbative approaches, such as Monte Carlo (MC) simulations, the functional renormalization group also called the nonperturbative renormalization group (NPRG), and the conformal bootstrap (CB), have considerably expanded the available toolbox, perturbative RG continues to occupy a unique place in both statistical physics and quantum field theory. Beyond its enduring practical relevance, it often remains the only analytically tractable method for addressing a wide range of problems. Assessing its accuracy and convergence properties therefore remains a question of fundamental methodological importance.

The calculation of critical exponents in $d$-dimensional O$(n)$ models has long served as a benchmark for evaluating the performance of perturbative approaches. Mainly, it concerns three-dimensional (that means space dimension $d=3$) models. The reason is that these models can also be studied by the alternative methods mentioned above -- MC, NPRG and CB -- as well as high temperature expansions which provide estimates of critical quantities with, for some of them, an accuracy far exceeding that of present perturbative calculations. As a consequence, they offer a unique opportunity to quantitatively assess the reliability and limitations of perturbative RG predictions.

The prevailing view in the literature, including many pedagogical accounts, is that perturbative RG methods lead to $\epsilon=4-d$ expansions of the O($n$) critical exponents whose resummations {\it \`a la} Padé--Borel  progressively approach the ``exact'' critical exponents -- those obtained with the CB for instance -- as additional perturbative orders are incorporated. This picture is largely supported by low-order calculations, which display an apparent convergence and provide quantitatively successful predictions. Yet the large order behavior of the $\epsilon$-expansion remains incompletely understood. In particular, several authors have pointed out that nonanalytic contributions, invisible to conventional perturbation theory, could affect the convergence of resummed series, see Sec.~\ref{review} below. Determining whether such effects become relevant at high orders therefore remains an important open question. The recent evaluation of all seven-loop contributions and of a substantial subset of eight-loop diagrams by Schnetz \cite{schnetz2018,schnetz2023} offers an unprecedented opportunity to revisit this issue, precisely in the regime where a straightforward convergence toward the exact exponents ceases to be evident.

We focus in the following on the $3d$ Ising model corresponding to the $n=1$ case. We compute three independent critical exponents $\eta,\nu$ and the correction to scaling exponent $\omega$ using the $\epsilon$-expansion at the highest available loop orders. We apply a well-tested Borel resummation framework enhanced by conformal mapping, further stabilized by a homographic transformation of the series to assure the most accurate up-to-date error estimation, with the aim of clarifying the origin of the discrepancies with the exact results and assessing the convergence and reliability of the perturbative estimates. 

The remainder of this paper is organized as follows. In Sec.~\ref{review}, we present a comprehensive overview of the established perturbative and nonperturbative approaches used to determine critical exponents  in  the three-dimensional Ising universality class. As a prerequisite to the subsequent analysis, we briefly review the main methods that have yielded high-precision estimates of critical exponents, including field-theoretical renormalization-group techniques combined with resummation procedures \cite{leguillou1985, guida1998, kompaniets2017, kompaniets2020}, high-temperature expansions \cite{campostrini2002}, Monte Carlo simulations \cite{hasenbusch2010, hasenbusch2021}, the nonperturbative renormalization group (NPRG) \cite{depolsi2020, balog2019}, and conformal-bootstrap methods \cite{elshowk2014, kos2016, simmons_duffin2017, bootstrap_3d}. A detailed comparison of these approaches and of their quantitative predictions for the Ising universality class is provided. 
In Sec.~\ref{Slow_conv}, we discuss the potential slow convergence of the perturbative RG series and the
reliability of high-order perturbative estimates.
Sec.~\ref{RG_and_resummation} describes the  RG framework employed in this work. We first discuss the highest-order perturbative expansions currently available for the RG functions and then present the Borel--conformal resummation procedure used throughout our analysis. Sec.~\ref{convergence} is devoted to the assessment of convergence properties and to the methods used to quantify the reliability of the resummed estimates.
Our results are presented and discussed in Sec.~\ref{results}, where we analyze the critical exponents obtained from high-order perturbative expansions and examine their convergence properties. Finally, Sec.~\ref{conclusion} summarizes our main findings and discusses possible directions for future research. Additional technical details are collected in the Appendixes.

    \section{A review of the different methods and their results}
    \label{review}
	
Perturbative  RG approaches to the determination of critical exponents in the three-dimen\-sional Ising universality class are formulated within the continuum
scalar $\phi^4$ field theory, which provides a convenient framework for computing the RG functions as perturbative series in an expansion parameter. Two complementary perturbative schemes have been extensively employed. The first is the $\epsilon=4-d$ expansion, in which all quantities are expressed as power series in $\epsilon$ and evaluated at $\epsilon=1$. The second is the fixed-dimension approach, where the RG functions are calculated directly in three dimensions as series in the renormalized coupling constant.

The leading critical exponents $\eta$ and $\nu$ are obtained from the RG functions $\beta$, $\gamma_{\phi}$, and $\gamma_{\phi^{2}}$, where $\gamma_{\phi}$ and $\gamma_{\phi^{2}}$ denote the anomalous dimensions of the field $\phi$ and of the composite operator $\phi^2$, respectively. These quantities are evaluated at the stable fixed-point coupling $g^\ast$, defined as the nontrivial zero of the beta function, $\beta(g^\ast)=0$. The leading correction-to-scaling exponent $\omega$ characterizes the velocity of the RG flow  along the first irrelevant direction, corresponding to the $\phi^4$ operator, and is given by the derivative of the beta function evaluated at the fixed point.

Since the perturbative expansions of the RG functions $\beta(g)$, $\gamma_{\phi}(g)$, and $\gamma_{\phi^{2}}(g)$ are asymptotic and exhibit factorial growth at large orders, meaningful numerical predictions require appropriate resummation procedures. This has motivated the development of a variety of techniques, ranging from simple Pad\'e approximants to more sophisticated methods designed to control the singularity structure of truncated perturbative series.

\begin{figure}
		\begin{center}
			\subfloat[]{\includegraphics[width=.5\textwidth]{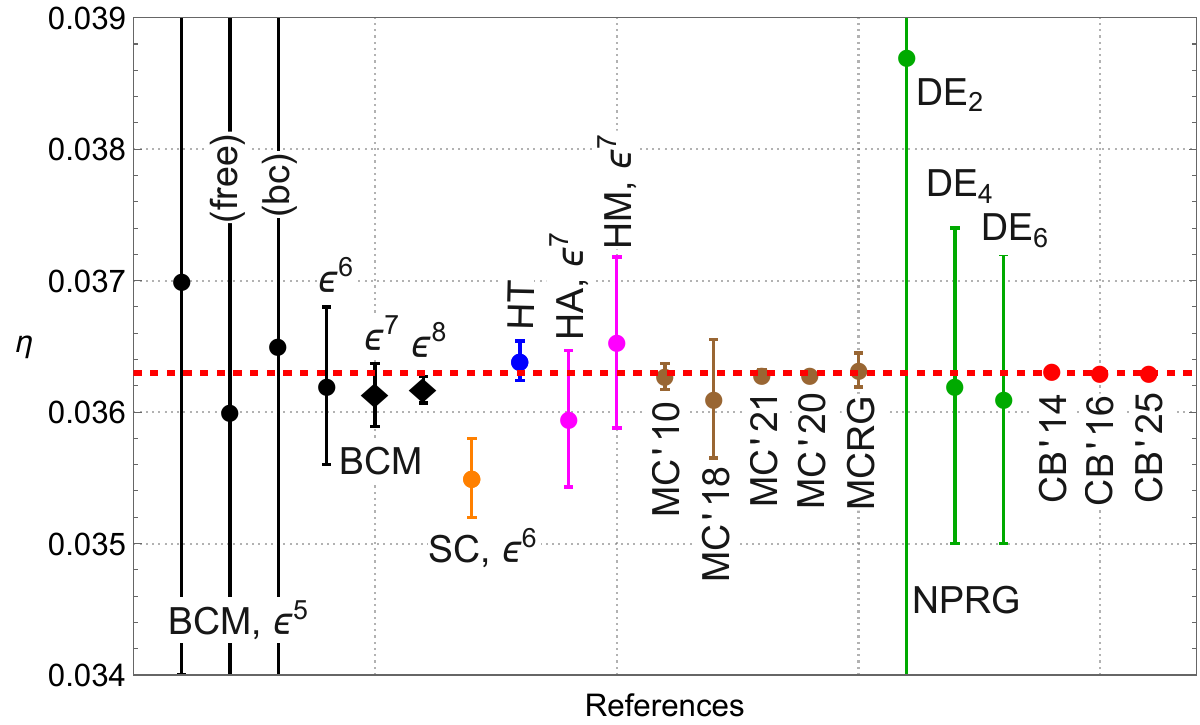}}\hfill
			\subfloat[]{\includegraphics[width=.5\textwidth]{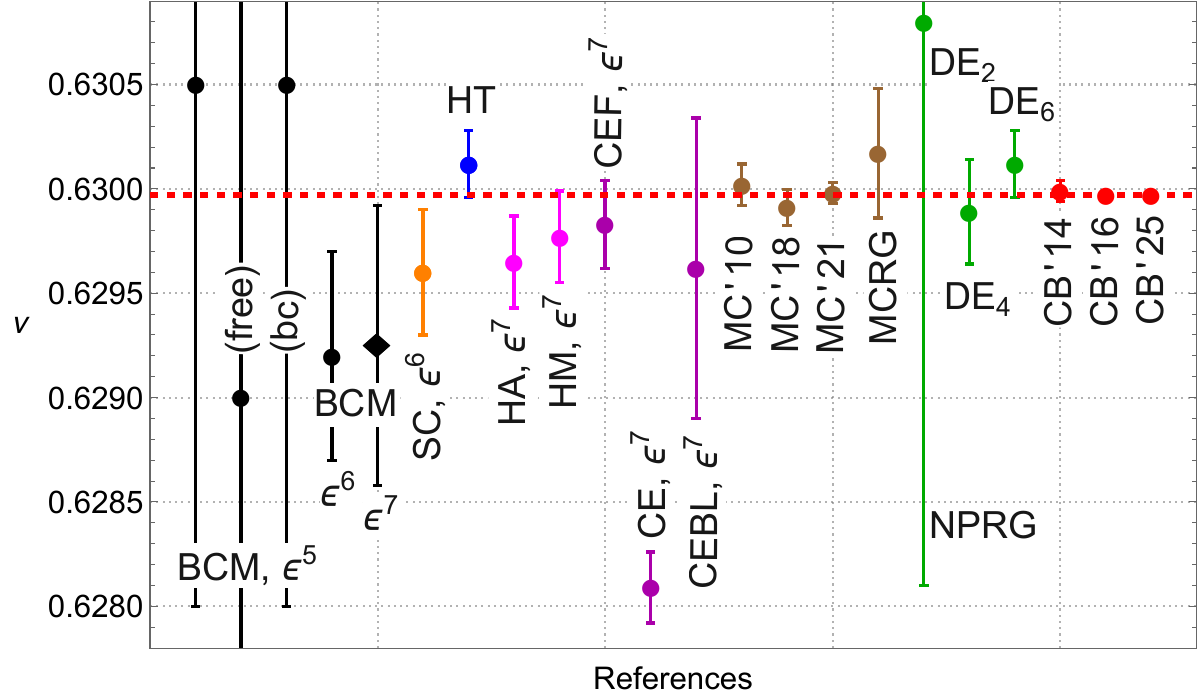}}	
		\end{center}
		\begin{center}
			\subfloat[]{\includegraphics[width=.5\textwidth]{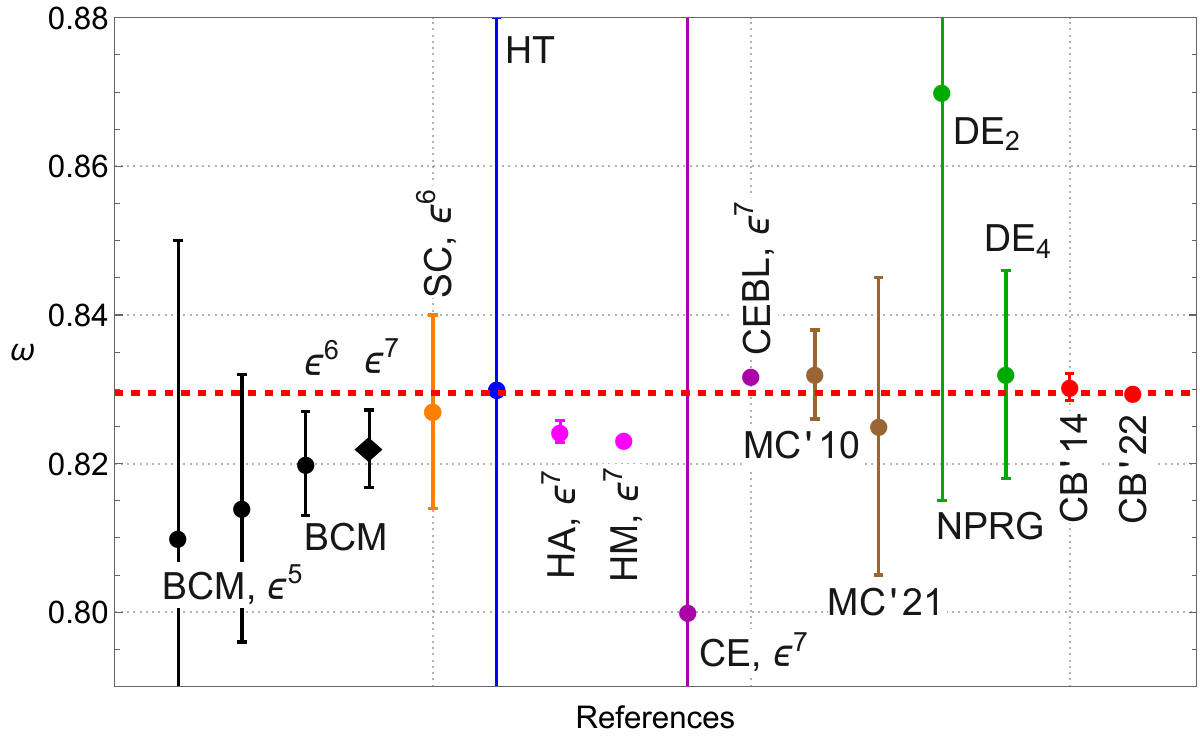}}
		\end{center}
		\caption{Comparison of estimates of the critical exponents $\eta$ [(a)], $\nu$ [(b)], and $\omega$ [(c)] for the three-dimensional Ising universality class obtained using a variety of analytical and numerical approaches. The estimates are categorized by method, within which they are presented in chronological order. Black symbols correspond to Borel resummation with conformal mapping (BCM) at different perturbative orders: \emph{free} denotes the standard $\epsilon$-expansion resummation, while \emph{bc} incorporates the exact two-dimensional value as a boundary condition \cite{leguillou1985, guida1998, kompaniets2017}. Orange symbols correspond to the self-consistent resummation procedure (SC) \cite{kompaniets2020}. Blue symbols denote high-temperature (HT) expansions \cite{campostrini2002}. Pink symbols represent hypergeometric approximants (HA) and hypergeometric--Meijer approximants (HM) \cite{shalaby2021, shalaby2021b}. Purple symbols correspond to the continued exponential fraction (CEF), continued exponential (CE), and continued exponential with Borel--Leroy transformation (CEBL) approaches \cite{abhignan2023}. Brown symbols indicate Monte Carlo estimates, including MC$'10$ \cite{hasenbusch2010}, MC$'18$ \cite{ferrenberg2018}, MC$'21$ \cite{hasenbusch2021}, MC$'20$ \cite{gori2020} and Monte Carlo renormalization-group (MCRG) results \cite{kaupuzs2023}. Green symbols show nonperturbative renormalization-group (NPRG) estimates obtained within the derivative expansion at orders $\partial^2$ (DE$_2$), $\partial^4$ (DE$_4$), and $\partial^6$ (DE$_6$) \cite{depolsi2020, balog2019}. Red symbols correspond to conformal-bootstrap (CB) determinations, including CB$'14$ \cite{elshowk2014}, CB$'16$ \cite{kos2016, simmons_duffin2017}, CB$'22$ \cite{reehorst2022}, and CB$'25$ \cite{bootstrap_3d}. The horizontal dashed lines indicate the currently most precise estimates available for the three-dimensional Ising universality class: Ref.~\cite{bootstrap_3d} for $\eta$ and $\nu$, and Ref.~\cite{reehorst2022} for $\omega$. Black diamonds denote the estimates obtained in the present work (see Table~\ref{tab:final_results}).}
		\label{eta_nu_omega_ref}
	\end{figure}

A particularly influential approach is based on a combination of Borel transformation and conformal mapping (BCM) \cite{leguillou1980}.
The method constructs the Borel transform of a divergent perturbative series and subsequently performs an analytic continuation through a conformal mapping of the Borel plane. By relocating the dominant singularities to the boundary of the convergence domain, this procedure enables accurate numerical reconstruction of the original function. Applied to the Ising universality class, BCM yielded critical exponents of significantly higher accuracy than those obtained from direct truncation of the perturbative series.
Moreover, its successful agreement with exact results in two dimensions provided strong evidence for the reliability of the method (see Fig.~\ref{eta_nu_omega_ref} and Table~\ref{tab:brief_review}).

The BCM approachis widely used in  the fixed-dimension formulation, where seven-loop perturbative expansions became available for the anomalous dimensions and the RG function associated with the $\phi^2$ operator \cite{murraynickel, guida1998} (see Fig.~\ref{eta_nu_omega_ref} and Table~\ref{tab:brief_review}). Several important refinements were also introduced. In particular, a homographic transformation of the coupling constant was employed to reduce the influence of spurious singularities in the complex coupling plane, thereby improving the convergence properties of the resummed series (see Sec.~\ref{convergence}) \cite{leguillou1985}.

More recently, the complete six-loop RG functions of the O$(n)$ models, including the scalar ($n=1$) theory relevant to the Ising universality class, were computed within the $\epsilon$-expansion framework \cite{kompaniets2017}. The resulting perturbative series were then resummed using BCM together with several optimization procedures, one of which will be discussed in Sec.~\ref{convergence}. The corresponding estimates of the critical exponents, along with the resummed $\epsilon^5$ results \cite{leguillou1985},  are shown as [BCM, $\epsilon^5$],  [BCM, $\epsilon^6$] in Fig.~\ref{eta_nu_omega_ref} and are summarized in Table~\ref{tab:brief_review}.

Notably, the six-loop estimates differ only marginally from the corresponding five-loop results, illustrating the slow convergence of the perturbative expansions even after resummation. Building on this work, Ref.~\cite{kompaniets2020} extended the six-loop $\epsilon$-expansion analysis to the calculation of fractal dimensions associated with critical curves in O($n$)-symmetric field theories. By introducing additional constraints into the resummation procedure, the critical exponents were evaluated via the self-consistent resummation technique. Shown as [SC, $\epsilon^6$] in Fig.~\ref{eta_nu_omega_ref}, the corresponding values provide an additional estimate of the critical exponents (see also Table~\ref{tab:brief_review}). 

High-temperature (HT) series expansions constitute an independent and complementary route to the determination of critical exponents. They are based on expansions of thermodynamic observables, such as the susceptibility and specific heat, in powers of the inverse temperature within the disordered phase \cite{campostrini2002}. After suitable extrapolation towards the critical point, these series yield remarkably precise estimates of critical quantities. The resulting values provide an important point of comparison with field-theoretical, Monte Carlo, and other nonperturbative approaches (see Table~\ref{tab:brief_review} and the entries labeled [HT] in Fig.~\ref{eta_nu_omega_ref}). 

Alongside the more established techniques for resumming divergent asymptotic series, several alternative approaches have recently been proposed. One such method relies on generalized hypergeometric approximants \cite{shalaby2021, shalaby2021b}. The central idea is to construct approximants that simultaneously incorporate information from the weak-coupling expansion, the strong-coupling regime, and the known large-order behavior of the perturbative coefficients. In the context of O$(n)$-symmetric field theories, critical exponents are represented by generalized hypergeometric functions whose parameters are adjusted to reproduce these different asymptotic constraints. Applied to the Ising universality class, this approach yields estimates that are broadly consistent with those obtained from Borel resummation (see the entries [HA, $\epsilon^7$] and [HM, $\epsilon^7$] in Fig.~\ref{eta_nu_omega_ref}). It therefore provides an interesting alternative framework for analyzing high-order perturbative expansions.

Another recently proposed strategy combines hypergeometric extrapolation, continued fractions, and the Borel--Leroy transformation \cite{abhignan2023}. By exploiting information from both weak- and strong-coupling regimes, this method aims to infer the behavior of higher-order contributions without requiring their explicit computation. For scalar field theory, the resulting estimates appear less sensitive to the perturbative truncation order than those obtained with conventional resummation procedures, suggesting a potentially improved stability as the loop order increases. However, when applied to the Ising universality class, the corresponding predictions differ noticeably from those obtained using more established approaches (see the entries [CEF, $\epsilon^7$], [CE, $\epsilon^7$], and [CEBL, $\epsilon^7$] in Fig.~\ref{eta_nu_omega_ref}, as well as Table~\ref{tab:brief_review}). These discrepancies underline the need for further investigation of the systematic uncertainties associated with such alternative resummation schemes.

	\begin{table}[h!t]
		\centering
		\caption{Comparison of estimates of the critical exponents $\eta$, $\nu$, and $\omega$ for the three-dimensional Ising universality class obtained using various theoretical and numerical approaches. BCM denotes Borel resummation with conformal mapping. Within this framework, \emph{free} refers to the standard $\epsilon$-expansion resummation, \emph{bc} incorporates the exact two-dimensional value by resumming the series $(f(\epsilon)-f(2))/(2-\epsilon)$, where $f(\epsilon)$ denotes the corresponding critical exponent, and \emph{SC} denotes a self-consistent resummation procedure. The superscript $^{a}$ indicates an estimate derived from the perturbative expansion of $\nu^{-3}$. HT: high-temperature expansion; HA: hypergeometric approximants; HM: hypergeometric--Meijer approximants; CEF: continued exponential fraction transformation; CE: continued exponential transformation; CEBL: continued exponential transformation combined with the Borel--Leroy procedure; MC: Monte Carlo simulation; MCRG: Monte Carlo renormalization group; NPRG: nonperturbative renormalization group estimates obtained within the derivative expansion at orders $\partial^2$, $\partial^4$, and $\partial^6$; CB: conformal bootstrap.}
		\label{tab:brief_review}
	\begin{tabular}{l | l l l} 
			\hline 
			\hline 
			Methods & \qquad $\eta$ & \qquad $\nu$ & \qquad $\omega$ \\ 
			\hline 
			BCM, $\epsilon^{5}$ {\cite{leguillou1985}} & 0.037(3) & 0.6305(25) &  0.81(4)\\ 
			BCM, $\epsilon^{5}$ (free) {\cite{guida1998}} & 0.0360(50) & 0.6290(25) & 0.814(18) \\ 
			BCM, $\epsilon^{5}$ (bc) {\cite{guida1998}} & 0.0365(50) & 0.6305(25) & 0.814(18) \\ 
			BCM, $\epsilon^{5}$ {\cite{kompaniets2017}} & 0.0366(11) & 0.6290(20) & 0.818(8) \\ 
			BCM, $\epsilon^{6}$ {\cite{kompaniets2017}} & 0.0362(6) & 0.6292(5) & 0.820(7) \\ 
			SC, $\epsilon^{6}$ {\cite{kompaniets2020}} & 0.0355(3) & 0.6296(3)$^{a}$ & 0.827(13) \\
			High T (25th-order) {\cite{campostrini2002}} & 0.03639(15) & 0.63012(16) & 0.83(5) \\
			HA, $\epsilon^{7}$ {\cite{shalaby2021}} & 0.03595(52) & 0.62965(22) &  0.8243(15) \\ 
			HM, $\epsilon^{7}$ {\cite{shalaby2021b}} & 0.03653(65) & 0.62977(22) & 0.82311(50)\\ 
			CEF, $\epsilon^{7}$ {\cite{abhignan2023}} &  & 0.62983(21) &  \\ 
			CE, $\epsilon^{7}$ {\cite{abhignan2023}} &  & 0.62809(17) &  0.80(15) \\ 
			CEBL, $\epsilon^{7}$ {\cite{abhignan2023}} &  & 0.62962(72) & 0.83176(2) \\ 
			MC {\cite{hasenbusch2010}}  & 0.03627(10) & 0.63002(10) & 0.832(6) \\ 
			MC {\cite{ferrenberg2018}}  & 0.03610(45) & 0.629912(86) &  \\ 
			MC {\cite{hasenbusch2021}}  & 0.036284(40) & 0.62998(5) & 0.825(20) \\ 
			MC {\cite{gori2020}}  & 0.036284(16) &  &  \\
			MCRG {\cite{kaupuzs2023}}  & 0.03632(13) & 0.63017(31) &  \\ 
			NPRG, $\partial^{2}$ {\cite{balog2019, depolsi2020}} & 0.0387(55) & 0.6308(27) & 0.870(55) \\ 
			NPRG, $\partial^{4}$ {\cite{balog2019, depolsi2020}}& 0.0362(12) & 0.62989(25) & 0.832(14) \\ 
			NPRG, $\partial^{6}$ {\cite{balog2019, depolsi2020}} & 0.0361(11) & 0.63012(16) & \\ 
			CB {\cite{elshowk2014}} & 0.03631(3) & 0.62999(5) & 0.8303(18) \\ 
			CB {\cite{kos2016, simmons_duffin2017}}& 0.0362978(20) & 0.629971(4) &  \\ 
			CB {\cite{reehorst2022}}&  &  &  0.82951(61)\\ 
			CB {\cite{bootstrap_3d}}& 0.036297612(48) & 0.62997097(12)  &  \\ 
			\hline 
			\hline 
		\end{tabular}
	\end{table}

	Monte Carlo simulations provide one of the most important independent approaches for determining critical exponents and testing scaling properties in the Ising universality class. State-of-the-art studies have combined high-statistics simulations with sophisticated finite-size scaling analyses and improved lattice models designed to suppress leading scaling corrections \cite{hasenbusch2010, hasenbusch2021}. These developments have enabled estimates of critical exponents whose precision is comparable to that achieved by the most advanced field-theoretical methods, using lattice systems with linear sizes of up to approximately 200 (see Table~\ref{tab:brief_review} and the entries [MC$'10$] and [MC$'21$] in Fig.~\ref{eta_nu_omega_ref}). 

    Independent high-precision Monte Carlo estimates were also reported in Ref.~\cite{ferrenberg2018} (see Table~\ref{tab:brief_review} and the entry [MC$'18$] in Fig.~\ref{eta_nu_omega_ref}). In addition, a high-precision value of $\eta$ was obtained in Ref.~\cite{gori2020} via Monte Carlo simulations of the Blume-Capel model (see Table~\ref{tab:brief_review} and entry [MC$'20$] in Fig.~\ref{eta_nu_omega_ref} (a)). More recently, Monte Carlo results have been systematically compared with Monte Carlo renormalization-group (MCRG) calculations \cite{kaupuzs2023}. Such comparisons provide insight into the role of finite-size corrections and allow the consistency of different numerical approaches to be assessed (see Table~\ref{tab:brief_review} and the entry [MCRG] in Fig.~\ref{eta_nu_omega_ref}).

The nonperturbative (or functional) renormalization group (NPRG) provides an alternative framework for the determination of critical exponents, based on a systematic truncation of the effective average action within a derivative expansion. This expansion presents several advantages: it does not rely on a Taylor expansion in a small physical parameter of the model, and it is expected to exhibit a finite radius of convergence and a small expansion parameter \cite{balog2019, depolsi2020}. High-precision estimates of the critical exponents have been obtained up to sixth order in the derivative expansion, yielding results with small quoted uncertainties; these values are reproduced in Table~\ref{tab:brief_review} and shown by the entries [NPRG] in Fig.~\ref{eta_nu_omega_ref}.

The conformal bootstrap (CB) constitutes one of the most powerful nonperturbative approaches for the determination of critical exponents in conformal field theories, in particular for O$(n)$-symmetric models \cite{bootstrap_3d, elshowk2014, kos2016, simmons_duffin2017}. In the case of the three-dimensional Ising universality class, it provides the most precise determinations currently available. Extensions and refinements of the method continue to be actively developed, including higher-precision numerical implementations and studies of broader classes of CFTs (see, e.g., Refs.~\cite{elshowk2014, kos2016, simmons_duffin2017, reehorst2022}). The corresponding results are listed in Table~\ref{tab:brief_review} and shown as [CB$'16$], [CB$'22$], and [CB$'25$] in Fig.~\ref{eta_nu_omega_ref}. Beyond this benchmark case, the method has been successfully extended to isolate islands for other universality classes, including the three-dimensional {$\mathcal{N} = 1$} SUSY Ising model~\cite{atanasov2022}, as well as $O(2)$~\cite{chester2020} and $O(3)$~\cite{chester2021} models and Gross-Neveu-Yukawa theories~\cite{erramilli2023}.
		
The close agreement between a variety of theoretical and numerical approaches—including field-theoretic resummations, Monte Carlo simulations, the functional renormalization group, and the conformal bootstrap—leading to consistent estimates of critical exponents within small relative uncertainties, reflects the overall internal consistency of the theoretical framework.  The observed agreement between different methods thus constitutes a nontrivial cross-validation of these frameworks in statistical physics. The complementarity of these approaches provides a robust and multifaceted description not only of the Ising universality class, but also of the broader class of O$(n)$ models.

Recent efforts have also aimed at constructing rigorous or computer-assisted bounds for critical exponents, thereby partially bridging the gap between high-precision numerical estimates and mathematically controlled results (see, e.g., Ref.~\cite{kai2026}). 

    \section{A possible slow convergence of the perturbative series and an underestimate of the error bars}\label{Slow_conv}
	
A nonperturbative method such as CB (and possibly also NPRG) is able to produce reliable error bars for the determination of critical exponents. As for MC and HT, the presence of corrections to scaling must be taken into account with great care and reduced as much as possible to reach the same kind of reliability on the error bars (these are the so-called improved models for which the leading correction to scaling almost vanishes). 

A priori, perturbative field theoretic methods do not suffer from this problem because the corrections to scaling do not interfere with the determination of the leading critical exponents. However, the possible existence of nonanalyticities right at the fixed point can be a source of underestimation of error bars -- first pointed out by Nickel \cite{Nickel}-- and the two-dimensional Ising model is a test case of this phenomenon \cite{Caselle2001,pellisetto2002}.

Sokal \cite{Sokal1994,Sokal1995} has explained the reason for the presence of these nonanalyticities and the fact that they are unavoidable. We reproduce his argument below for the sake of completeness, formulating it in a slightly different and maybe more modern way.

On the one hand, the perturbative RG flow at criticality takes place in a one-dimensional space of coupling constant (for the Ising model) corresponding to the $\phi^4$ coupling.
On the other hand, the exact RG flow, that is, the Wilsonian flow, takes place in the infinite dimensional space of all possible $\mathbb{Z}_2$-invariant couplings.  On the critical (hyper-)surface, a particular one-dimensional RG trajectory emanates from the Gaussian fixed point and ends at the Wilson-Fisher fixed point. It has been called the large river \cite{Bagnuls} because the RG flow is slow along this trajectory: it looks like a large river in a valley. In the vicinity of the fixed point, all eigendirections lying within the critical surface are contracting, corresponding to irrelevant scaling operators. Consequently, a generic renormalization-group trajectory starting on the critical surface rapidly converges onto the least irrelevant direction, that is, the large river.  All RG trajectories except the large river are like small rivers in the mountains that flow very rapidly towards the large and slow river in the valley. The perturbative RG flow is nothing but the projection of the flow along the large river onto the $\phi^4$ axis which is a priori possible because they are both one-dimensional. However, barring miracles, the $\phi^4$ direction is not an eigendirection of the RG flow at the Wilson-Fisher fixed point and the approach to this fixed point, which is really along the direction of the large river at the Wilson-Fisher fixed point, has components along all $\phi^{2n}$ directions. Reciprocally, the flow along the $\phi^4$ direction has projections onto all the irrelevant eigendirections. Nonanalytic terms are therefore unavoidable in all perturbative RG functions in the vicinity of the Wilson-Fisher fixed point. Notice that nonanalyticities of the beta-function were also found by Parisi \cite{parisi1973, parisi1980} in his fixed dimension approach where the perturbative expansion is performed in the coupling constant directly in $d=3$.

A more refined analysis predicts that the beta-function $\beta(g)$ exhibits nonanalytic corrections at $g = g^\ast$, involving powers of $(g^\ast-g)^{\Delta_i/\Delta}$, where $\Delta_i$ and $\Delta$ denote subleading correction-to-scaling exponents associated with the irrelevant eigenvalues of the linearized RG flow around the Wilson--Fisher fixed point. Such nonanalytic contributions can, in principle, be incorporated within the $\epsilon$-expansion framework. However, as discussed in Refs.~\cite{Caselle2001,pellisetto2002}, they may significantly affect the convergence of resummed perturbative series.

This issue has been investigated in detail in two dimensions, where exact results are available and where the corresponding nonanalyticities are particularly pronounced. In three dimensions, these effects are expected to be considerably weaker, which is consistent with the fact that already five-loop calculations yield reasonably accurate estimates of the critical exponents. Nevertheless, the improvement observed when going from five to six loops is relatively modest, suggesting that the convergence may already be slowing down at these orders.

Sokal's geometric interpretation of the RG flow provides a natural framework for understanding this behavior. Since different observables probe different directions in the vicinity of the fixed point, they need not converge at the same rate. This expectation is indeed reflected in the distinct convergence patterns observed for the critical exponents $\eta$, $\nu$, and $\omega$. Assessing the impact of these effects and their implications for the reliability of high-order perturbative estimates constitutes one of the main motivations for the present study of the available seven- and eight-loop series.

	\section{RG functions and resummation procedure}\label{RG_and_resummation}
	
	\subsection{RG approach for the Ising model} \label{RG_Ising}
	Because of universality, the critical exponents of the Ising model can be computed from the $\phi^4$ field theory
		\begin{eqnarray}
			\label{ham_eff}
			{\cal H} = \int d^d x \left\{ \frac{1}{2}  (\nabla \phi)^2 +\frac{1}{2}  m^2 \phi^2  + \frac{g_0}{4!} \phi^4 \right\}, 
	\end{eqnarray}
where close to criticality $m^2$ is proportional to the deviation from the critical temperature: $m^2 \propto T - T_c + O\left((T - T_c)^2\right)$  and $g_0 > 0$ is the unrenormalized (``bare'') coupling.	
	
Different renormalization schemes are used in RG approaches to eliminate the ultraviolet divergences that are a characteristic of field theory \cite{amit2005}. The renormalization factors $Z_{i}$ [$i = g, \phi, \phi^{2}, m^{2}$, for the coupling $g$, for the field $\phi$, for the $\phi^2$-insertion, for the mass term $m^{2}$: $Z_{m^{2}} = Z_{\phi^2} Z_{\phi}^{-1}$] allow to transfer these divergences into ``bare'' quantities so that the renormalized physical quantities are finite. These quantities are defined at a certain scale $\mu$ and the RG flow  measures how the renormalized parameters and factors $Z_{i}$ change with  $\mu$. It is expressed by the RG functions:
	\begin{eqnarray}
		\label{RG_functions}
		\beta = \frac{\partial g}{ \partial \ln{\mu}}\Big{|}_{g_{0}}, \quad \gamma_{\phi} = \frac{\partial Z_{\phi}}{\partial \ln{\mu}}\Big{|}_{\phi_{0}}, \quad \gamma_{m^{2}} = \frac{\partial Z_{m^{2}}}{\partial \ln{\mu}}\Big{|}_{m_{0}} = \gamma_{\phi^{2}} - \gamma_{\phi}, 
	\end{eqnarray}
	where the derivatives are taken at fixed ``bare'' parameters. Within perturbation theory the RG functions are  expressed as  series in a putatively small parameter.

In particular, these RG functions (\ref{RG_functions}) enable the calculation of the universal characteristics of critical behavior. This behavior is quantified by critical exponents; here, we focus on $\eta$ and $\nu$. Through familiar scaling relations, knowing these two exponents allows the remaining ones to be uniquely determined. The  critical exponent $\eta$ governs the power-law decay of correlations between two points exactly at the critical temperature $T_c$:
	\begin{eqnarray}
		\langle \phi(x) \phi(0)\rangle \sim |x|^{2 - d - \eta}, 
	\end{eqnarray}	
	while the exponent $\nu$ describes the divergence of the correlation length $\xi$ as the temperature approaches $T_c$:
	\begin{eqnarray}
		\xi \sim |T - T_c|^{-\nu}.
	\end{eqnarray}	
	{These exponents characterize the leading asymptotic behavior in the vicinity of  the critical point. It is also of interest to examine exponents such as the correction-to-scaling exponent $\omega$. It accounts for deviations from power-law scaling in the vicinity of the transition, determining how fast the system converges to its asymptotic limit, e.g.
		\begin{eqnarray}
			\langle \phi(x) \phi(0)\rangle \sim |x|^{2 - d - \eta} \left(1 + A x^{-\omega}\right),
		\end{eqnarray}
		where $A$ is some constant.}
	
For the three-dimensional Ising universality class, the RG $\beta$- and $\gamma$-functions are currently known up to seventh order in the renormalized coupling, i.e., $\mathcal{O}(g^9)$  and
$\mathcal{O}(g^8)$ as reported in Refs.~\cite{schnetz2018,schnetz2023}. This unprecedented level of precision was achieved using  a novel computational framework that circumvents the principal bottlenecks encountered by conventional high-order perturbative calculations. By extending the perturbative series to an additional loop order, this approach provides valuable high-order information that can be combined with advanced resummation techniques, such as Borel resummation, to evaluate numerical values of the critical exponents.

	The critical exponents are determined by the values of the $\gamma-$functions, $\gamma_\phi$ and $\gamma_{m^2}$, at the nontrivial fixed point $g^{*}$ of the RG flow, where $\beta(g^{*}) = 0$:  $\eta = 2 \gamma_{\phi}(g^{*})$, and $\nu^{-1} = 2 + \gamma_{m^{2}}(g^{*})$. The available perturbative series are:\footnote{For detailed expressions and computational verification, see also the Maple package HyperlogProcedures, available on Oliver Schnetz's homepage: https://www.math.fau.de/person/oliver-schnetz/. For the sake of brevity, our findings are reported in a compact form; however, all numerical analyses were carried out to 20 decimal places utilizing high-precision results.}
	\begin{eqnarray}
		\label{betaSchnetz}
		\beta &\approx& - \epsilon g + 3 g^2 - 5.6666667 g^3 + 32.549683 g^4 - 271.60578 g^5  \nonumber \\ &+&  2848.5683 g^6 - 34776.131 g^7 +  474650.98 g^8 + {\cal{O}}\left(g^{9}\right),
	\end{eqnarray}
	
	\begin{eqnarray}
		\label{gamma}
		\gamma_{\phi}(g) &\approx& 0.083333333 g^2 - 0.062500000 g^3 + 0.33854167 g^4 - 1.9255850 g^5 \nonumber \\ &+&  14.383999 g^6 - 124.15855 g^7  +  1171.8757 g^8 +  {\cal{O}}\left(g^{9}\right),
	\end{eqnarray}
	and
	\begin{eqnarray}
		\label{gammam2}
		\gamma_{m^{2}}(g) &\approx& - g + 0.83333333 g^2 - 3.5000000 g^3 + 19.956305 g^4 - 150.75555 g^5 \nonumber \\ &+&  1354.6353 g^6  - 13759.782 g^7 + {\cal{O}}\left(g^{8}\right).
	\end{eqnarray}
	These functions were obtained in the minimal subtraction  RG scheme, and therefore the space dimension enters only the $\beta$-function and only in the term linear in $g$. 

To proceed, one has to find $g^*$ in the form of an expansion in a putatively  small parameter. In the $\epsilon$-expansion approach, we find:
	\begin{eqnarray}
		\label{gFPepsilon}
		g^* (\epsilon) &\approx& 0.33333333 \epsilon + 0.20987654 \epsilon^2 - 0.13755896 \epsilon^3  + 0.26865287  \epsilon^4  \nonumber \\ &-& 0.84368455 \epsilon^5 +  3.1543731 \epsilon^6 - 13.483134 \epsilon^7 + {\cal{O}}\left(\epsilon^{8}\right),
	\end{eqnarray}
 in agreement with \cite{kleinert2001}, \cite{kompaniets2017}  and  \cite{shalaby2021b}.

	Substituting Eq. (\ref{gFPepsilon}) in $\gamma_{\phi}(g)$ and $\gamma_{m^{2}}(g)$ \cite{schnetz2018}, we find for $\eta (\epsilon) = 2 \gamma_{\phi}(g^{*}(\epsilon))$:
	\begin{eqnarray}
		\label{etaepsilon}
		\eta (\epsilon) &\approx& 0.018518519 \epsilon^2 + 0.018689986 \epsilon^3 - 0.0083287703 \epsilon^4 + 0.025656451 \epsilon^5  \nonumber \\ &-& 0.081272643 \epsilon^6 + 0.31474941 \epsilon^7 - 1.3781866 \epsilon^8 + {\cal{O}}\left(\epsilon^{9}\right),
	\end{eqnarray}
	and for $\nu^{-1}(\epsilon) = 2 + \gamma_{m^{2}}(g^{*}(\epsilon))$:
	\begin{eqnarray}
		\label{nuepsilon}
		\nu^{-1}(\epsilon) &\approx& 2 - 0.33333333 \epsilon -0.11728395 \epsilon^2 + 0.12452741 \epsilon^3 - 0.30684954 \epsilon^4   \nonumber \\ &+& 0.95124031 \epsilon^5  - 3.5725821 \epsilon^6 + 15.286864 \epsilon^7 + {\cal{O}}\left(\epsilon^{8}\right).
	\end{eqnarray}
	The correction to scaling exponent is given by $\omega(\epsilon) = \partial_g \beta (g, \epsilon)|_{g = g^*(\epsilon)}$, that is:
	\begin{eqnarray}
		\label{omegaepsilon}
		\omega(\epsilon) &\approx& \epsilon - 0.62962963 \epsilon^2 + 1.6182207 \epsilon^3 - 5.2351360  \epsilon^4 + 20.749836 \epsilon^5  \nonumber \\ &-&  93.111286 \epsilon^6 + 458.74239 \epsilon^7  + {\cal{O}}\left(\epsilon^8\right).
	\end{eqnarray}	
	
	As anticipated, these series exhibit no obvious signs of convergence \cite{eckmann1975, zinnjustin2002, kleinert2001, leguillou1985}. Resummations are therefore required which is the subject of the next sections.

	\subsection{Borel transformation with conformal mapping} \label{BCM_section}
	
	An asymptotic series is an expansion of a function in terms of a small parameter, say $z$, such that truncating the series at a finite order provides a useful approximation, even though the full infinite series does not necessarily converge. In many cases, the coefficients of an asymptotic expansion grow factorially, $a_{k} \sim (-1)^{k} k!$, which ensures divergence when the series is summed to all orders. A series $\sum_{k} a_{k} z^{k}$ is said to be asymptotic to a function $f(z)$ if, for fixed $L$ and sufficiently small $z$,
	\begin{equation}
		\left| f(z) - \sum_{k=0}^{L} a_{k} z^{k} \right| \le a_{L+1} z^{L+1}.
	\end{equation}
	This condition ensures that the truncated expansion approximates $f(z)$ with increasing accuracy as $z \to 0$, although it does not imply convergence of the full series. It is well-known \cite{hardy1948}, that asymptotic expansions typically approach the exact value up to an optimal number of terms $N_{0}$, after which additional terms worsen the approximation and drive the partial sums away from the correct result. A classical example is the Stirling series for the factorial. This is why resummations are required \cite{amit2005, kleinert2001}.
	
Despite the systematic nature of these resummation procedures, extracting accurate results from perturbative expansions in $d=3$ remains a nontrivial task. The convergence properties of the resummed series are strongly influenced by the singularity structure of the underlying functions, and the choice of resummation scheme can have a significant impact on the accuracy and reliability of the resulting estimates of critical exponents \cite{kleinert2001,pellisetto2002}.

One of the most widely used approaches is the Pad{\'e} approximation, in which a truncated power series is replaced by a rational function whose numerator and denominator are chosen such that its Taylor expansion reproduces the known coefficients of the original series \cite{baker2009}. By providing an analytic continuation beyond the radius of convergence of the perturbative expansion, Pad{\'e} approximants often improve the behavior of truncated series and can capture aspects of the nonperturbative structure of the underlying theory.

A more powerful method for dealing with factorially divergent perturbative expansions is Pad{\'e}--Borel resummation \cite{baker1976,baker1978}. In this approach, the original series is first converted into its Borel transform, which generally possesses more favorable convergence properties. A Pad{\'e} approximant is then constructed for the Borel-transformed series, and the original function is recovered through an inverse Borel transform. This procedure can substantially improve the stability and accuracy of perturbative predictions. Nevertheless, Pad{\'e}--Borel resummation has important limitations. In particular, when singularities of the Borel transform lie on the integration contour of the inverse Borel transform, the latter becomes ambiguous, potentially compromising the effectiveness of the resummation.

Although the Borel transform reorganizes a factorially divergent perturbative series into a form that is more amenable to summation, we employ a resummation procedure supplemented by a conformal mapping, which provides the required analytic continuation while significantly accelerating the convergence of the truncated resummed series by mapping the singularities away from the expansion domain \cite{leguillou1980,leguillou1985,frustrated_b}. To obtain reliable estimates of the critical exponents from the asymptotic expansions (\ref{etaepsilon})--(\ref{omegaepsilon}), we proceed as follows.

Consider the truncated perturbative expansion
\begin{eqnarray}
\label{series}
f(z) = \sum_{k = 0}^{L} a_{k} z^{k},
\end{eqnarray}
where $L$ denotes the highest available order of the expansion ($L=7$ or $8$ in the present work). The coefficients $a_k$ exhibit factorial growth at large orders and obey the asymptotic behavior
\begin{eqnarray}
\label{constants_ab_e}
a_{k} = k! (- a)^k k^{b^{\prime}} \left(1 + {\cal{O}}\left(\frac{1}{k}\right)\right)  \quad \text{as} \quad  k \to \infty,
\end{eqnarray}
where the constants $a$ and $b^{\prime}$ characterize the leading large-order divergence of the series. Their values can be inferred from the asymptotic behavior of the perturbative coefficients. According to Refs.~\cite{brezin1977,guida1998},
	\begin{eqnarray}
		\label{constants_ab_g}
		a = \frac{1}{3}, \quad b^{\prime} = 
		\begin{cases}
			3 + 1/2 \, \, \text{for} \, \, \eta,\\
			4 + 1/2 \, \, \text{for} \, \,  \nu^{-1}, \\
			5 + 1/2 \, \, \text{for} \, \,  \omega.
		\end{cases}
	\end{eqnarray}

The Borel-Leroy image associated with the function $f(z)$ is defined by:
	\begin{eqnarray}
		\label{borel_gamma}
		B(z) = \sum_{k = 0}^{L} \frac{a_{k}}{\Gamma{\left(k + b + 1\right)}} z^{k}, 
	\end{eqnarray}
	where  $\Gamma(k + b + 1) = (k + b)!$ is the Euler Gamma-function used to suppress the factorial growth of the series coefficients and $b$ is a fit parameter. 
		
	 As usual, the radius of convergence is given by the singularity closest to the origin in the complex plane of $z$, that is, for $|z| < 1/a$ which is the first singularity of the Borel-Leroy transform.  To reconstruct the function $f(z)$ from (the resummation of) $B(z)$ it will be mandatory to analytically continue $B(z)$ such that the convergence then holds for the entire positive real axis and we now explain how this is performed.
	
	Using the integral representation $\Gamma(z) = \int_{0}^{\infty} d t e^{- t} t^{z - 1}$, we first rewrite Eq.~(\ref{series}) as
	\begin{eqnarray}
		f(z) = \sum_{k = 0}^{L} \frac{a_{k}}{\Gamma{\left(k + b + 1\right)}} z^{k} \int_{0}^{\infty} d t e^{- t} t^{k + b}
	\end{eqnarray}
	and by changing the order of summation and integration the  function $f(z)$ can be defined as 
	\begin{eqnarray}
		\label{Borel_trans}
		f(z) = \int_{0}^{\infty} d t e^{-t} t^{b} B(z t). 
	\end{eqnarray}
	
	\begin{figure}[h!] 
		\includegraphics[width=0.75\textwidth]{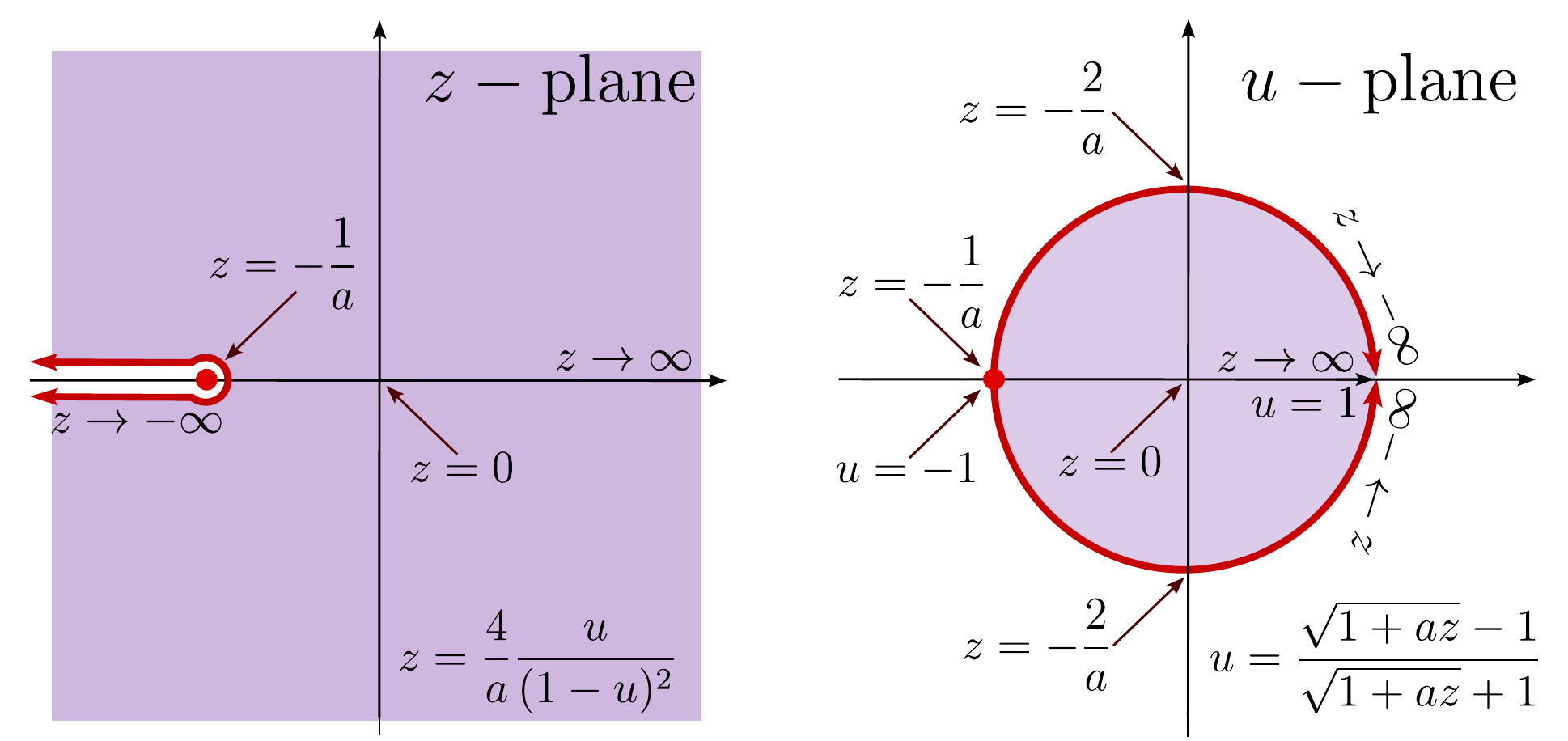}
		\centering
		\caption{Conformal mapping $u(z)$ of the $z$-plane, cut along the negative real axis from $-1/a$ to $-\infty$ (left), onto the interior of the unit disk in the $u$-plane (right). The analytic domain (shaded light purple) is mapped to the interior $|u| < 1$. Red arrows illustrate how the upper and lower edges of the branch cut are mapped onto the boundary $|u| = 1$, meeting at the point $u = 1$ which corresponds to $z \to \infty$.}
		\label{cm}
	\end{figure}
	
	To evaluate the integral in Eq.~(\ref{Borel_trans}), one must determine the analytic continuation of $B(z)$. This can be achieved in several ways, such as  employing Pad{\'e} approximants \cite{baker1976, baker1978, baker2009} or conformal mapping \cite{leguillou1985, guida1998, kompaniets2017, frustrated_c}.  In this work, we focus on the latter technique. We assume that all singularities of $B(z)$ lie on the negative real axis, and thus, the Borel-Leroy sum is analytic in the entire complex plane except for the cut $\left.\left(-\infty, -1/ a\right.\right]$ (see Fig.~\ref{cm}). Next, we perform a change of variables, namely \textit{the conformal transformation}:
	\begin{eqnarray}
		\label{conformal_trans}
		u(z) = \frac{\sqrt{1 + a z} - 1}{\sqrt{1 + a z} + 1} \quad \text{and} \quad z(u) = \frac{4}{a} \frac{u}{(1 - u)^2}.
	\end{eqnarray}
	This mapping transforms the cut $z$-plane onto the interior of the unit disk in the $u$-plane such that the singularities of $B(z(u))$ are relocated to the boundary $|u| = 1$ (see Fig.~\ref{cm}). Here, the mapping parameter $a$ is fixed by the large-order behavior, as given in Eq.~\eqref{constants_ab_g}. Thus, the leading singularity $z = - 1/a$ is mapped onto the boundary point $ u = -1$. Consequently, the corresponding expansion of $B(z(u))$ in powers of $u$ converges throughout the entire unit disk  $|u| < 1$, effectively covering the full integration domain of the Borel-Leroy sum:
	\begin{eqnarray}
		\label{Borel_cm}
		B(z(u)) =  \sum_{k = 0}^{L} d_{k}(a, b) [u(z)]^{k},
	\end{eqnarray}
	where $d_{k}(a, b)$ are the coefficients of the re-expansion of the Borel–Leroy image of $f(z)$. Once the analytic continuation of $B(z)$ is obtained via Eq.~(\ref{Borel_cm}), the resummed expression for the series $f(z)$ in Eq.~(\ref{series}) can be determined by the inverse Borel transform:
	\begin{eqnarray}
		\label{series_res}
		f_{R}(z) = \sum_{k = 0}^{L} d_{k}(a, b) \int_{0}^\infty d t \, e^{-t} \, t^{ b} \, [u(z t)]^{k}.
	\end{eqnarray}
	To incorporate the strong-coupling behavior of the series, $f(z \to \infty) \sim z^{\alpha/2}$ \cite{frustrated_b}, one often employs a generalized version of Eq.~(\ref{series_res}) by introducing a factor $1 = [1 - u(z t)]^{\alpha}/[1 - u(z t)]^{\alpha}$ into the integrand:
	\begin{eqnarray}
		\label{series_res__gen}
		f_{R}(z) = \sum_{k = 0}^{L} d_{k}(\alpha, a, b) \int_{0}^\infty d t \, e^{-t} \, t^{ b} \, \frac{[u(z t)]^{k}}{[1 - u(z t)]^{\alpha}}, 
	\end{eqnarray}
	where the coefficients $d_{k}(\alpha, a, b)$ are computed to match the original power series  Eq.~(\ref{series}) upon re-expansion in $z$. This yields our final resummation expression for the resummation of asymptotic series Eqs.~(\ref{etaepsilon}) -- (\ref{omegaepsilon}). 
	
	In principle, the exact function $f_{R}(z)$,  Eq.~(\ref{series_res__gen}), must be independent of the choice of auxiliary parameters $a$, $b$ and $\alpha$.  However, in any approximation based on a truncated series, a dependence persists. This parametric sensitivity can be exploited to optimize the numerical results and, as discussed in the next section, provides a systematic framework for error estimation via the criteria of minimal sensitivity and fastest apparent convergence.

	To conclude this section, let us note that a closely related approach based on the Borel transform with conformal mapping (BCM) exists \cite{kompaniets2017}, where additional resummation parameter, $\lambda$, is introduced in a slightly different way, but plays the same role as $\alpha$ in the method presented above \cite{kleinert2001}.  Both approaches, while structurally distinct, are mathematically equivalent as they share the same underlying conformal mapping and Borel–Leroy kernel. We apply a prefactor $(1 - u (x))^{-\alpha}$ [see Eq.~(\ref{series_res__gen})] to the transformed series, whereas the authors of Ref.~\cite{kompaniets2017} absorb this factor into the integrand through the power $(a x /u(x))^\lambda$ with the corresponding redefinition of the coefficients\footnote{From Eq. (\ref{conformal_trans})  one obtains $(1 - u)^{-\alpha} = 2^{-\alpha} (1+ \sqrt{1 + a x})^{\alpha}$. Simultaneously, the same conformal transformation implies $\frac{1}{u} = \frac{(1 + \sqrt{1 + a x})^2}{a x}$, which leads to $\left(\frac{a x}{u}\right)^{\lambda} = (1 + \sqrt{1 + a x})^{2 \lambda }$. Under the condition $\alpha = 2 \lambda$ the two expressions become proportional. The remaining constant factor $2^{-2 \lambda}$ is absorbed into the redefined coefficients of the series expansion in \cite{kompaniets2017}.}. Consequently, both methods yield identical resummed values, formally linked by the relation $\alpha = 2 \lambda$. In the next Section, we will present the principles that are used to resum divergent $\epsilon$-expansions, as well as provide two improved methods of precise numeric evaluation of the critical exponents.

	\section{Principles of convergence} \label{convergence}

	\subsection{Principle of minimal sensitivity (PMS) and Principle of fastest apparent convergence (PFAC)} \label{two_param_sensivity}

	A physical observable $\mathcal{O}$ should, by construction, be independent of the auxiliary parameters $a$, $b$, and $\alpha$ introduced in the previous section. In practical perturbative calculations truncated at a finite loop order $L$, however, all observables acquire an artificial dependence on these parameters. While the parameter $a$ can be fixed unambiguously from the large-order asymptotic behavior of the series (see Section~\ref{RG_and_resummation}), the residual dependence on $b$ and $\alpha$ persists at any finite order and represents a systematic ambiguity of the resummation procedure.
	
	To control this spurious dependence and to extract reliable estimates for physical quantities, we adopt the optimization strategy proposed in Ref.~\cite{frustrated_b}, which combines the principle of minimal sensitivity (PMS) with the principle of fastest apparent convergence (PFAC). Within PMS, the optimal values of the parameters $b$ and $\alpha$ at a given loop order $L$ are determined by requiring the observable $\mathcal{O}^{(L)}(b,\alpha)$ to be stationary with respect to variations of these parameters,
	\begin{eqnarray}
		\label{PMS}
		\left.\frac{\partial \mathcal{O}^{(L)}(b,\alpha)}{\partial b}\right|_{b^{(L)}_{opt},\,\alpha^{(L)}_{opt}}
		=
		\left.\frac{\partial \mathcal{O}^{(L)}(b,\alpha)}{\partial \alpha}\right|_{b^{(L)}_{opt},\,\alpha^{(L)}_{opt}}
		=0 .
	\end{eqnarray}
	The solutions $\left(b^{(L)}_{opt},\alpha^{(L)}_{opt}\right)$, which generally depend on the loop order $L$, correspond to stationary points of $\mathcal{O}^{(L)}$ and define the optimized estimate $\mathcal{O}^{(L)}_{opt}=\mathcal{O}^{(L)}\left(b^{(L)}_{opt}, \alpha^{(L)}_{opt}\right)$.
	
	In addition, PFAC provides a complementary criterion by comparing results for successive loop orders. Specifically, the optimal parameters are chosen such that the difference between the estimates at orders $L$ and $L+1$ is minimized:
	\begin{eqnarray}
		\label{PFAC}
		\mathcal{O}^{(L+1)}\!\left(b^{(L+1)},\alpha^{(L+1)}\right)
		-
		\mathcal{O}^{(L)}\!\left(b^{(L)},\alpha^{(L)}\right)
		=
		\delta^{(L+1,L)},
	\end{eqnarray}
	where $\delta^{(L+1,L)}$ is required to be as small as possible. This condition ensures that the optimized sequence of approximants exhibits the fastest apparent convergence with increasing loop order.
	
	To ensure the stability of the resummation, $b$ and $\alpha$ are chosen to yield the highest convergence and the least dependence on the parameter choice. We map the ($b, \alpha$) parameter space to locate a stability region where the optimal values,  ($b_{opt}, \alpha_{opt}$) are identified via the simultaneous application of PMS (\ref{PMS}) and PFAC (\ref{PFAC}) criteria. To this end, we implement an algorithm that minimizes the following quantity to find a sensible region ($b_{opt}, \alpha_{opt}$) \cite{serone2018}:
	\begin{eqnarray}
		\label{func_serone}
		\Delta \mathcal{O} = \left( \frac{\partial 	\mathcal{O}^{(L)}}{\partial b}\right)^{2} \left|_{b^{(L)}_{opt},  \alpha^{(L)}_{opt}}\right. + \left( \frac{\partial 	\mathcal{O}^{(L)}}{\partial \alpha}\right)^{2} \left|_{b^{(L)}_{opt},  \alpha^{(L)}_{opt}}\right. + \left(\left| \delta^{(L, L-1)}\right|  - \left| \delta^{(L - 1, L-2)}\right|  \right)^{2},
	\end{eqnarray}
	where the first two terms represent the sensitivity of the observable to the resummation parameters $b$ and $\alpha$, while the last term reflects the convergence rate of the series. In other words, by minimizing the functional (\ref{func_serone}), the algorithm seeks a point where the gradient of $\mathcal{O}(b, \alpha)$ vanishes (a stationary point). Simultaneously, the condition $\left| \delta^{(L, L-1)}\right|  \approx \left| \delta^{(L - 1, L-2)} \right|$ ensures that the resummation has entered a stable linear or better convergence mode, characterized by a consistent approach to a limit at a predictable rate, rather than chaotic oscillations.
	
	In accordance with \cite{serone2018}, the error of the resummed result is estimated using a combined approach is defined as:
	\begin{eqnarray}
		\label{error_serone}
		{\rm Err} \,  \mathcal{O}^{(L)} = \left(\frac{1}{\Delta b} + \frac{1}{\Delta \alpha}\right) \frac{1}{N} \sum_{n = 1}^{N} \left| \mathcal{O}_{n}^{(L)} - \mathcal{O}^{(L)} \right|  +  \left| \mathcal{O}^{(L)} - \mathcal{O}^{(L-1)} \right|.
	\end{eqnarray}		
	The first term quantifies the sensitivity of the observable to the choice of the resummation parameters ($b, \alpha$). It accounts for the uncertainty stemming from the imprecise knowledge of these parameters by averaging the deviation of $\mathcal{O}_{n}^{(L)}$ within a stability window $[b_{opt} \pm \Delta b, \alpha_{opt} \pm  \Delta \alpha]$. The weighting factor $\left(1 / \Delta b + 1 / \Delta \alpha\right)$ effectively penalizes regions where the stability plateau is too narrow. The second term in (\ref{error_serone}), $\left| \mathcal{O}^{(L)} - \mathcal{O}^{(L-1)} \right|$, represents the internal convergence of the series, providing an estimate of the truncation error for asymptotic expansions. 
	
\begin{figure}[h!] 
			\subfloat[]{\includegraphics[width=0.35\textwidth]{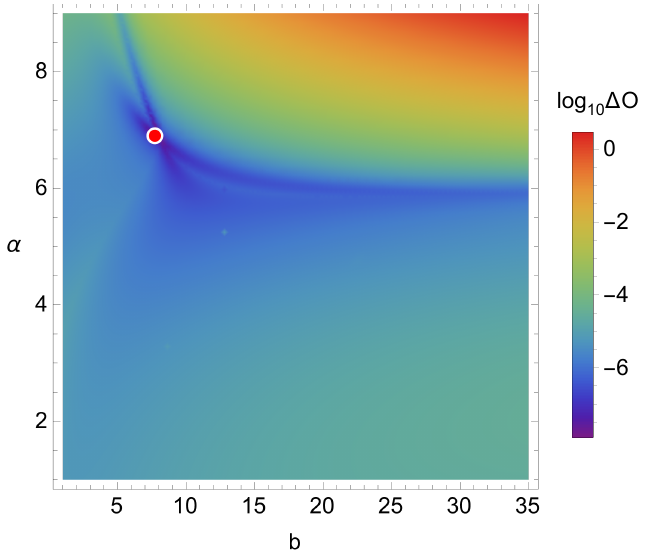}}\hfill
			\subfloat[]{\includegraphics[width=0.32\textwidth]{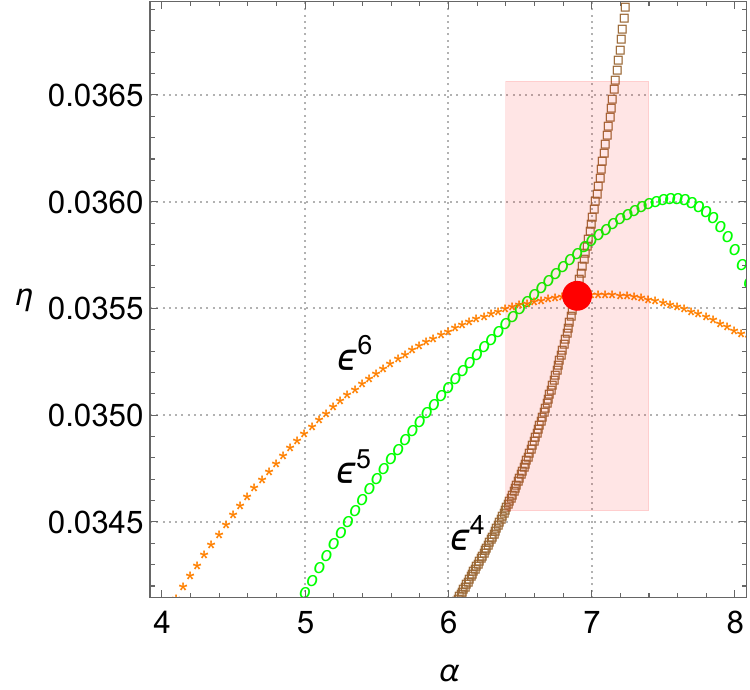}}\hfill
			\subfloat[]{\includegraphics[width=0.32\textwidth]{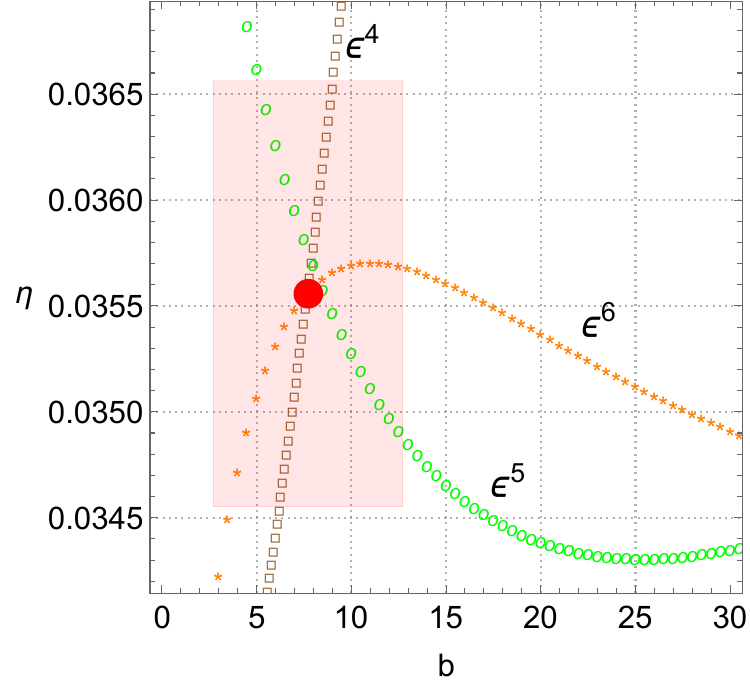}}\
		\centering
		\caption{(a): Density plot of $\log_{10} \Delta \mathcal{O}$ [Eq.~(\ref{func_serone})] as a function of the resummation parameters $b$ and $\alpha$ for the sixth-order expansion for critical exponent $\eta$. The color scale indicates the degree of stability, with dark blue regions representing the highest stability (minimal $ \Delta \mathcal{O}$). The red dot in the figures marks the identified optimal point  ($b_{opt}, \alpha_{opt}$) = (7.72, 6.9), where the PMS and PFAC criteria are simultaneously satisfied. (b) and (c): sensitivity of the critical exponent $\eta$ to the resummation parameters $\alpha$ (middle panel) and $b$ (right panel) around the point of the optimal values $(b_{opt}, \alpha_{opt})$ based on the sixth-order perturbative calculation. The shaded red rectangle illustrates the stability window $[\alpha_{opt} \pm \Delta \alpha]$ (central panel) and $[b_{opt} \pm \Delta b]$ (right panel) along the horizontal axis and the total error estimate ${\rm Err} \, \eta_{\epsilon^6}$ along the vertical axis. }
		\label{serone_two_param}
	\end{figure}

\begin{figure}[h!] 
			\subfloat[]{\includegraphics[width=0.49\textwidth]{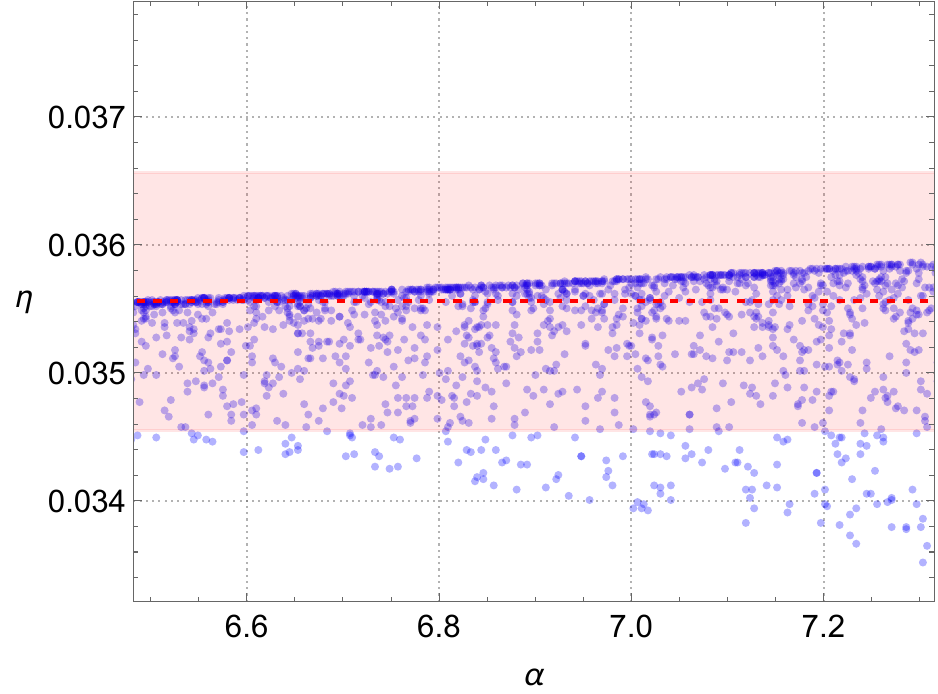}}\hfill
			\subfloat[]{\includegraphics[width=0.49\textwidth]{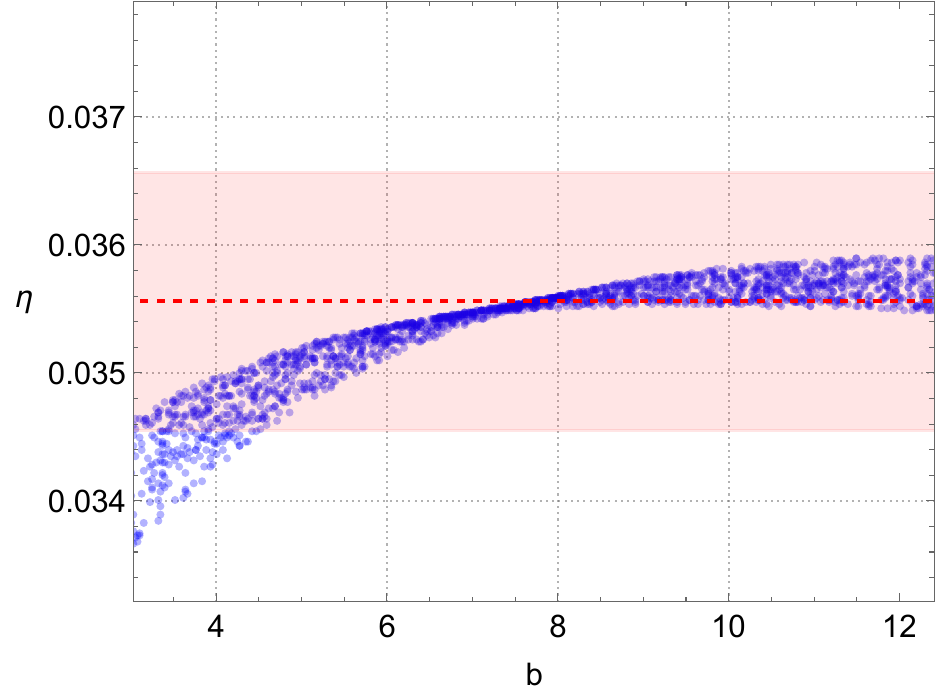}}
		\centering
		\caption{Estimate of the error in the resummation of the critical exponent $\eta$ based on the sixth-order perturbative expansion, Eq.~\ref{etaepsilon}, within the stability window $[b_{opt} \pm \Delta b, \alpha_{opt} \pm  \Delta \alpha]$. The scatter plots (blue dots) represent $N = 2000$ random samples from the parameter space. Panel (a) shows the sensitivity to  $\alpha$, while panel (b) illustrates the dependence on  $b$. The central dashed red line shows the final estimate $\eta_{\epsilon^6} = 0.0356(10)$, and the shaded red area represents the total uncertainty ${\rm Err} \, \eta_{\epsilon^6}$ calculated according to Eq. (\ref{error_serone}).}
		\label{errors_ala_serone}
	\end{figure}

	We begin by inspecting this approach for the critical exponent $\eta$ in the six-loop approximation [see Eq.~(\ref{etaepsilon})] and compare it with the numerical strategy proposed in \cite{kompaniets2017}. The choice of the stability window sizes, $\Delta b = 5.0$ and $\Delta \alpha = 0.5$, is determined by the landscape of the stability functional (\ref{func_serone}) shown in Fig.~\ref{serone_two_param} [panel (a)] \footnote{While variations in series coefficients across different exponents and loop orders suggest the need for adaptive windows, exploring alternative intervals yields inconclusive results without a clear, overarching benefit for all estimates (see Appendix~\ref{app__two_parameter} for further analysis).}. These intervals are selected to cover the plateau region (the dark blue area), where $\Delta \mathcal{O}$ remains consistently small. Defining a window that is too narrow would fail to capture the parameter sensitivity, while an overly large window would include regions where the resummation starts to diverge, visible as sharp color gradients at the edges of the stability valley in Fig.~\ref{serone_two_param} [panel (a)]. The robustness of this choice is further illustrated in panels (b) and (c) of Fig.~\ref{serone_two_param}, where we fix one parameter at its optimal value and vary the other. In these panels, the optimal result lies within a characteristic plateau, and the shaded red rectangles, defined by our chosen stability windows, successfully encompass the regions where the loop orders exhibit minimal mutual sensitivity. Within these bounds, the resummed value of $\eta$ exhibits minimal variation, as confirmed by the dense clustering of points in the scatter plots (see Fig.~\ref{errors_ala_serone}). There, we present the distribution of Borel integral values around the optimal point ($b_{opt}, \alpha_{opt}$) with respect to resummation parameters $\alpha$ and $b$. The high density of samples within the shaded uncertainty region in Fig.~\ref{errors_ala_serone} demonstrates that the combined uncertainty (\ref{error_serone}) provides a conservative and reliable estimate of the truncation and parameter-induced errors.
	
	 The exponents $\nu$ and $\omega$ are calculated by a similar procedure, now based on the expressions (\ref{nuepsilon}) and (\ref{omegaepsilon}), respectively. The final results for the critical exponents $\eta$, $\nu$, and $\omega$ obtained in the sixth-loop approximation using the stability criteria (\ref{func_serone}) and the error estimation (\ref{error_serone}) are summarized in Table~\ref{tab:comparison}. As can be seen from the table, the current approach based on the stability criteria (\ref{func_serone}) yields more conservative estimates with slightly larger uncertainties compared to the numerical results reported in \cite{kompaniets2017}. It is important to note that both studies utilize the same perturbative series. However, the strategy in \cite{kompaniets2017} achieves higher precision by incorporating an additional homographic transformation of the Borel plane, which significantly accelerates the convergence. Our current estimates remain in a good agreement with their values, confirming the reliability of the chosen stability functional, albeit providing a relatively rougher approximation at this stage, see Table~\ref{tab:comparison}. In Appendix~\ref{app__two_parameter} we provide also numerical results for the evaluation  for the critical exponents $\eta$, $\nu$ and $\omega$ up to the seven(eight)-loop order.
	
	While the current results are robust, our tests of the two-parameter resummation reveal that it merely shifts the central values while leaving the error margins effectively unchanged (see Table~\ref{tab:appendix_results} in Appendix~\ref{app__two_parameter}). Given that this approach fails to reduce the overall uncertainty, we proceed to a three-parameter resummation scheme. In the following section, we extend the present formalism by introducing an additional parameter through a homographic transformation. We will demonstrate how this extension improves the convergence properties of the investigated asymptotic series, allowing us to refine the precision of the critical exponents and bring our estimates closer to the current leading estimates.
	
	\begin{table}[h!]
		\centering
		\caption{Comparison of the six-loop critical exponents for the 3D Ising model. The confidence intervals are determined from Eq.~(\ref{error_serone}). }
		\label{tab:comparison}
		\begin{tabular}{lccc}
			\hline\hline
			Method & $\eta$ & $\nu$ & $\omega$ \\
			\hline
			This work, Eq.~(\ref{error_serone}) & 0.0356(10) & 0.6266(43) & 0.817(14) \\
			Kompaniets \& Panzer \cite{kompaniets2017} & 0.0362(6) & 0.6292(5) & 0.820(7) \\
			\hline\hline
		\end{tabular}
	\end{table}

	\subsection{Homographic transformation} \label{homographic_section}
	
	Here we introduce an additional resummation parameter $q$ by employing a homographic transformation (also known as a M{\" o}bius transformation) of the original expansion parameter $\epsilon$ \cite{leguillou1985, kompaniets2017}. This technique is essential for improving the convergence and reliability of perturbative expansions since it helps to change the position of singularities in the Borel plane (e.g. the critical exponents for O$(n)$-symmetric $\phi^4$ theory \cite{leguillou1985, kompaniets2017}, critical exponents of Dirac fermions \cite{ihrig2018}, critical exponents in perovskites with a structural phase transition \cite{aharony2022}). Homographic transformation relates the original expansion parameter $\epsilon$ to a shifted parameter $\epsilon^{\prime}$ as:
	\begin{equation}
		\label{homographic_trans} 
		\epsilon = \frac{\epsilon^{\prime}}{1 + q \epsilon^{\prime}}, \quad \text{and} \quad \epsilon^\prime = \frac{\epsilon}{1 - q \epsilon}, 
	\end{equation}
	where $q$ is a tuning parameter. In our implementation, this is equivalent to replacing the argument in the resummed expression with an effective parameter $z_{q} = \frac{\epsilon}{1 - q  \epsilon}$.
	
	To incorporate the required strong-coupling asymptotics, we consider the function $(1 - u)^{\alpha} B(z(u))$ and expand it into a Taylor series in $u$, obtaining the coefficients $d_{k}(\alpha, a, b)$. The homographic transformation is then applied by substituting the effective parameter $z_q$ into the resulting integral. The final resummed expression, which we use instead of Eq.~(\ref{series_res__gen}) for numerical integration, is then given by:
	\begin{equation} 
		\label{series_res_final_correct} 
		f_{R}(\epsilon) = \sum_{k = 0}^{L} d_{k}(\alpha, a, b) \int_{0}^\infty dt \, e^{-t} t^{b} \frac{[u(z_q t)]^{k}}{[1 - u(z_q t)]^{\alpha}}.
	\end{equation}
	
	Now, the resummation procedure relies on several adjustable parameters to improve the convergence and stability of the perturbative series:
	\begin{enumerate}
		\item the parameter $a$ is typically fixed by the large-order behavior of the series [see Eq.~(\ref{constants_ab_g})];
		\item the Borel-Leroy parameter $b$ is used to suppress the factorially growing coefficients, often by introducing a slightly stronger factorial dependence in the Borel integral [see Eq.~(\ref{borel_gamma})];
		\item the parameter $\alpha$ is chosen to match the expected strong-coupling power-law asymptotics of the function [see Eq.~(\ref{series_res__gen})];
		\item finally, the homographic parameter $q$ optimizes the convergence in the vicinity of $\epsilon \sim 1$; the transformation (\ref{homographic_trans}) effectively shifts the singularities of the Borel function away from the physical region of the integration path, further accelerating the convergence of the resummed series [see Eq.~(\ref{series_res_final_correct})].
	\end{enumerate}
	
	\begin{figure}
			\subfloat[]{\includegraphics[width=0.49\textwidth]{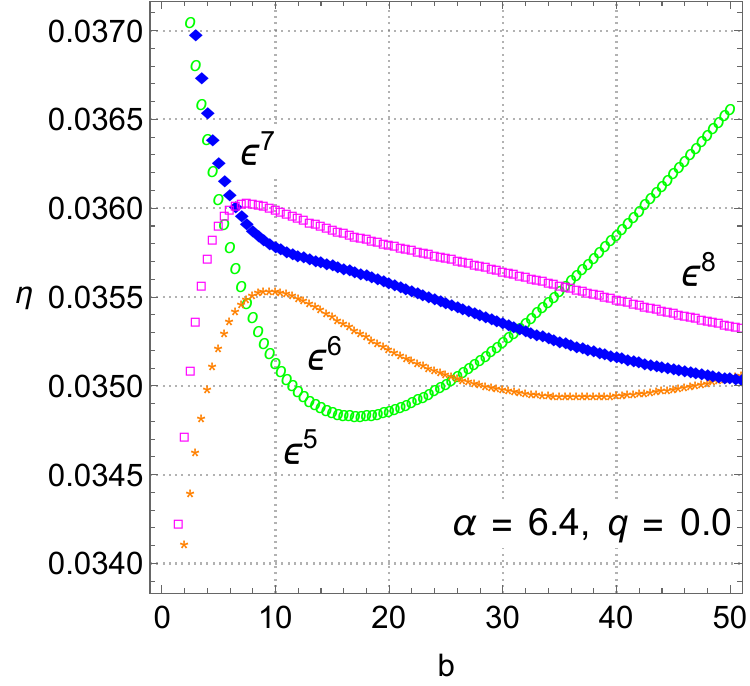}}\hfill
			\subfloat[]{\includegraphics[width=0.49\textwidth]{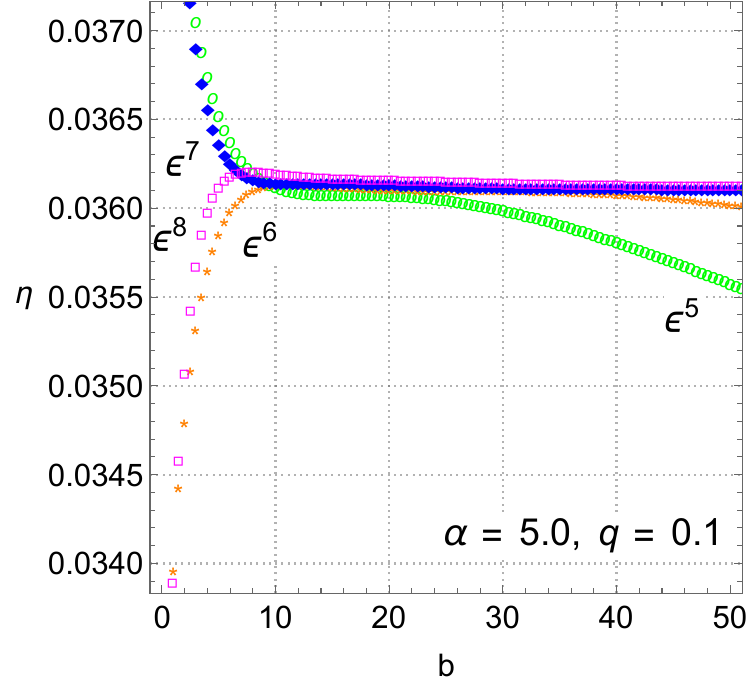}}\hfill
			\subfloat[]{\includegraphics[width=0.49\textwidth]{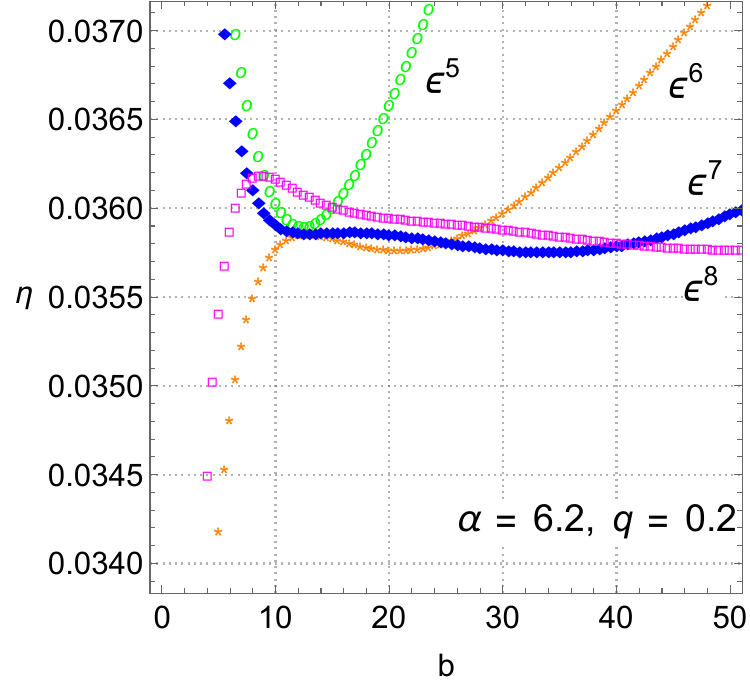}}
			\subfloat[]{\includegraphics[width=0.49\textwidth]{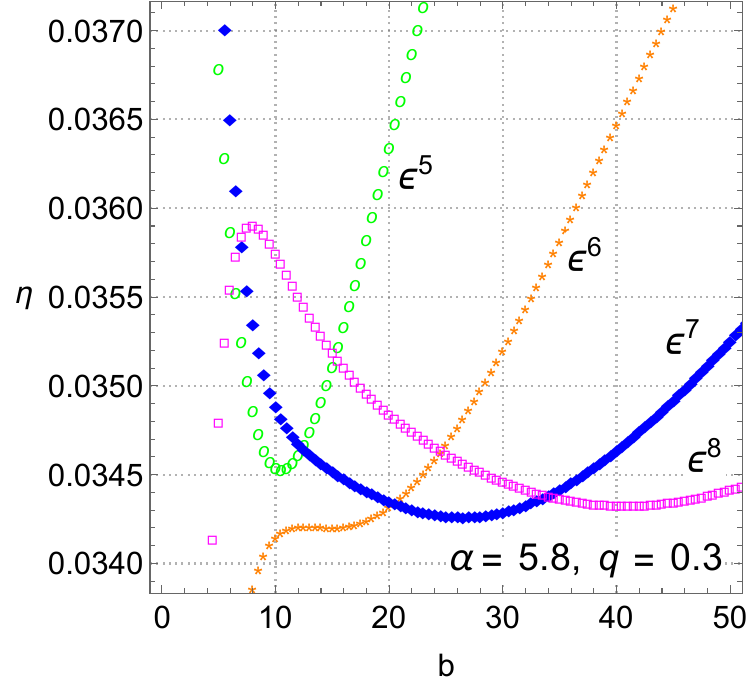}}
		\centering
		\caption{Dependence of $\eta$ on $b$ at different orders
			of the $\epsilon$-expansion at fixed
			$\alpha = 6.4$, $q = 0.0$ [(a)],
			$\alpha = 5.0$, $q = 0.1$ [(b)],
			$\alpha = 6.2$, $q = 0.2$ [(c)], and
			$\alpha = 5.8$, $q = 0.3$ [(d)], respectively.}
				\label{HT_eta}
	\end{figure}

	The introduction of the third resummation parameter, $q$, which controls the homographic mapping, significantly enhances the flexibility of the method but also exposes the limitations of the standard stability criteria. Specifically, the functional $\Delta \mathcal{O}$ defined in Eq.~(\ref{func_serone}) becomes increasingly problematic in a multiparameter setting. While one could formally extend Eq.~(\ref{func_serone}) by adding a third derivative term $(\partial \mathcal{O}/\partial q)^2$, such a straightforward modification implicitly assumes a uniform metric in the parameter space. In our case, the resummation parameters exhibit strongly disparate sensitivity scales: the observable $\mathcal{O}$ is globally robust with respect to $b$ but extremely sensitive to even minor variations in $q$ (see Fig.~\ref{HT_eta}).
	
	Assigning the same weight to all parameters is misleading because their sensitivities differ. The high sensitivity of $q$ can overwhelm the functional, forcing the solution to the edge of the parameter space, i.e. the functional simply ``rolls down'' the steepest gradient until it reaches the boundary, obscuring the actual physical plateau. Furthermore, the error estimate ${\rm Err}\, \mathcal{O}^{(L)}$ in Eq.~(\ref{error_serone}) relies on an average variation within a sampling window. However, local derivatives (and by extension, local averages) probe stability only infinitesimally close to a given point. They fail to capture the global robustness required to ensure that the resummation has entered a stable convergence mode over physically relevant parameter intervals.  Therefore, a reliable multiparameter optimization necessitates a transition to a stability criterion that effectively balances these disparate sensitivity scales. Instead of relying on local gradients, such a criterion should ensure that the resummed result remains consistent over physically meaningful parameter ranges. Let us introduce two resummation algorithms that implement the homographic transformation. In what follows below, we will refer to them as "Method I" (Section~\ref{error_estimations}) and "Method II" (Section~\ref{rms_error_estimations}).
	
	\subsection{Method I. Quantifying error and reliability. The combined PFAC and PMS method}  \label{error_estimations}
	
	To address the limitations of gradient-based functionals and the resulting boundary biases, let us formulate a stability criterion based on optimization strategy developed in \cite{kompaniets2017} that prioritizes global robustness over local derivatives. To this end, we replace infinitesimal variations with a method that accounts for the intrinsic scales of the parameters, in turn, this allows us to define a total estimated error, $E_{total}(b, \alpha, q)$, across the entire $(b, \alpha, q)$ space. The minimization of this functional allows us to simultaneously identify the optimal parameter set  $(b_{opt}, \alpha_{opt}, q_{opt})$ and provide a reliable estimate of the associated uncertainty. The core goal of this approach is thus to determine both the central value of the physical observable $\mathcal{O}^{(L)}$ and its total estimated error. 
	
	The overall total error, $E_{total}$, is calculated as the sum of the PFAC component and the PMS sensitivity components ($S_x$):
	\begin{eqnarray}
		\label{error}
		E_{\mathrm{total}}  &=& \underbrace{\max\left\{|\mathcal{O}^{(L)}(b, \alpha, q) - \mathcal{O}^{(L-1)}(b, \alpha, q)|, |\mathcal{O}^{(L)}(b, \alpha, q) - \mathcal{O}^{(L-2)}(b, \alpha, q)|\right\}}_{\text{PFAC}}  \nonumber \\ &+& \underbrace{\max(S_b[\mathcal{O}^{(L)}(b, \alpha, q) ], S_b[\mathcal{O}^{(L-1)}(b, \alpha, q) ]) + S_{\alpha}[\mathcal{O}^{(L)}(b, \alpha, q) ] + S_q[\mathcal{O}^{(L)}(b, \alpha, q) ]}_{\text{PMS}}.
	\end{eqnarray}
	Note, that in contrast to Eq.~(\ref{error_serone}), here PFAC quantifies the model uncertainty or the current convergence status of  the observable $\mathcal{O}^{(L)}(b, \alpha, q)$ against two previous approximations ($\mathcal{O}^{(L-1)}(b, \alpha, q),\mathcal{O}^{(L-2)}(b, \alpha, q)$), while  the sensitivity of the observable $\mathcal{O}(x)$ to a specific parameter $x$ (one of $b, \alpha, q$) is calculated using \textit{the min-max sensitivity method} \cite{kompaniets2017, ihrig2018}. Within this method,the sensitivity $S_x$ at each point $x_0$ is defined as the minimum of the maximum of local variations within a window of width $\Delta_x$:
	\begin{equation}
		\label{sensivity}
		S_{x}(\mathcal{O}(x_{0})) = \min_{x: x_0 \in [x, x + \Delta_x]} \left\{ \max_{x^{\prime} \in [x, x + \Delta_x]} |\mathcal{O}(x^{\prime}) - \mathcal{O}(x)| \right\}.
	\end{equation}
	Here, $\Delta_x$ represents a physical perturbation scale rather than a numerical grid step. It acts as a diagnostic tool to test the robustness of the results: a large window (e.g., $\Delta_b = 20$) is employed for parameters where stability is expected over a broad range, while a narrow window (e.g., $\Delta_q = 0.01$) is used for parameters known to be highly sensitive. The $\min$ operation in the Eq. (\ref{sensivity}) ensures that one finds the most stable (flattest) plateau of width $\Delta_x$ that the point $x_{0}$ belongs to. If $x_{0}$ is a part of at least one such stable region, the resulting $S_{x}$ is be small.
	
	We perform a detailed analysis of the error (\ref{error}) by first calculating the critical exponents via the resummation formula (\ref{series_res_final_correct}) across a comprehensive parameter space. We scan the parameter space for $(b, \alpha, q)$ in the range $[0, 40] \times [0, 9] \times [0, 0.4]$, using fine steps of $\delta_b = 0.5, \delta_\alpha = 0.01$, and $\delta_q = 0.01$. The analysis requires determining the local sensitivity for each parameter, necessitating the definition of varying perturbation windows, $\Delta_x$, based on the parameter's nature. Specifically, we set a larger perturbation $\Delta_{b} = 20$ due to the wide plateaus in $b$; a smaller perturbation $\Delta_{\alpha} = 2$ is chosen as the function is more sensitive to $\alpha$; and the smallest $\Delta_{q} = 0.01$ is used due to the strong dependence on $q$.
	
	\subsubsection{Algorithmic optimization}
	
	The initial, straightforward calculation of the sensitivity component $S_x$ necessitates a cubic time complexity (at least), $\mathcal{O}(N^3)$, due to the required all-to-all comparisons within the perturbation window $\Delta_x$, making the analysis computationally prohibitive for large datasets. We overcome this performance bottleneck by redesigning the core algorithm to achieve a quasilinear $\mathcal{O}(N \log N)$ complexity. The optimization relies on first sorting the data by the coordinate $x$ in $\mathcal{O}(N \log N)$ time, followed by replacing the $\mathcal{O}(N^2)$ window search  with a linear $\mathcal{O}(N)$ scan, achieved by utilizing the monotonic deque technique (a standard approach for the sliding window maximum/minimum problem) to retrieve the necessary running extreme values in amortized $\mathcal{O}(1)$ time \cite{cormen2022, eppsteinCS261}. The process is implemented in \texttt{Cython} to leverage static typing and achieve near-native C performance, with the overall workload being further accelerated through parallelization of independent data groups using \texttt{OpenMP}.
	
	\subsubsection{Two key points to the error estimation} 
	
	Two key principles are considered when estimating the error \cite{kompaniets2017, ihrig2018}. Firstly, the three-sigma rule (or three standard deviations rule) is applied: if the error has an approximately Gaussian (normal) distribution, the interval $\pm 3\sigma$ covers $\sim 99.7 \%$ of all possible values (for comparison, $\pm 2\sigma$ covers $\sim 95 \%$, and $\pm \sigma$ covers $\sim 68 \%$). Based on this principle, we typically consider $3\overline{E}$, where $\overline{E}$ is the minimal error. Secondly, if sets of parameters $(b, \alpha, q)$ yield an equally good minimum error, the spread of real values is found to be wider than $3\bar{E}$. In this scenario, the final error is determined as $\overline{E} + 2\sigma$, where $\sigma$ is the standard deviation of a weighted set of values whose error $E_i$ does not exceed $3\overline{E}$. To determine the central value, the weighted mean of all points in the parameter space that satisfy the relative error condition $E_i / \overline{{E}} < (E_i / \overline{{E}})^{*}$ is used, calculated according to \cite{kompaniets2017}:
	\begin{equation}
		\label{weight}
		\overline{{f}} = \frac{\sum_{i \in \{E_i / \overline{{E}} < (E_i / \overline{{E}})^{*}\}} \omega_{i} f(b_{i}, \alpha_{i}, q_{i})}{\sum_{i \in \{E_i / \overline{{E}} < (E_i / \overline{{E}})^{*}\}} \omega_{i}},
	\end{equation}
	where the weights $\omega_{i}$ are defined as $\omega_{i} = 1/E_{i}^{2}$.

	\begin{figure}[h!] 
			\subfloat[]{\includegraphics[width=0.32\textwidth]{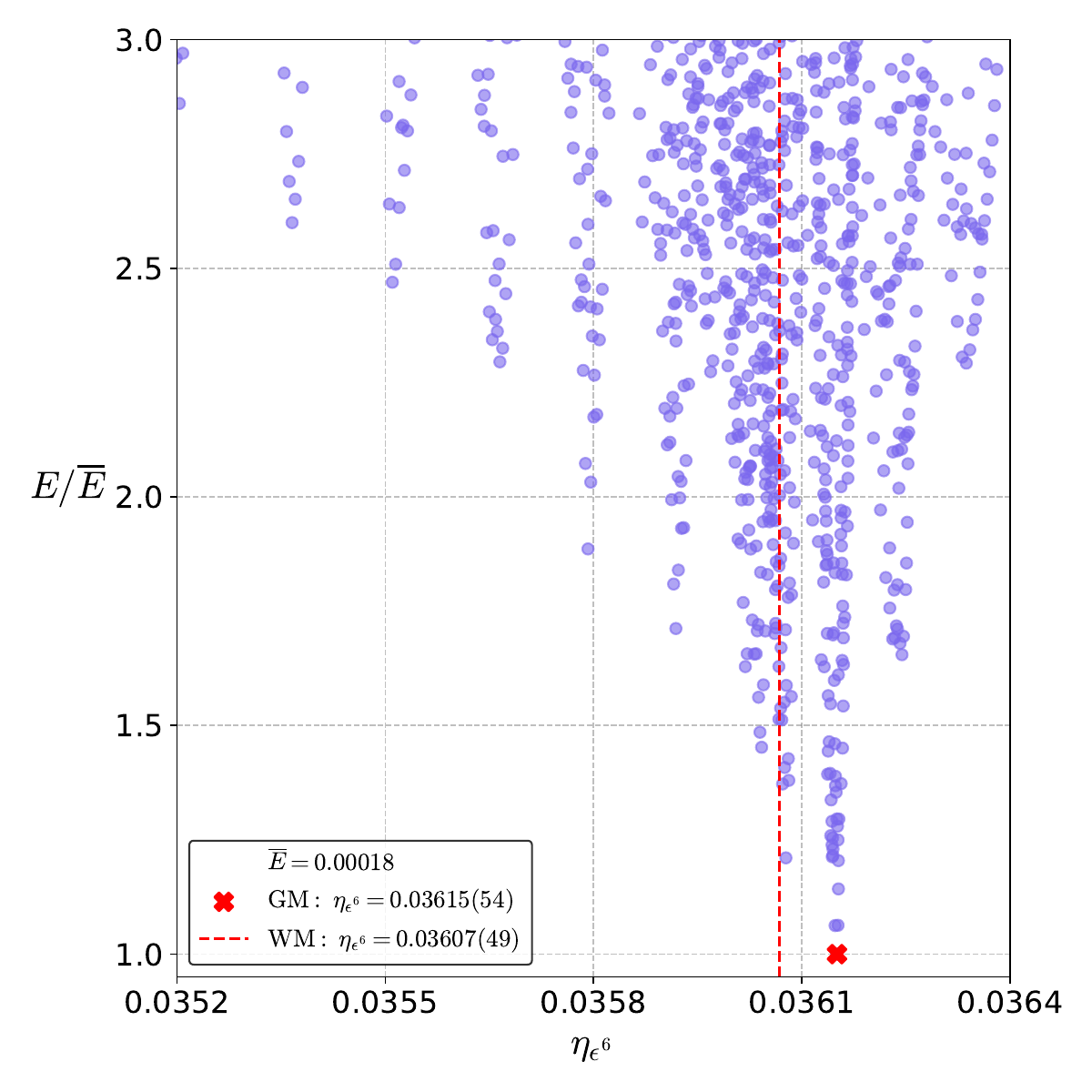}}\hfill
			\subfloat[]{\includegraphics[width=0.32\textwidth]{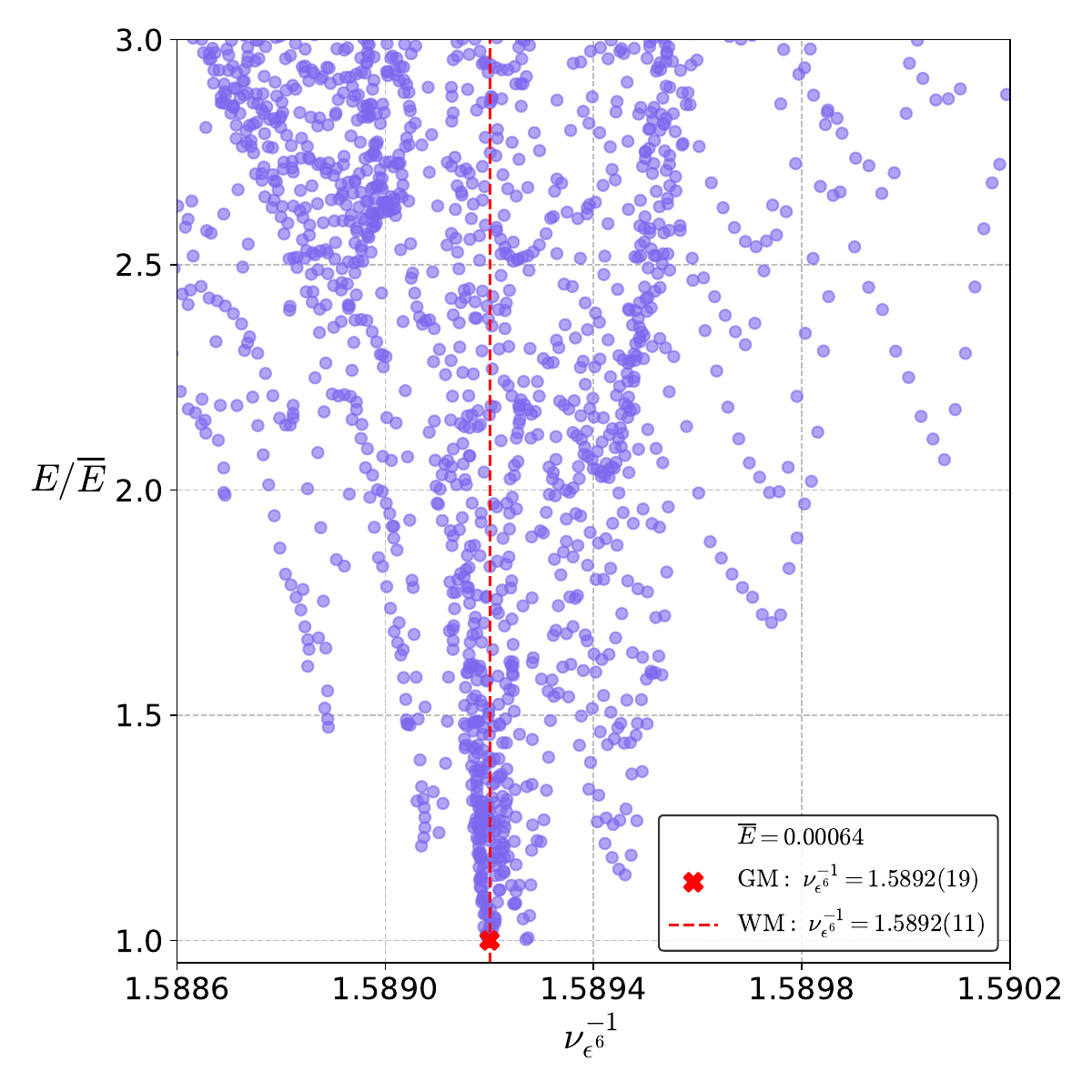}}\hfill
			\subfloat[]{\includegraphics[width=0.32\textwidth]{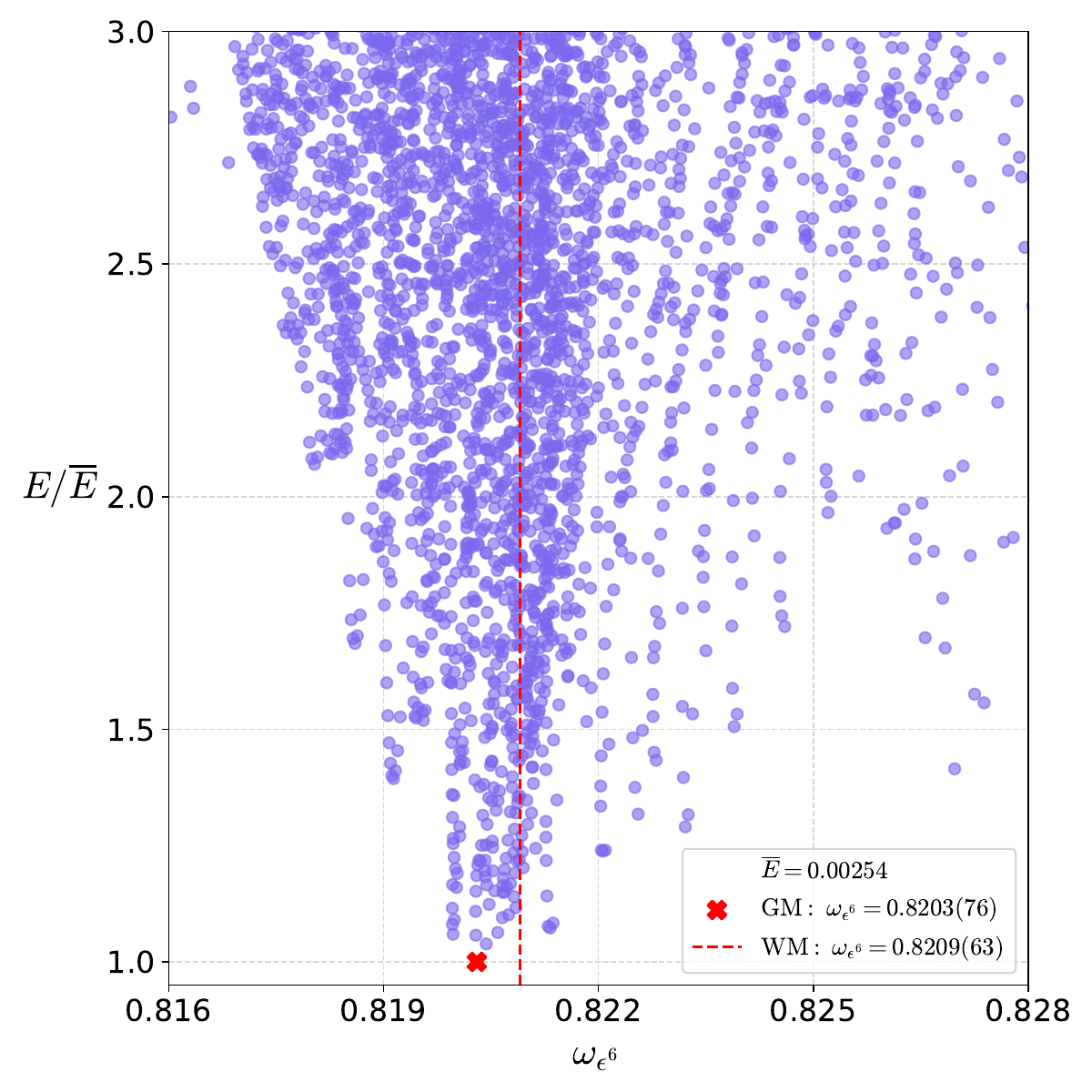}}
		\centering
		\caption{Error distribution and statistical analysis of the critical exponents $\eta$ [(a)], $\nu$ [(b)], and $\omega$ [(c)] at sixth-order. The scatter plots show the normalized error $E/\overline{E}$ as a function of the exponent values calculated across the parameter space $(b, \alpha, q)$, where $\overline{E}$ denotes the global minimum error. The red cross, GM, represents the global minimum point with an associated uncertainty of $3\overline{E}$ (indicated in parentheses). The dashed vertical line, WM, marks the weighted mean calculated over the candidate set within the plateau region $E \leq 3\overline{E}$ [see Eq.~(\ref{weight})]. Its corresponding uncertainty in parentheses is defined as $\overline{E} + 2\sigma_w$, where $\sigma_w$ is the weighted standard deviation. This dual representation provides a robust estimation of the exponent values by accounting for both local stability and the statistical spread within the optimal parameter region.
		}
		\label{minmax_six_loop}
	\end{figure}
	
	To validate this approach, we first applied the resummation algorithm to the six-loop expansions of the critical exponents $\eta$ and $\nu$. Our results are in excellent agreement with those reported in~\cite{kompaniets2017} (see Table~\ref{tab:comparison__two_methods}), with minor deviations arising from the numerical handling of the stability criteria (as discussed in the summary of this section). The statistical distribution of these results is presented in Fig.~\ref{minmax_six_loop}. The use of a dual representation, combining the global minimum (GM) with a weighted mean (WM) over the stability plateau, provides a robust estimation. By accounting for the statistical spread within the region $E \leq 3\overline{E}$, we ensure that the final exponent values are not biased by isolated local extrema but represent the most stable zone of the parameter space.
	
	\begin{figure}[h!] 
			\subfloat[]{\includegraphics[width=0.33\textwidth]{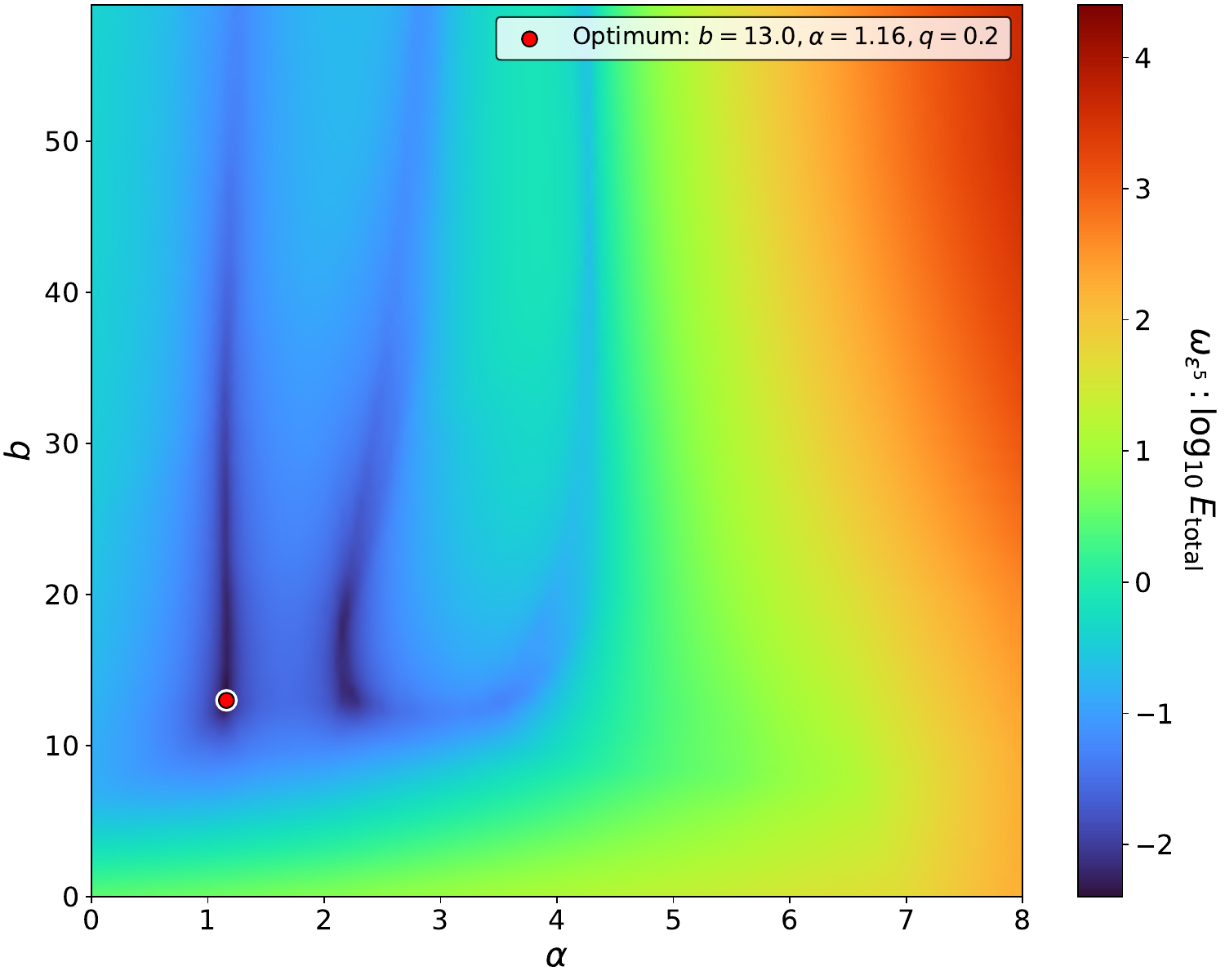}}\hfill
			\subfloat[]{\includegraphics[width=0.33\textwidth]{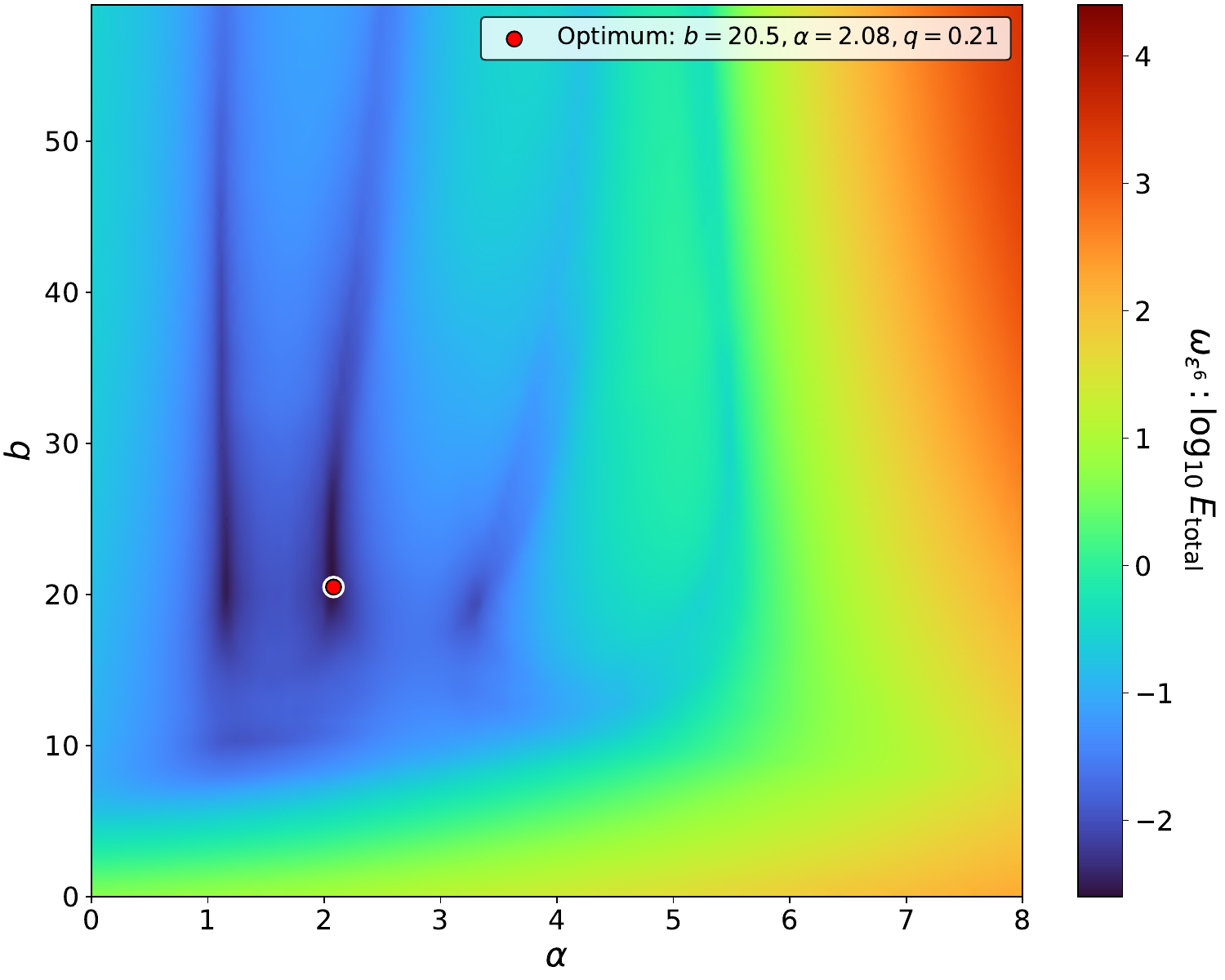}}\hfill
			\subfloat[]{\includegraphics[width=0.33\textwidth]{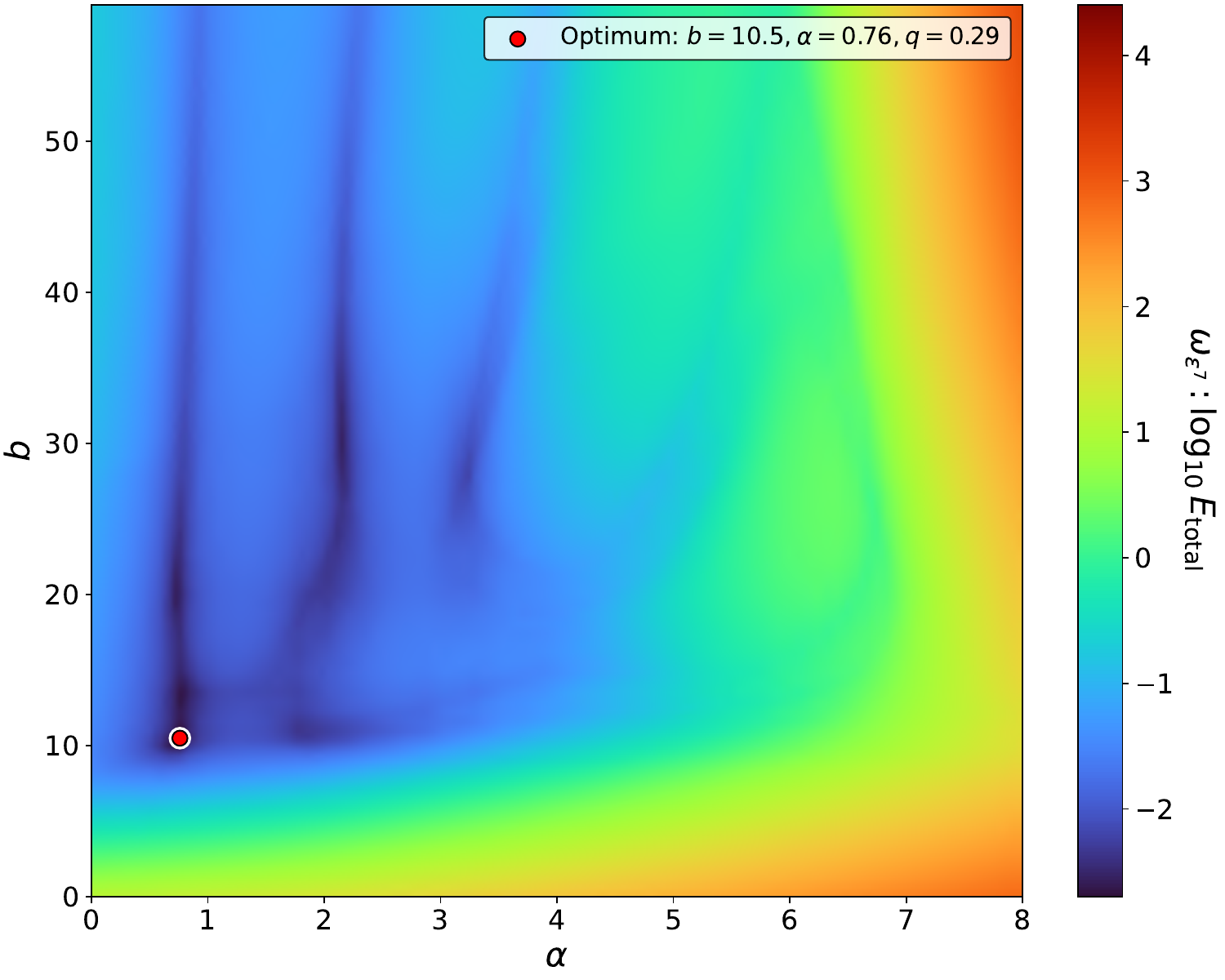}}
		\centering
		\caption{The total error landscape $E_{\mathrm{total}}(b, \alpha, q)$ [see Eq.~(\ref{error})] for the correction-to-scaling exponent $\omega$ at the resummation orders $\epsilon^5$ [(a)], $\epsilon^6$ [(b)], and $\epsilon^7$ [(c)].  In the lower orders, $\omega_{\epsilon^5}$ and $\omega_{\epsilon^6}$, the optimum settles at higher $b$ values, following the extended vertical `canyons' of mutual stability between successive approximations. As we reach the seventh order, $\omega_{\epsilon^7}$ , the structural refinement of the series causes the optimum to shift toward a more localized region. This shift may demonstrate how higher-order terms actively filter the parameter space, identifying a specific zone where convergence (PFAC) and stability (PMS) reach their most robust compromise.}
		\label{minmax_error_landscape__omega}
	\end{figure}
	
	For the critical exponent $\eta$, our results closely reproduce those of \cite{kompaniets2017} (see Table~\ref{tab:comparison__two_methods}). For the correlation length exponent $\nu$, we likewise find excellent agreement at the level of the central value. However, the corresponding error distribution exhibits a relatively well-localized minimum (see (b) in Fig.~\ref{minmax_six_loop}). In this situation, both the GM and WM lead to nearly identical estimates, indicating a high degree of stability of the resummation: $\left( \nu_{\epsilon^{6}}^{(\mathrm{GM})} \right)^{-1} = 1.5892(19)$ and $\left( \nu_{\epsilon^{6}}^{(\mathrm{WM})} \right)^{-1} = 1.5892(11)$.  A noticeable difference arises in the associated uncertainty. Namely, the WM, being based on an average over the stability region, yields a somewhat narrower error bar, while the GM reflects the local balance between convergence (PFAC) and stability (PMS) at the optimal point. In this approach, we adopt the latter as a more conservative estimate of the uncertainty, while noting that both prescriptions remain fully consistent.

	A more complex behavior is observed for the correction-to-scaling exponent $\omega$. Despite this,  our five- and six-loop estimates closely reproduce those of \cite{kompaniets2017} (see Fig.~\ref{omega5__n1} in Appendix~\ref{figures_app} for $\omega_{\epsilon^5}$ and Table~\ref{tab:comparison__two_methods} with Fig.~\ref{omega6__n1} in Appendix~\ref{figures_app} for $\omega_{\epsilon^6}$). As shown in Fig.~\ref{minmax_error_landscape__omega}, the optimal parameters $(b, \alpha, q)$ undergo a significant shift when transitioning from lower orders  $\omega_{\epsilon^5}$, $\omega_{\epsilon^6}$ to the seventh-order approximation $\omega_{\epsilon^7}$. In the fifth and sixth orders, the error landscape is characterized by extended vertical ``canyons'' of stability, where the functional remains relatively flat over a wide range of $b$ values. It is important to note that at these orders, our methodology reproduces the established literature values for critical exponents $\eta$, $\nu$, and $\omega$ with the same high precision as in \cite{kompaniets2017}.  However, upon reaching the seventh order, we observe a significant refinement for the correction-to-scaling exponent $\omega$. The restrictive nature of the convergence criterion (PFAC) effectively collapses these broad stability `canyons', relocating the global minimum toward a more localized and robust zone at $b \approx 10.5$ and $\alpha \approx 0.76$. This transition suggests that higher-order terms act as a numerical filter, narrowing the parameter space to a region where physical consistency and mathematical stability finally converge. While the physical stability plateau remains present, the convergence criterion (PFAC) becomes significantly more stringent at higher orders and begins to dominate over the stability measure (PMS). Consequently, the algorithm does not simply select the center of a broad plateau but rather identifies the specific point where successive approximations intersect most precisely. In this sense, the procedure is remarkably `honest' -- although a region at higher $b$ values might exhibit lower local sensitivity (PMS), the slightly better convergence (PFAC) leads to a smaller total error $E_{\mathrm{total}}$. To minimize the combined functional, the algorithm is prepared to sacrifice the `width' of the stability plateau in favor of the `depth' of perturbative convergence, resulting in the observed localization of the optimal result (more figures are presented in Appendix~\ref{figures_app}).
	
	To conclude this section, we address the potential reasons for the slight discrepancies between our results and those reported in \cite{kompaniets2017}. While the core algorithm and the resummation values $\mathcal{O}$ are consistent, small variations in the final $(b, \alpha, q)$ coordinates are expected due to several technical factors:
	\begin{itemize}
		\item[(i)] Both this work and \cite{kompaniets2017} employ discrete grids. The ``min-max'' operation [Eq.~(\ref{sensivity})] is performed over these discrete points. Since the error landscape is often a ``flat valley'' rather than a sharp peak, the global minimum can easily shift between neighboring grid points due to infinitesimal differences in the underlying values;
		\item[(ii)] Even if the integrated values look close, small differences in numerical integration may introduce minor fluctuations. The discrete nature of the max operator makes it highly sensitive to this numerical resolution.
		\item[(iii)] The fact that we obtain identical optimal coordinates $(b, \alpha, q)$ confirms the consistency of the core methodology. The minor discrepancies in the final value of $E_{\mathrm{total}}$ are purely technical, resulting from how the sliding window boundaries interact with the specific discrete sampling of the parameter space.
	\end{itemize}	
	
	\subsection{Method II. Root-mean-square (rms) stability and convergence selection in the three-parameter space}
	\label{rms_error_estimations}
	
	As we have seen from the previous sections, the introduction of the third parameter $q$ (the homographic transformation) significantly expands the available parameter space, often creating extended `canyons' of stability where the PMS criterion alone cannot distinguish between different results. Our analysis by {\it Method I} (Section~\ref{error_estimations} \cite{kompaniets2017}) indicates that although these regions may appear locally stable, they do not all satisfy the convergence requirements equally. To address this, we employ a complementary approach that combines PFAC-based control with an alternative sensitivity analysis based on the rms variation \cite{taylor1996_book}. This allows us to double-check the results and ensure they represent the most robust compromise between stability and convergence.
	
	The rms-based analysis introduced here is used as a complementary diagnostic tool rather than a replacement of the established `min--max' error estimation framework \cite{kompaniets2017}. While the latter provides a conservative global uncertainty estimate, the former allows us to resolve the local structure of stability plateaus at high orders. The agreement between the regions selected by these two complementary criteria serves as an additional consistency check of our results.
	
	For a fixed-loop order $L$, we quantify the local numerical stability of the observable $\mathcal{O}^{(L)}$ by evaluating its rms variation within a finite neighborhood of each point in the auxiliary parameter space. Similarly to Section~\ref{error_estimations}, for each point $(b_0,\alpha_0,q_0)$ we define a neighborhood
	\begin{equation}
		\mathcal{N}(b_0,\alpha_0,q_0) = \left\{(b,\alpha,q)\;\big|\; |b-b_0|\le\Delta_b,\; |\alpha-\alpha_0|\le\Delta_\alpha,\; |q-q_0|\le\Delta_q \right\},
	\end{equation}
	where $\Delta_b$, $\Delta_\alpha$ and $\Delta_q$ denote the fixed variation scales as in \textit{Method I}.
	
	In contrast to the `min--max' criterion employed in Section~\ref{error_estimations}, which probes global variations across the full parameter domain, {\it global worst-case sensitivity}, the rms-based analysis is designed to detect {\it local stability}, capturing stationarity within parameter neighborhoods. In practice, the rms sensitivity is evaluated separately along each parameter direction using finite windows, and combined into an effective PMS stability estimator. For each direction, we compute the local standard deviation of $\mathcal{O}^{(L)}$ within the corresponding neighborhood slice, yielding the directional contributions $S_b$, $S_\alpha$, and $S_q$. These are then aggregated into a single measure of local stability,
	\begin{equation}
		\label{rms_sensitivity}
		E_{\mathrm{pms}}(b_0,\alpha_0,q_0) =	\max\left(S_b^{(L)}, S_b^{(L-1)}\right) + S_\alpha^{(L)} + S_q^{(L)},
	\end{equation}
	where $S_b$, $S_\alpha$, and $S_q$ denote the local standard deviations evaluated along the corresponding parameter directions. The use of two consecutive loop orders in the $b$-direction improves the robustness of the stability estimate.
	
	At sufficiently high orders, numerical stability typically manifests itself as an extended plateau rather than a sharply localized extremum. In such situations, the rms-based criterion provides a statistically robust measure of stability, as it captures the overall density of consistent solutions and effectively suppresses numerical noise by accounting for the spread of values within local neighborhoods.
	
	To control the convergence of the perturbative series, we additionally impose a PFAC-based consistency condition, as defined in Eq.~(\ref{error}). Within the PMS-stable domain, we retain only those points for which the discrepancy between different loop orders remains small [see Eq.~(\ref{PFAC})]. This requirement ensures that local numerical stability is accompanied by a consistent convergence pattern of the resummed series.
	
	The combined application of PMS stability and PFAC consistency typically yields a compact but nontrivial region in the auxiliary parameter space. While the `min--max' approach provides a conservative global error estimate and remains a useful benchmark \cite{kompaniets2017}, the present rms strategy improves the local resolution inside the most physically reliable region by explicitly targeting stationary plateaus. At the selected optimal point, the total uncertainty is decomposed into two complementary contributions. The truncation error is estimated from the PFAC discrepancy, $E_{\mathrm{pfac}}$, while the residual sensitivity to auxiliary parameters is quantified by the local rms-based stability measure, $E_{\mathrm{pms}}$. The total error is then obtained as a quadratic combination,
	\begin{equation}
		\label{hybrid_error}
		E_{\mathrm{total}} = \sqrt{E_{\mathrm{pfac}}^{\,2} + E_{\mathrm{pms}}^{\,2}} .
	\end{equation}
	Compared to the {\it Method I}, the rms-based estimate yields a more conservative and statistically meaningful uncertainty, as it reflects the combined effect of convergence and local stability.
	
	The selection of the optimal resummation parameters can naturally be formulated as a multi-objective optimization problem, where the goals of minimizing the perturbative discrepancy $E_{\mathrm{pfac}}$ and minimizing the local instability $E_{\mathrm{pms}}$ are generally competing. Within this framework, the log-normalized scoring function $E_{\mathrm{balanced}}$ provides a practical way to select a representative optimal point by balancing these two contributions on a comparable scale. The resulting optimal value of $\mathcal{O}$ is therefore associated with a stable solution embedded in a consistent convergence regime.
	
\begin{figure}[h!] 
			\subfloat[]{\includegraphics[width=0.33\textwidth]{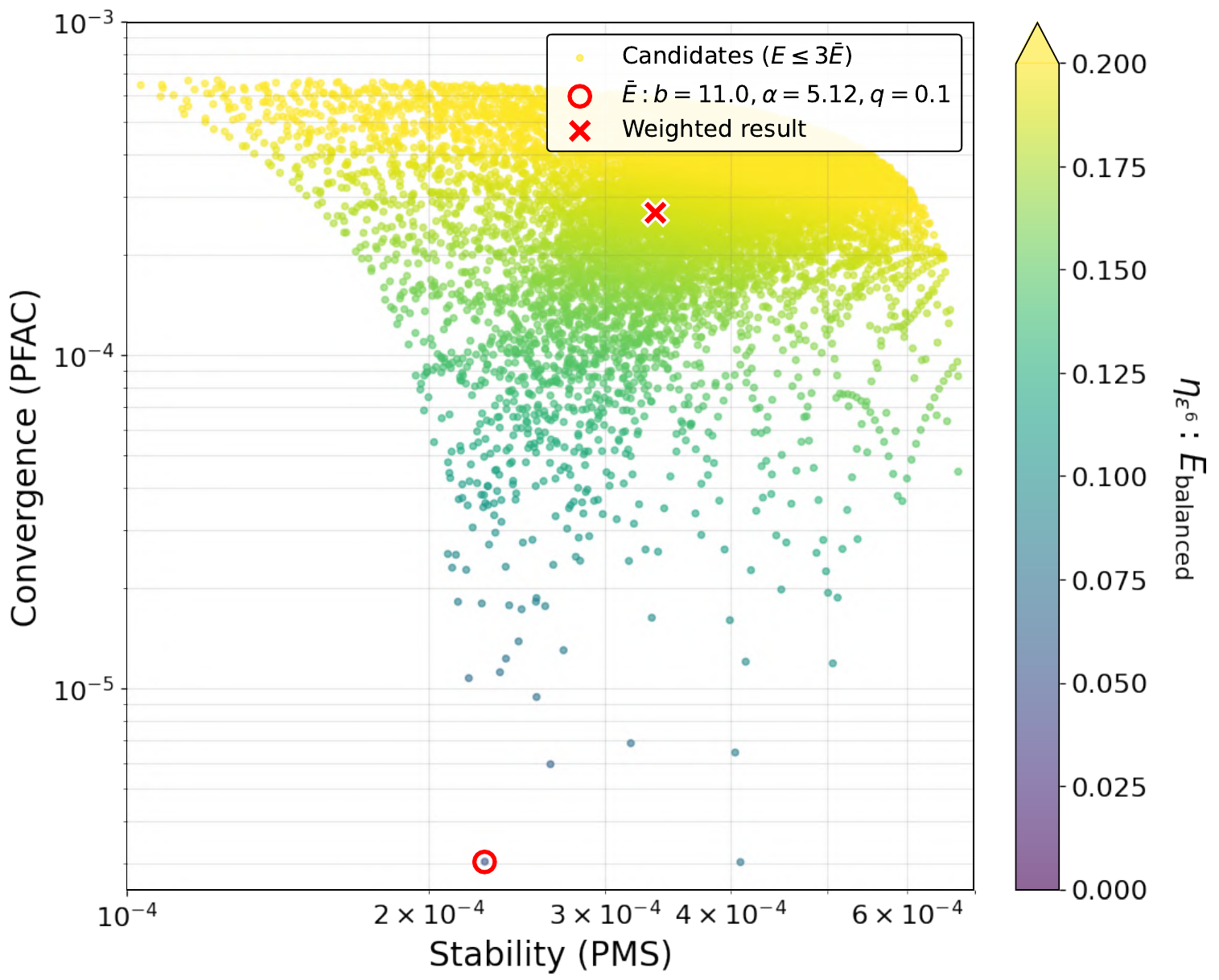}}\hfill
			\subfloat[]{\includegraphics[width=0.33\textwidth]{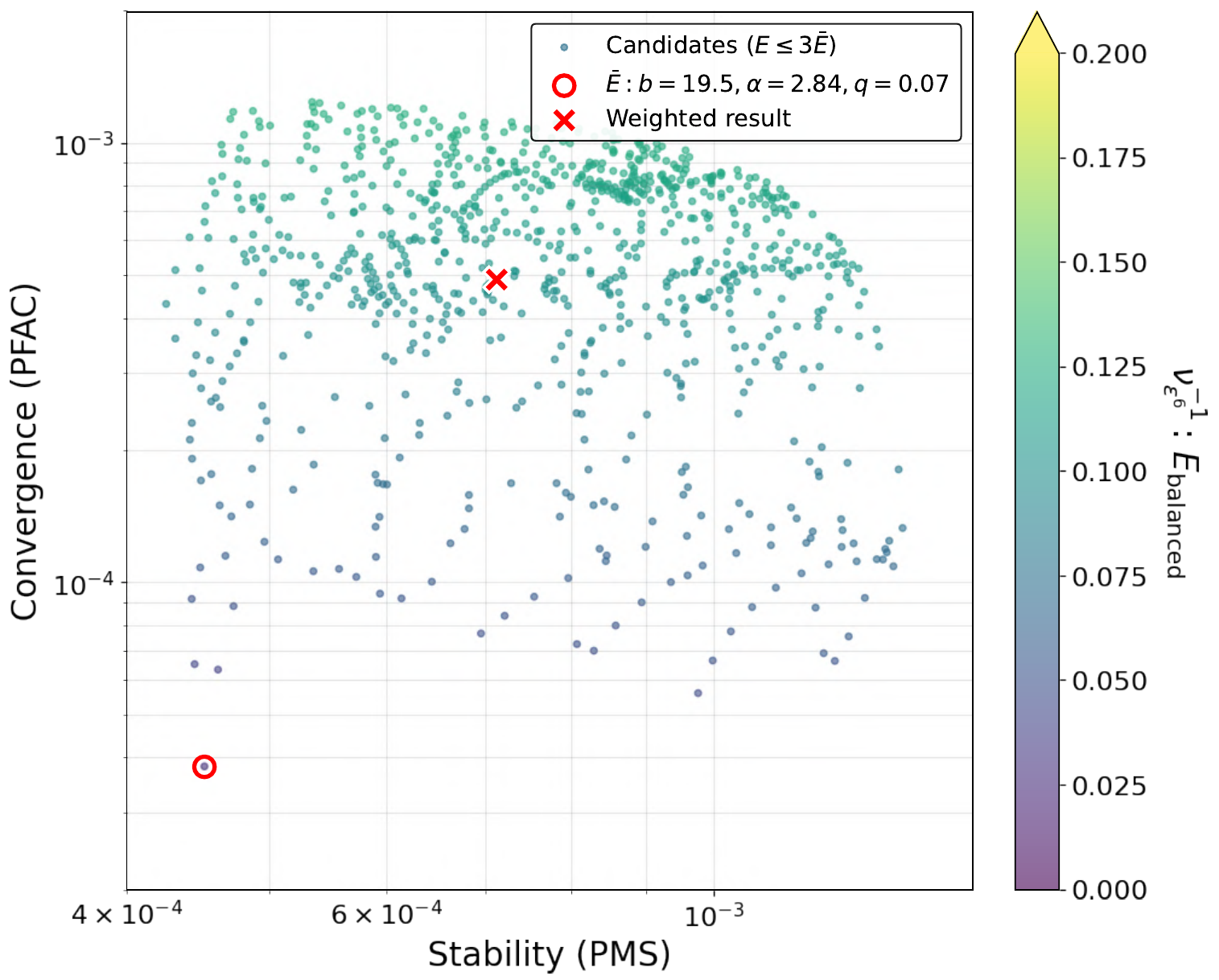}}\hfill
			\subfloat[]{\includegraphics[width=0.33\textwidth]{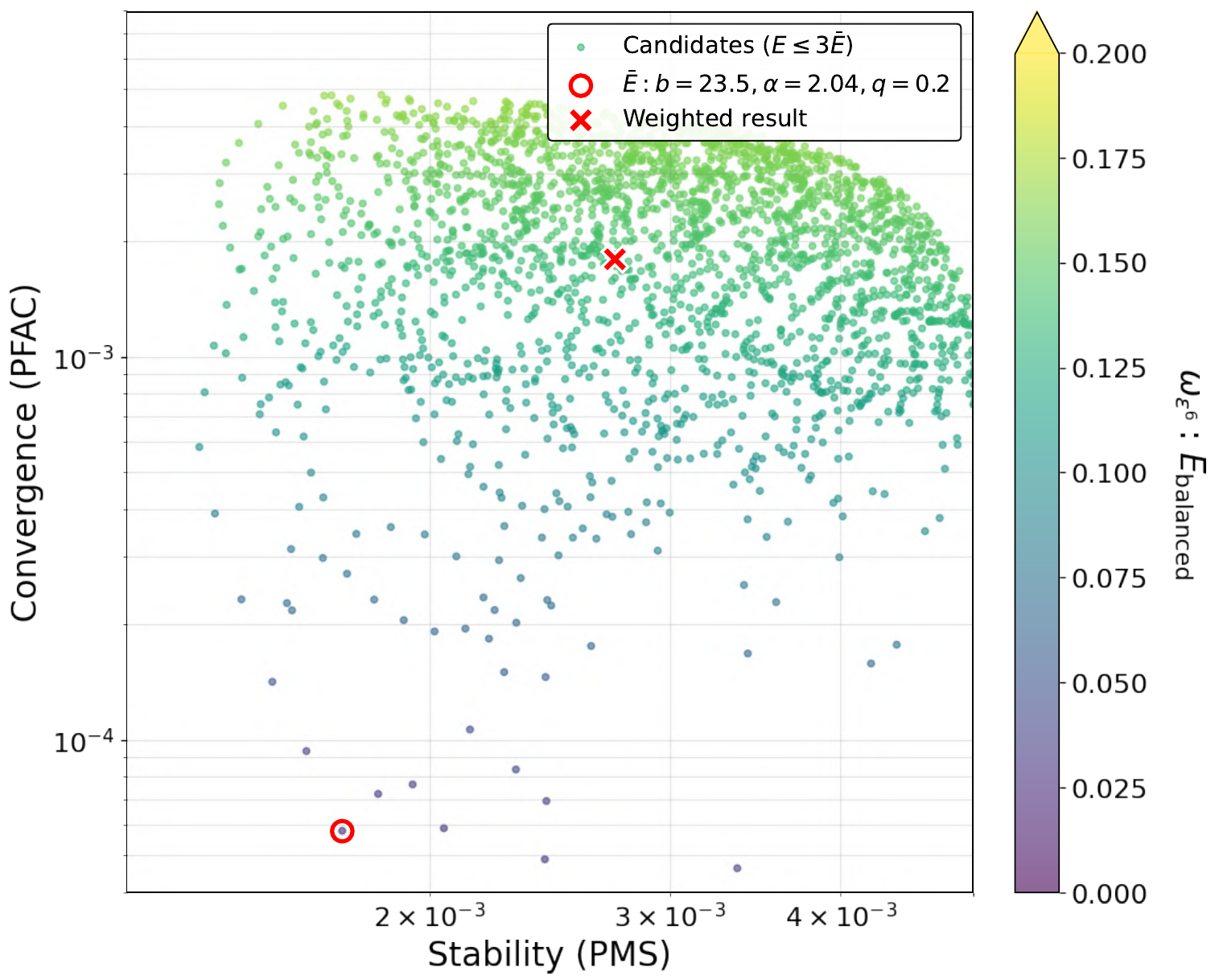}}\hfill
			\subfloat[]{\includegraphics[width=0.33\textwidth]{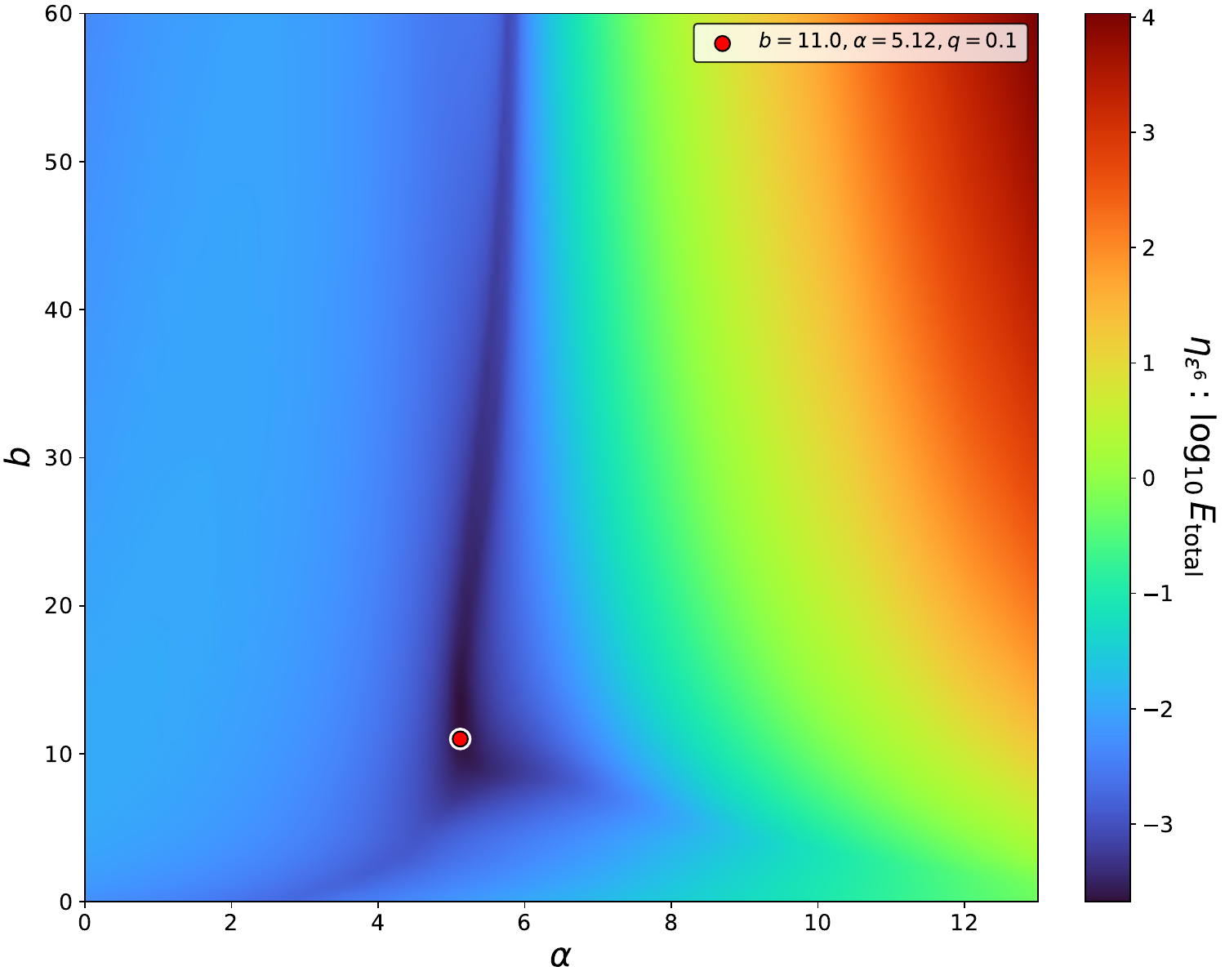}}\hfill
			\subfloat[]{\includegraphics[width=0.33\textwidth]{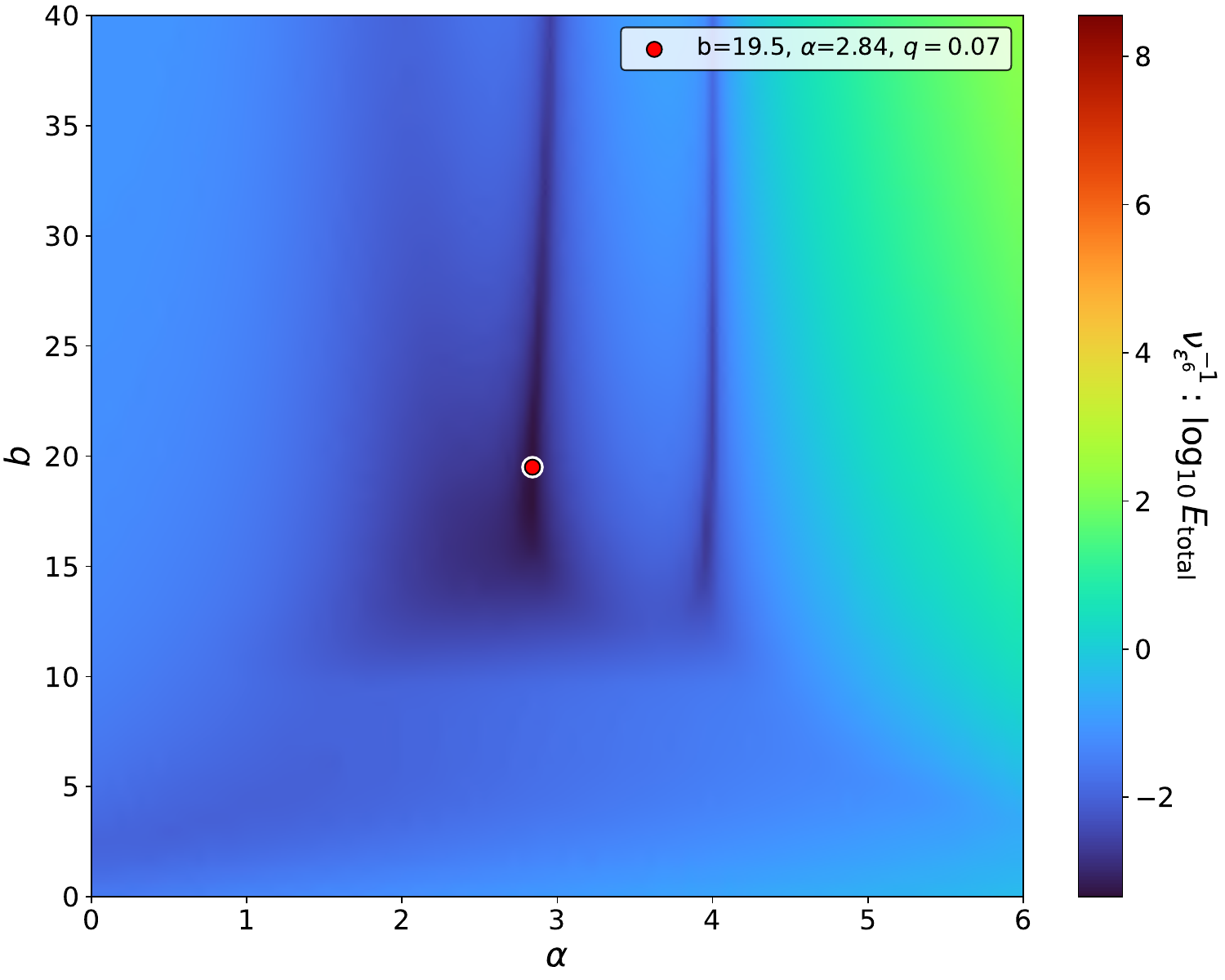}}\hfill
			\subfloat[]{\includegraphics[width=0.33\textwidth]{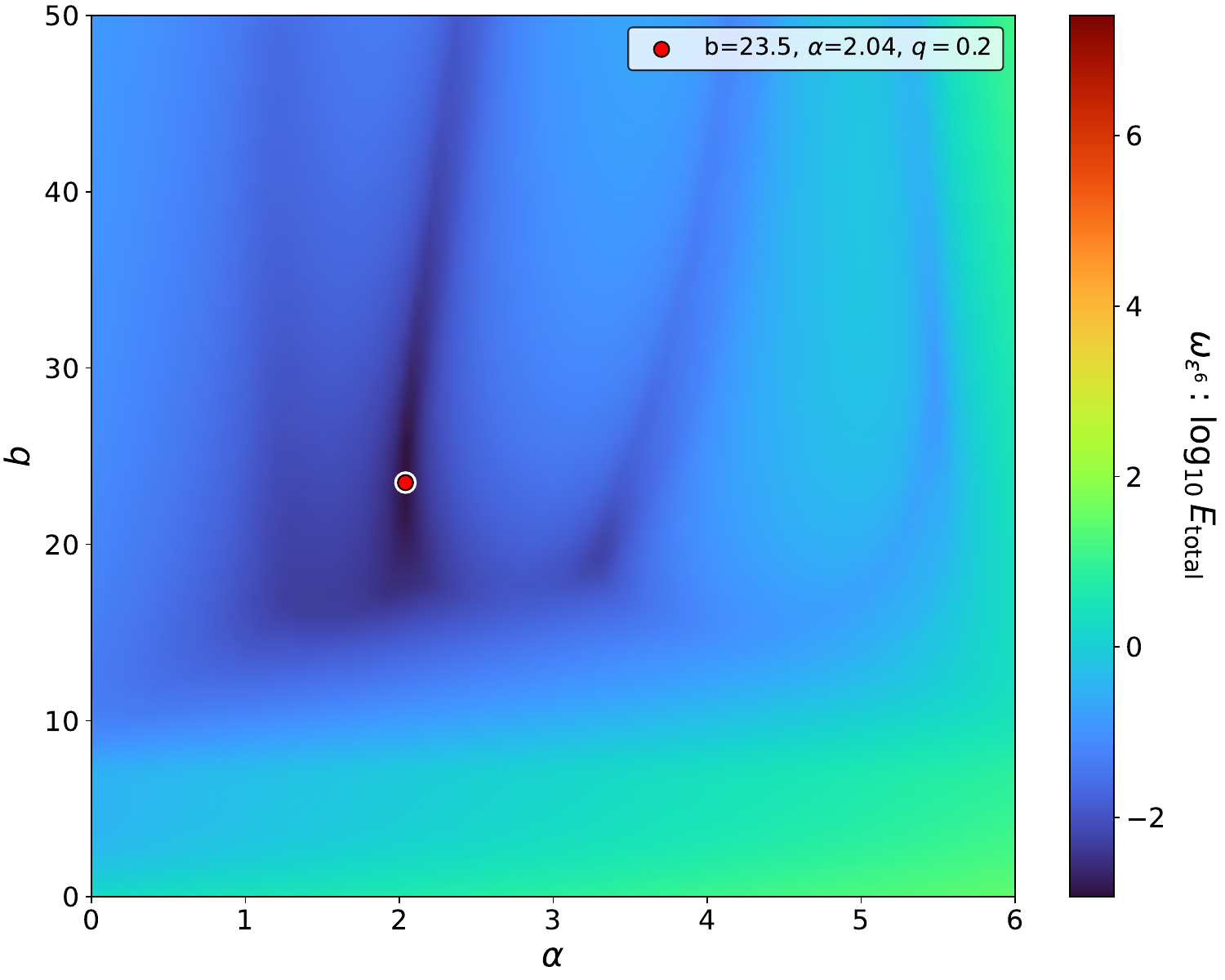}}
		\centering
		\caption{Analysis of the optimization procedure and the error landscape for the 3D Ising critical exponents $\eta$ [(a), (d)] , $\nu$ [(b), (e)] , and $\omega$ [(c), (f)] at  the six-loop perturbative expansion. Panels (a)-(c): the distribution of the calculated points in the plane of stability ($E_{\mathrm{pms}}$) and convergence ($E_{\mathrm{pfac}}$). The color mapping represents the balanced log-normalized score $E_{\mathrm{balanced}}$, which accounts for the different orders of magnitude between the criteria [see Eqs.~(\ref{balanced_error}) and (\ref{balanced_error_final})] and serves as a diagnostic tool for visualizing regions where stability and convergence are simultaneously optimized. The red open circle identifies the global minimum of $E_{\mathrm{balanced}}$, corresponding to the selected optimal point. The candidate set are defined by $E_{\mathrm{total}} \leq 3\,E_{\min}^{\mathrm{total}}$, consistent with the plateau criterion used in the analysis. For cases with significant plateau spread, the weighted mean result is indicated by a red cross. Panels (d)-(f): the error landscapes in the $(b, \alpha)$ plane at the optimal $q$ value. The contours show the total physical error $\log_{10}{E_{\mathrm{total}}}$ calculated according to Eq.~(\ref{hybrid_error}). The optimal configuration is marked by a red dot, situated within a deep ``valley'' of stability that represents the best compromise between convergence and sensitivity to the resummation parameters.}
		\label{pareto_six_loop}
	\end{figure}

	Therefore, in order to balance the relative contributions of the convergence (PFAC) and stability (PMS) criteria, which typically differ by several orders of magnitude, we introduce a log-normalized scoring procedure. Each error component is mapped to a dimensionless scale via
	\begin{equation}
		\label{balanced_error}
		\hat{E}_i = \frac{\log_{10} E_i - \min(\log_{10} E_i)} {\max(\log_{10} E_i) - \min(\log_{10} E_i)} ,
	\end{equation}
	where the extrema are taken over the full parameter grid. The optimal region is identified by minimizing the distance to the origin in this normalized log-space:
	\begin{equation}
		\label{balanced_error_final}
		E_{\mathrm{balanced}} = \sqrt{ \hat{E}_{\mathrm{pfac}}^{\,2} + \hat{E}_{\mathrm{pms}}^{\,2}}.
	\end{equation}
	In addition to its role in selecting the optimal point, the balanced score $E_{\mathrm{balanced}}$ is used as a diagnostic quantity to visualize the structure of the parameter space. In particular, in Fig.~\ref{pareto_six_loop} (the first row), the color coding represents the value of $E_{\mathrm{balanced}}$, allowing one to identify regions where stability and convergence are simultaneously optimized. This provides an additional qualitative criterion for assessing the consistency of the selected solutions and for identifying extended plateaus in the multiparameter space.
	
	The choice between the global minimum (GM) and the weighted mean (WM) is governed by the topology of the error landscape $E_{\mathrm{total}}(b, \alpha, q)$ (see the second row in Fig.~\ref{pareto_six_loop}). The GM corresponds to the point where the combined error $E_{\mathrm{total}}$ reaches its absolute minimum, thus providing the optimal compromise between perturbative convergence (PFAC) and local stability (PMS). As such, it represents the primary estimate of the observable. For the exponent $\eta$ at the sixth-loop order, the error landscape is not sharply localized but instead forms a relatively broad minimum region (see (d) in Fig.~\ref{pareto_six_loop}).  Although the GM is well-defined and provides a stable estimate, the surrounding plateau contains many nearby points with similar error values. As a result, WM exhibits a noticeably larger spread:  $\eta_{\epsilon^{6}}^{(\mathrm{GM})} = 0.0362(7)$ and  $\eta_{\epsilon^{6}}^{(\mathrm{WM})} = 0.0361(14)$. As we will report in the next Section this situation improves at higher loop orders. At the seven- and eight-loop orders, the plateau becomes much more localized, and the GM and WM results nearly coincide, indicating a well-defined and stable minimum (see Section~\ref{final_critical_exponents} and second rows in Figs.~\ref{eta7__n1} and \ref{eta8__n1}  in Appendix~\ref{figures_app}).
	
	For $\nu$ and $\omega$ at the six-loop order, the error landscape develops a more complex structure with multiple narrow stability regions (see (e) and (f) in Fig.~\ref{pareto_six_loop}). These regions correspond to distinct areas in parameter space where the balance between convergence and stability is achieved. Despite this fragmented structure, the resulting estimates remain highly consistent, namely the GM and WM values are in close agreement and lie well within the corresponding uncertainties: $\left( \nu_{\epsilon^{6}}^{(\mathrm{GM})} \right)^{-1} = 1.5895(14)$ and $\left( \nu_{\epsilon^{6}}^{(\mathrm{WM})} \right)^{-1} = 1.5893(15)$, while $\omega_{\epsilon^{6}}^{(\mathrm{GM})} = 0.8200(52)$ and  $\omega_{\epsilon^{6}}^{(\mathrm{WM})} = 0.8194(49)$. This indicates that, although the optimal parameters are not unique, the physical observable itself is stable across different candidate regions. In this situation, the GM continues to provide the final estimate, while the WM confirms that the result is not sensitive to the particular choice of the optimal point within the landscape.

	In summary, the analysis shows that, although the structure of the error landscape varies significantly between different exponents and loop orders, GM consistently provides a stable and well-defined estimate. WM, in turn, serves as a complementary diagnostic, reflecting the spread of acceptable solutions within the stability region. In all cases, we therefore adopt the GM as the final estimate, while the WM is reported as a supplementary measure characterizing the structure of the error landscape and the associated uncertainty. The resulting six-loop estimates are summarized in Table~\ref{tab:comparison__two_methods}. As can be seen, both methods -- "Method I" (Section~\ref{error_estimations}) and "Method II" (Section~\ref{rms_error_estimations}) -- lead to mutually consistent results that are in excellent agreement with the benchmark values of ~\cite{kompaniets2017}, confirming the reliability and robustness of the proposed approach.
	
	\begin{table}[h!]
		\centering
		\caption{Comparison of the six-loop estimates of the critical exponents for the 3D Ising model.}
		\label{tab:comparison__two_methods}
		\begin{tabular}{lccc}
			\hline\hline
			Method & $\eta$ & $\nu$ & $\omega$ \\
			\hline
			Method I, Eqs.~(\ref{error}), (\ref{sensivity}) & 0.03615(54) & 0.62925(76) & 0.8203(76) \\
			Method II, Eqs.~(\ref{rms_sensitivity}), (\ref{hybrid_error}) & 0.03615(68) & 0.62912(67) & 0.8200(52) \\
			Kompaniets \& Panzer \cite{kompaniets2017} & 0.0362(6) & 0.6292(5) & 0.820(7) \\
			\hline\hline
		\end{tabular}
	\end{table}

	\section{Results and discussion} \label{results}
	
	\subsection{Critical exponents at high perturbation orders} \label{final_critical_exponents}
	
	In this section, we present the numerical results for the critical exponents $\eta$, $\nu$, and $\omega$ obtained from the seven- and eight-loop perturbative expansions. To ensure the reliability of our estimates, we perform a detailed comparative analysis using two independent strategies: the conservative global `min-max' approach (\textit{Method I} \cite{kompaniets2017}, see Section~\ref{error_estimations}) and the local stability rms-based analysis (\textit{Method II}, see Section~\ref{rms_error_estimations}).
	
	The correlation function critical exponent $\eta$ exhibits the most stable behavior among the investigated exponents. For the seven-loop expansion, \textit{Method I} identifies an optimal value at $(b, \alpha, q) = (10.0, 5.04, 0.10)$, yielding $\eta = 0.03613(24)$. \textit{Method II}, focusing on local stability plateaus, finds the optimum at $(b, \alpha, q) = (13.5, 5.76, 0.09)$ with a consistent value of $\eta = 0.03611(14)$ (see Table~\ref{tab:final_results} and Fig.~\ref{eta7__n1}).  The inclusion of the eight-loop term further refines these estimates and significantly reduces the associated uncertainties. At this order, the results from both methods are remarkably congruent: \textit{Method I} gives $\eta = 0.03617(10)$ at  $(b, \alpha, q) = (17.5, 6.08, 0.11)$, while \textit{Method II} yields $\eta = 0.03616(7)$ at  $(b, \alpha, q) = (18.5, 5.88, 0.10)$. As illustrated in the error distribution plots (see Table~\ref{tab:final_results} and Figs.~\ref{eta7__n1} and \ref{eta8__n1}  in Appendix~\ref{figures_app}), the global minimum (GM) and the weighted mean (WM) practically coincide for $\eta$. This reflects a single, deep, and well-localized stability region in the parameter space where the statistical spread is minimal.
	
	The seven-loop analysis for $\nu$ reveals a noticeable shift of the optimal parameters between the two methods (see Fig.~\ref{nu7__n1}), similar to the behavior already observed at six-loop order (see Section~\ref{error_estimations}). However, as discussed in Section~\ref{error_estimations}, such variations in the optimal coordinates do not necessarily translate into a significant spread of the physical observable. Indeed, despite the apparent relocation in the parameter space, the resulting values of $\nu$ remain highly stable. The corresponding error distribution is characterized by a well-defined minimum, for which the GM and WM estimates are fully consistent within uncertainties. In line with our general strategy, we therefore adopt the GM as the final estimate, while using the WM as a supplementary measure of the spread within the stability region. The final results, $\nu = 0.62925(67)$ and $\nu = 0.62914(54)$ for \textit{Method I} and \textit{Method II}, respectively (see Table~\ref{tab:final_results}), demonstrate good agreement between the two optimization procedures.
	
	A qualitatively different behavior is observed for the correction-to-scaling exponent $\omega$ at the seven-loop order. While the five- and six-loop results show a consistent pattern and are in excellent agreement with Ref.~\cite{kompaniets2017}, the seven-loop estimates exhibit a noticeable upward shift in the central value, accompanied by a reduced overlap with lower-order results (as discussed in Section~\ref{error_estimations}).
	
	Within \textit{Method I}, the optimization procedure identifies a well-defined global minimum at $(b, \alpha, q) = (10.5, 0.76, 0.29)$, yielding $\omega = 0.8257(42)$. Although this point satisfies the combined PFAC and PMS criteria, the resulting value lies  above the five- and six-loop estimates [see the inset in Fig.~\ref{final_trend}(c)]. 
	As illustrated in Fig.~\ref{minmax_error_landscape__omega}(c), this behavior arises from the fact that the stability landscape consists of multiple `canyons.' In contrast, within \textit{Method II} (see Fig.~\ref{omega7__n1} in Appendix~\ref{figures_app}), the optimal parameters are located in a distinct region of the parameter space, $(b, \alpha, q) = (30.5, 2.08, 0.24)$, and lead to the result $\omega = 0.8220(46)$. This suggests that the observed shift is not an artifact of a particular optimization scheme, but rather reflects an intrinsic structural feature of the resummed series for $\omega$ [see Eq.~(\ref{omegaepsilon})].

	To further probe this behavior, we specifically initialized the optimization in \textit{Method I} towards the regime of larger $b$, seeking alternative stable regions that might align with the broader plateaus observed at lower orders. This procedure identified a local optimum at $(b, \alpha, q) = (28.5, 2.04, 0.23)$, yielding $\omega = 0.8220(52)$. This value is in remarkable agreement with the result found by \textit{Method II}, confirming that the upward shift persists across distinct stability domains and is not due to isolated numerical instabilities.
	
	Overall, the seven-loop analysis suggests that the resummation of the $\omega$ series is less stable than for $\eta$ and $\nu$. The lack of overlap with lower-order estimates points to a slower convergence or a higher sensitivity of $\omega$ to the details of the resummation procedure. This may be related to the definition of $\omega$ as the derivative of the $\beta$-function at the stable fixed point.
	
\begin{table}[h]
	\centering
	\caption{Summary of the seven- and eight-loop estimates of the critical exponents for the 3D Ising model.}
	\label{tab:final_results}
	\begin{tabular}{lcccc}
		\hline\hline
		Exponent & Order & Method I & Method II & Conformal bootstrap \\ \hline
		$\eta$ & $\epsilon^{7}$ & $0.03613(24)$ & $0.03611(14)$ &   \\
		& $\epsilon^{8}$ & $0.03617(10)$ & $0.03616(7)$  & $0.036297612(48)$ \cite{bootstrap_3d} \\
		$\nu$  & $\epsilon^{7}$ & $0.62925(67)$ & $0.62914(54)$ & $0.62997097(12)$ \cite{bootstrap_3d}  \\
		$\omega$ & $\epsilon^{7}$ & $0.8257(42)^{(a)}$  &   & \\
		&   & $0.8220(52)^{(b)}$  & $0.8220(46)$ & $0.82951(61)$ \cite{reehorst2022}  \\
		\hline\hline
		\multicolumn{5}{l}{\footnotesize $^{(a)}$ Global minimum obtained with default initialization.}\\
		\multicolumn{5}{l}{\footnotesize $^{(b)}$ Result obtained by initializing the search towards the larger-$b$ regime.}
	\end{tabular}
\end{table}
	
	\subsection{Convergence trends of high-loop exponents} \label{convergence_trend_Ising}

\begin{figure}
		\begin{center}
				\subfloat[]{\includegraphics[width=.5\textwidth]{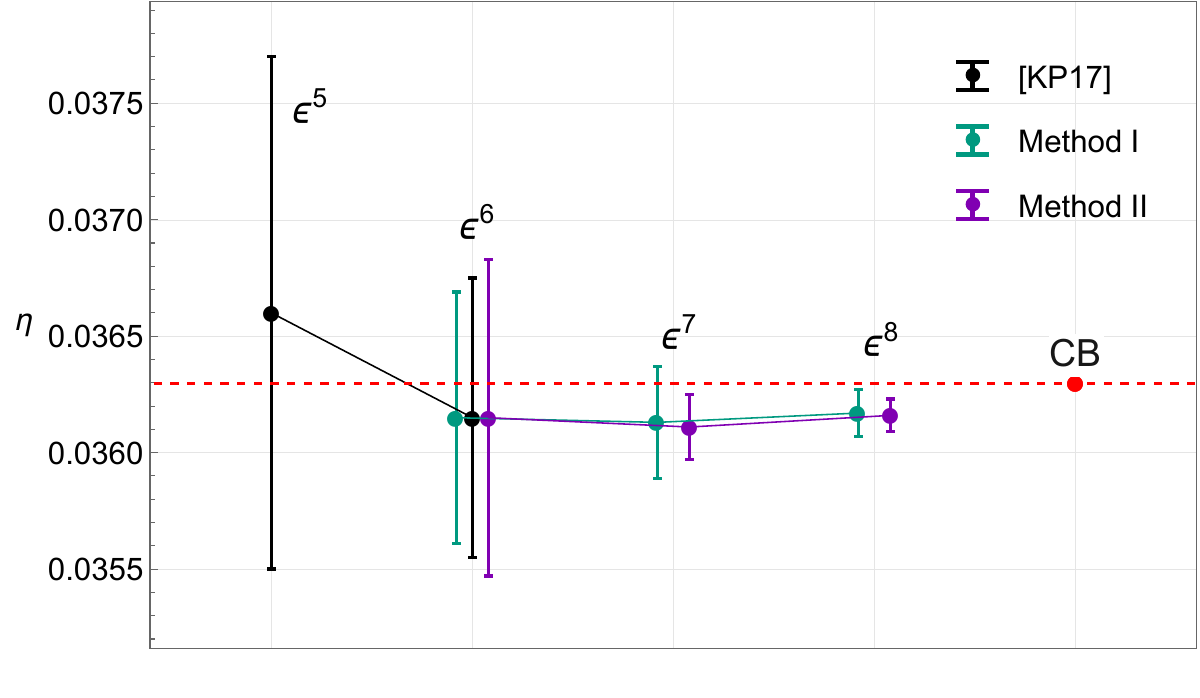}}\hfill
				\subfloat[]{\includegraphics[width=.5\textwidth]{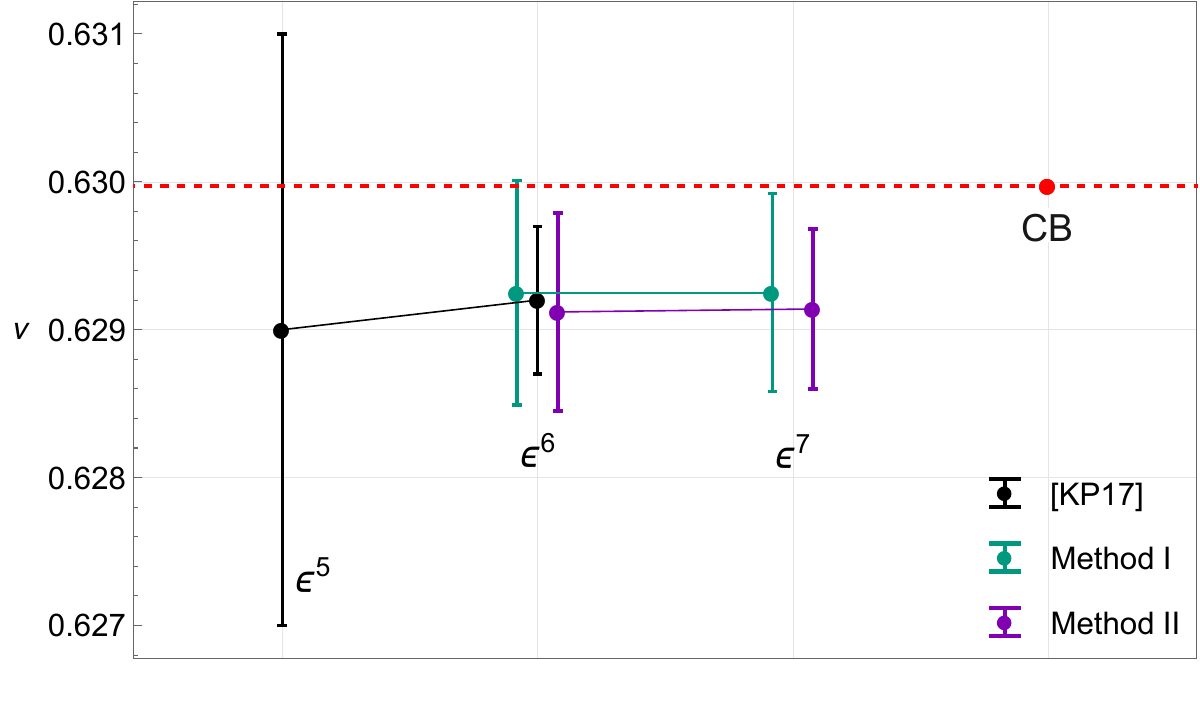}}	
		\end{center}
		\begin{center}
				\subfloat[]{\includegraphics[width=.5\textwidth]{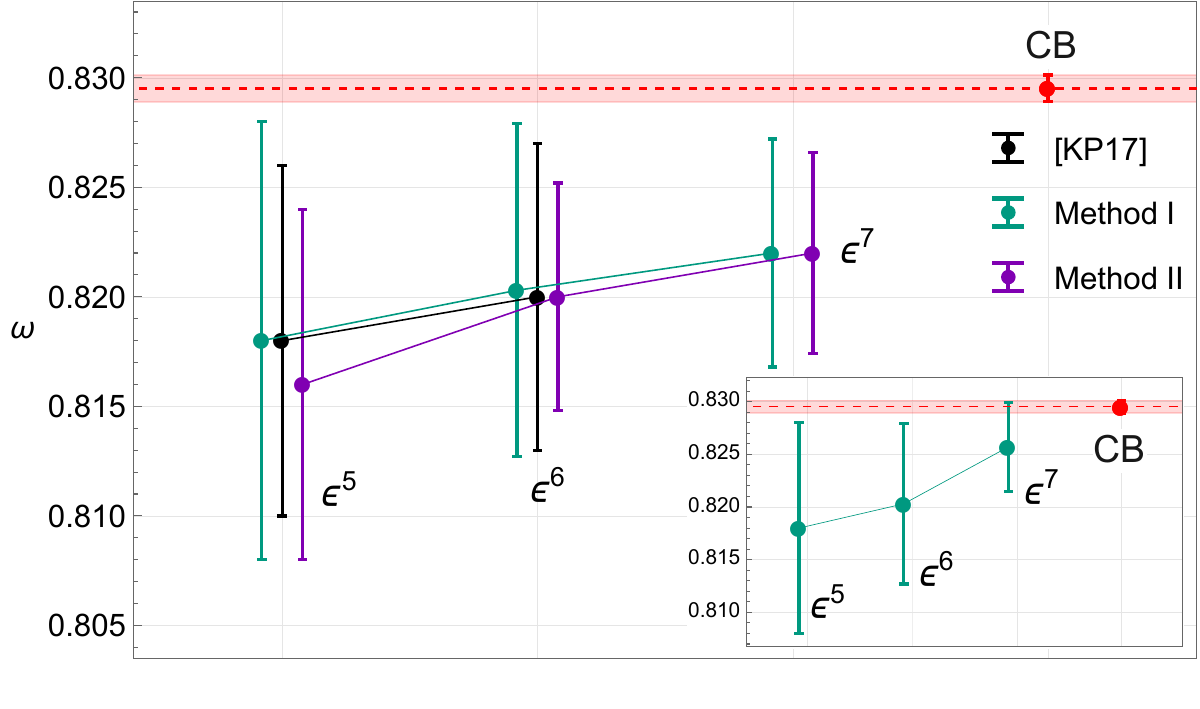}}
		\end{center}
		\caption{Convergence trends of the high-loop exponents $\eta$ [(a)], $\nu$ [(b)], and $\omega$ [(c)] compared to the CB values \cite{bootstrap_3d,kos2016,simmons_duffin2017,reehorst2022}. Results from Ref.~\cite{kompaniets2017} at five and six loops are shown in black ([KP17]), while our estimates obtained from \textit{Method I} (including its alternative solution for $\omega$ in the inset; see Section~\ref{final_critical_exponents} for a detailed discussion) and \textit{Method II} are shown in green and purple, respectively. The CB benchmark is indicated by red dashed lines and filled red symbols. \cite{bootstrap_3d, kos2016, simmons_duffin2017, reehorst2022}. }
		\label{final_trend}
	\end{figure}

Having established accurate estimates of the critical exponents and their uncertainties in the previous subsection (see Table~\ref{tab:final_results}), we now examine their evolution with increasing order of the $\epsilon$-expansion. The six-, seven-, and eight-loop Borel--conformal resummation results are compared with the CB estimates \cite{bootstrap_3d,kos2016,simmons_duffin2017, reehorst2022}, which provide the most accurate nonperturbative reference currently available. Fig.~\ref{final_trend} summarizes the evolution of the resummed exponents from five to eight loops. By monitoring both the evolution of the central values and the reduction of the uncertainty intervals, we assess whether the perturbative estimates converge toward the CB values. Although all three exponents exhibit increasing numerical stability with perturbative order, their convergence toward the CB results remains inconclusive.

For the anomalous dimension $\eta$, the five-loop estimate lies above the CB value \cite{bootstrap_3d}, although its uncertainty interval fully overlaps with it. At six loops, the central value shifts below the CB estimate while remaining compatible with it. The seven-loop analysis further reduces the uncertainty. Within \textit{Method I}, which closely follows the optimization procedure of Ref.~\cite{kompaniets2017}, the confidence interval still marginally overlaps the CB value, whereas \textit{Method II} yields a narrower interval lying entirely below it. At eight loops, both methods produce estimates whose central values and uncertainty intervals remain below the conformal bootstrap benchmark. The currently available perturbative results therefore do not provide clear evidence that the resummed sequence converges toward the CB limit.

From six loops onward, the central value of $\eta$ varies only weakly with perturbative order, whereas the estimated uncertainty decreases rapidly [Fig.~\ref{final_trend}(a)]. In \textit{Method I}, for example, the uncertainty is reduced by approximately a factor of two at each successive loop order between five and eight loops. The sequence therefore becomes increasingly stable without exhibiting a comparable drift toward the conformal bootstrap estimate \cite{bootstrap_3d}.

The behavior of the correlation-length exponent $\nu$ is even more subtle. The central estimates at five, six, and seven loops remain systematically below the conformal bootstrap value \cite{bootstrap_3d}. At five loops, the uncertainty interval is sufficiently broad to include the CB result. At six loops, the weighted analysis of Ref.~\cite{kompaniets2017}, corresponding essentially to our \textit{Method I}, yielded a relatively narrow interval lying below the bootstrap estimate. In the present work, a more conservative treatment of the parameter sensitivity leads to a broader uncertainty interval that again marginally overlaps the CB value, whereas \textit{Method II} produces a narrower interval entirely below it. The seven-loop central estimates remain almost unchanged with respect to the six-loop values, with only a moderate reduction of the uncertainties. If convergence toward the CB limit is indeed present, it therefore appears considerably slower than for $\eta$.

The correction-to-scaling exponent $\omega$ exhibits the most intricate behavior [Fig.~\ref{final_trend}(c)]. Both optimization strategies reproduce the published five- and six-loop results \cite{kompaniets2017}. At seven loops, the central value continues its gradual upward evolution. The distinctive feature at this order is the persistence of multiple stability regions in the optimization landscape. In the main panel of Fig.~\ref{final_trend}(c), \textit{Method I} is initialized toward the large-$b$ region and converges to a solution $\omega = 0.8220(52)$ that closely agrees with the \textit{Method II} estimate in both central value and uncertainty $\omega = 0.8220(46)$ (see Table~\ref{tab:final_results}). The inset shows the alternative solution obtained from the default initialization of \textit{Method I}, which converges to a distinct stability region at smaller $b$, yielding $\omega = 0.8257(42)$ at seven loops (see Table~\ref{tab:final_results}). Although these two solutions differ systematically, both remain above the corresponding five- and six-loop estimates, indicating that the upward evolution is robust with respect to the choice of stability region. The existence of multiple competing optima appears to be an intrinsic feature of the seven-loop resummation rather than an artifact of the optimization procedure. At present, however, the perturbative results do not provide convincing evidence for convergence toward the conformal bootstrap value.

Overall, the convergence behavior depends strongly on the critical exponent. While all three exponents display improved numerical stability as the perturbative order increases, the evolution of their central values toward the conformal bootstrap estimates is much less systematic. For $\eta$, the uncertainty decreases rapidly whereas the central value changes only marginally from six loops onward. For $\nu$, the central estimates remain consistently below the bootstrap value with only a weak loop-order dependence. For $\omega$, the upward evolution persists through seven loops and is accompanied by the appearance of multiple competing stability regions. Taken together, these observations demonstrate that increasing perturbative order leads to highly stable resummed estimates but does not yet establish unambiguous convergence of the perturbative renormalization-group expansion toward the conformal bootstrap predictions.

\subsection{Systematic uncertainties and asymptotic convergence}
\label{Discussion_convergence}

The results presented in the previous section reveal a systematic reduction of the estimated uncertainties with increasing perturbative order, accompanied by a much weaker evolution of the corresponding central values. This behavior is particularly pronounced for the exponent $\eta$, whose uncertainty decreases by nearly a factor of two at each successive loop order between five and eight loops in Method I, whereas the central estimate remains almost unchanged from six loops onward. Such behavior indicates that the internal numerical stability of the resummation procedure improves rapidly with perturbative order. At the same time, it raises the question of whether this stability can be interpreted as evidence for equally rapid convergence toward the exact critical exponents.

As explained in Section~\ref{Slow_conv}, a possible  mechanism was outlined by Sokal \cite{Sokal1994,Sokal1995} and Caselle, Pelissetto, and Vicari \cite{Caselle2001}, who argued that the renormalization-group functions are generally nonanalytic in the vicinity of the infrared fixed point. Under these conditions, resummation procedures based on analytic approximations may exhibit excellent apparent convergence while approaching the exact result only slowly. In other words, the stability of the resummed sequence may substantially overestimate its actual proximity to the exact value. Although the present calculations neither confirm nor rule out this scenario, the observed evolution of the critical exponents is qualitatively consistent with such a mechanism.

The comparison between the two resummation strategies employed in the present work provides additional support for this interpretation. The simpler two-parameter optimization over the Borel-Leroy parameter $b$ and the strong-coupling parameter $\alpha$ (Sections~\ref{two_param_sensivity} and Appendix~\ref{app__two_parameter}) produces comparatively broad confidence intervals, but its central estimates evolve systematically toward the conformal bootstrap values as the perturbative order increases (Table~\ref{tab:appendix_results}). By contrast, introducing the homographic transformation parameter $q$ substantially reduces the estimated uncertainties (Table~\ref{tab:final_results}) while simultaneously making the central values almost insensitive to the perturbative order.

The homographic transformation is introduced to improve the convergence of the Borel integral by moving the singularities of the transformed Borel function away from the physical integration contour. Combined with the PMS and PFAC optimization criteria, it efficiently suppresses the residual dependence of the resummed estimates on the free resummation parameters. The resulting reduction of the uncertainty intervals demonstrates that the optimization procedure successfully stabilizes the truncated perturbative series. Nevertheless, comparison with the two-parameter analysis suggests that part of this stabilization may originate from the optimization procedure itself rather than exclusively from the convergence of the perturbative expansion.

These observations point to two possible sources of systematic uncertainty. The first is intrinsic and arises from the nonanalytic structure of the renormalization-group functions near the fixed point \cite{Caselle2001}. The second is methodological and reflects the tendency of increasingly sophisticated optimization procedures to suppress the apparent parameter dependence of truncated perturbative series. Although these two effects cannot be disentangled within the present analysis, they both act in the same direction: they reduce the apparent spread of the resummed estimates without necessarily producing a comparable reduction of the remaining systematic deviation from the exact result.

More generally, the markedly different convergence patterns observed for $\eta$, $\nu$, and $\omega$ indicate that the asymptotic behavior of the resummed perturbative series is strongly observable dependent. This conclusion is consistent with Sokal's discussion of nonanalytic corrections to scaling (see Section~\ref{Slow_conv}), according to which different observables may exhibit substantially different convergence rates even when derived from the same renormalization-group fixed point. It also remains possible that certain nonperturbative contributions are only partially captured within the present resummation framework, thereby limiting the apparent convergence even at the highest perturbative orders currently available.

Taken together, these observations suggest that the uncertainty associated with high-order resummation is not completely characterized by the residual dependence on the resummation parameters alone. While the two optimization strategies employed here provide important consistency checks, they may still underestimate the full systematic uncertainty of the procedure. Consequently, increasingly narrow confidence intervals should not automatically be interpreted as evidence that the exact asymptotic limit has been reached.

Further progress will require both longer perturbative series and a better understanding of the mathematical structure of the renormalization-group functions in the vicinity of the fixed point. Such developments should help clarify whether the slow evolution observed here reflects the intrinsic asymptotic behavior of perturbative field theory or residual limitations of the present resummation techniques.

	\section{Conclusion} \label{conclusion}

In this work, we investigated the critical exponents $\eta$, $\nu$, and $\omega$ of the three-dimensional Ising universality class using the highest-order perturbative $\epsilon$-expansion currently available, extended to seven and eight loops in Refs.~\cite{schnetz2018,schnetz2023}. The divergent perturbative series were resummed by means of the Borel--conformal mapping  technique, supplemented by a homographic transformation of the expansion parameter $\epsilon$ to improve the convergence properties of the Borel integral.

To obtain reliable estimates and assess the associated systematic uncertainties, we developed and applied two complementary optimization strategies. \textit{Method~I} combines a {\em global}  sensitivity criterion with the principle of fastest apparent convergence~\cite{kompaniets2017}, thereby probing the maximal variation of the resummed estimates over the optimization domain. \textit{Method~II} employs a complementary  {\em local} stability analysis to identify stationary plateaus in the three-dimensional parameter space $(b,\alpha,q)$. Both methods were first benchmarked against the published six-loop analysis of Ref.~\cite{kompaniets2017}, successfully reproducing its results and thereby validating the proposed resummation framework. The critical exponent estimates obtained in this work are displayed in Fig.~\ref{final_trend} and Table~\ref{tab:final_results}.
For convenience, we summarize our final estimates that are obtained via Method~I
at six-, seven- and eight-loop orders in Table~\ref{tab:final_table}. In the tables above, uncertainties are quoted with two significant digits. For the final estimates, we instead adopt the rounding convention recommended by the Particle Data Group~\cite{PRD_group}: the uncertainty is quoted with two significant digits if its first two significant digits form a number below 36, and with one significant digit otherwise. The central value is rounded to the same decimal place as the uncertainty.

\begin{table}[h!]
	\centering
	\caption{Our final estimates for the critical exponents obtained at successive orders of the resummed $\epsilon$-expansion, compared with the conformal bootstrap (CB) results used as a benchmark \cite{bootstrap_3d, kos2016, simmons_duffin2017, reehorst2022}.}
	\begin{tabular}{c|c|l}
		\hline
		\hline
		Exponent & Order & Estimates \\
		\hline
		$\eta$ & $\epsilon^{6}$ & $0.0362(5)$ \\
		& $\epsilon^{7}$  & $0.03613(24)$      \\
		& $\epsilon^{8}$  & $0.03617(10)$       \\
		& CB & $0.036297612(48)$ \\
		\hline
		$\nu$  & $\epsilon^{6}$  & $0.6293(8)$   \\
		& $\epsilon^{7}$  & $0.6293(7)$          \\
		& CB &$0.62997097(12)$ \\
		\hline
		$\omega$ & $\epsilon^{6}$  & $0.820(8)$ \\
		& $\epsilon^{7}$ & $0.822(5)$           \\
		& CB & $0.82951(61)$ \\
		\hline
		\hline
	\end{tabular}
	\label{tab:final_table}
\end{table}

For $\eta$, $\nu$, and $\omega$, the central values obtained from our optimal resummations show only a weak dependence on the perturbative order. Between the $\epsilon^6$ and $\epsilon^8$ approximations, the variation amounts to only about $0.1\%$ for $\eta$, while it is essentially negligible for $\nu$ between $\epsilon^6$ and $\epsilon^7$ and reaches $0.25\%$ for $\omega$. Surprisingly, the behavior of the corresponding error bars is markedly different and strongly depends on the exponent considered. For $\eta$, the uncertainty is reduced by approximately a factor of two at each successive order between $\epsilon^6$ and $\epsilon^8$, whereas it decreases by only $12.5\%$ and $37.5\%$ for $\nu$ and $\omega$, respectively, between $\epsilon^6$ and $\epsilon^7$. In absolute terms, the overall accuracy of our estimates remains high: the deviations of the central values from the benchmark conformal bootstrap estimates are only $0.36\%$ for $\eta$, $0.11\%$ for $\nu$, and $0.91\%$ for $\omega$.

The main difficulty with these results is that the remarkable stability of the central values, combined with the steady reduction of the error bars, eventually leads the highest-order estimates for all three exponents to become slightly incompatible with the CB results. Even more strikingly, at order $\epsilon^6$, the estimate of $\omega$ is already incompatible with the CB value, while that of $\nu$ is only marginally compatible within the quoted uncertainty.

The issue is therefore not the absolute accuracy of the estimates, but rather the convergence of the resummed perturbative series. At low perturbative orders, the resummed results become progressively closer to the CB values. However, this convergence slows down significantly from the six-loop order onward. One of the main outcomes of the present work is that an increasing perturbative order leads to rapidly improving numerical stability, but does not yet provide unambiguous evidence for convergence toward the conformal bootstrap values. Of course, it is not possible to draw definitive conclusions regarding the convergence of the perturbative series from the available results alone. Nevertheless, it is already apparent from our findings that the convergence to the exact values, if it occurs, is much slower than naively expected. Taking into account the  arguments concerning the nonanalyticity of RG functions in the vicinity of a fixed point \cite{Sokal1994, Sokal1995, Caselle2001, pellisetto2002}, we suggest that the convergence toward the conformal bootstrap benchmark proceeds at an anomalously slow rate. Moreover, one cannot exclude a potential underestimation of the error bars computed in this work. This could stem from the inherent properties of the homographic transformation itself, as well as from the sophisticated optimization procedure built upon it (see Section \ref{homographic_section}).
As explained in Section~\ref{Slow_conv}, this slow convergence could be the first signal of the impact of nonanalytic terms at $g^*$ of the $\beta(g)$ function. We think that a first step towards a full answer to this question is to broaden our horizons and to study the O($n$) models to see whether the Ising universality class is an exception or the rule.

	\section*{Acknowledgments}
	The work was supported by the National Research Foundation of Ukraine Project No.~2023.03/0099
	``Criticality of complex systems: fundamental aspects and applications''. B.D. thanks R. Guida for discussions. We thank Giacomo Gori, Giorgio Parisi, Slava Rychkov, and Oliver Schnetz for their comments and enlightening discussions during the preparation of the final version of this paper.
	
	\section*{Data availability}
	
	There are no publicly available research data or software supporting this manuscript. Requests for further information or data should be sent to the authors.
	
	
	\bibliographystyle{apsrev4-2}
	\bibliography{references}

	
	\appendix
	
	
	\newpage
	\section{Critical exponents within the two-parameter resummation scheme}
	\label{app__two_parameter}

This Appendix presents complementary results from the stability-based resummation procedure detailed in Section~\ref{two_param_sensivity}. We report critical exponent estimates across successive loop orders alongside their corresponding stability landscapes in the $(b, \alpha)$ parameter space.

\begin{figure}[h!] 
			\subfloat[]{\includegraphics[width=0.33\textwidth]{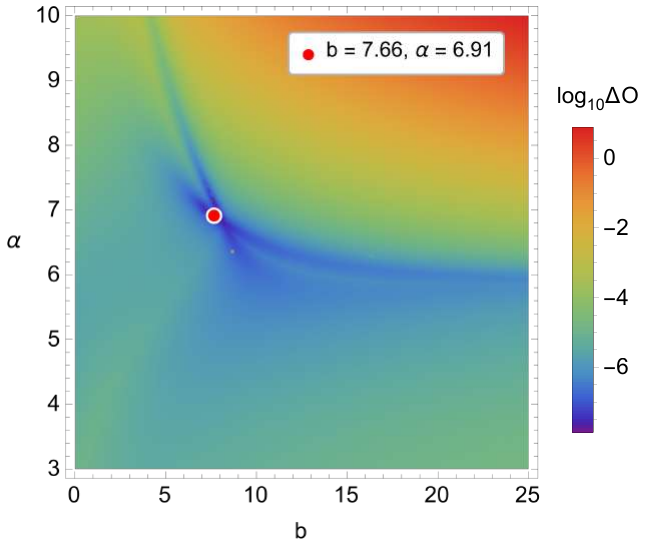}}\hfill
			\subfloat[]{\includegraphics[width=0.33\textwidth]{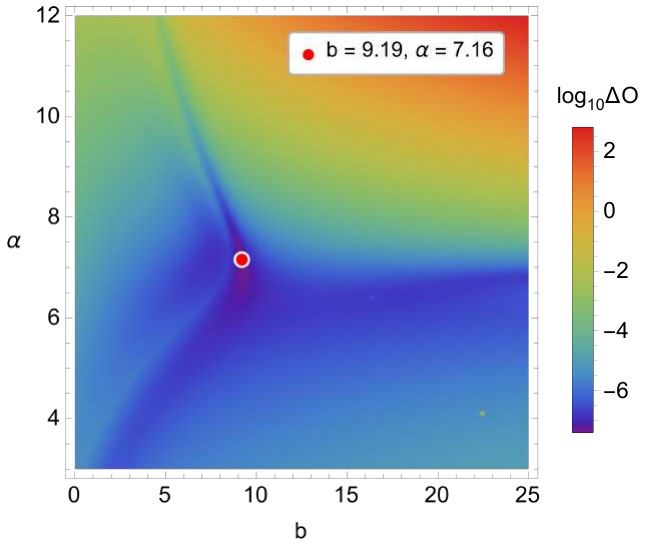}}\hfill
			\subfloat[]{\includegraphics[width=0.33\textwidth]{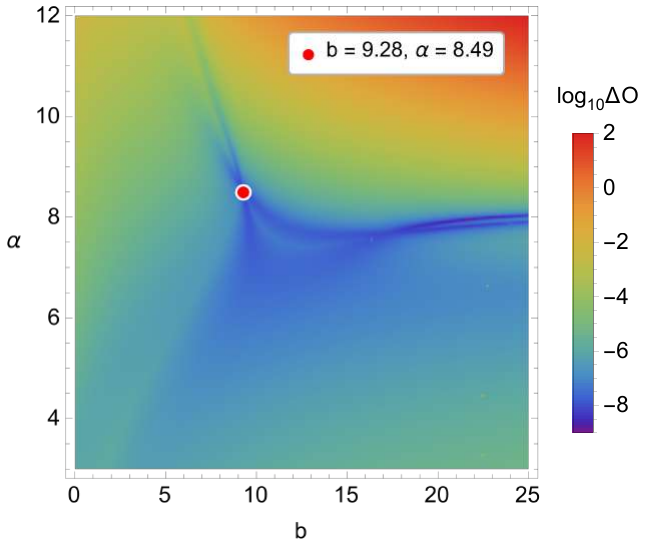}}\hfill
			\subfloat[]{\includegraphics[width=0.33\textwidth]{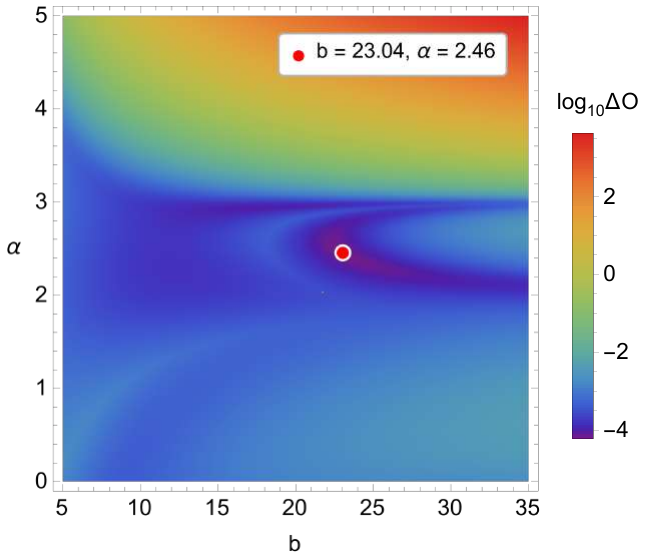}}\hfill
			\subfloat[]{\includegraphics[width=0.33\textwidth]{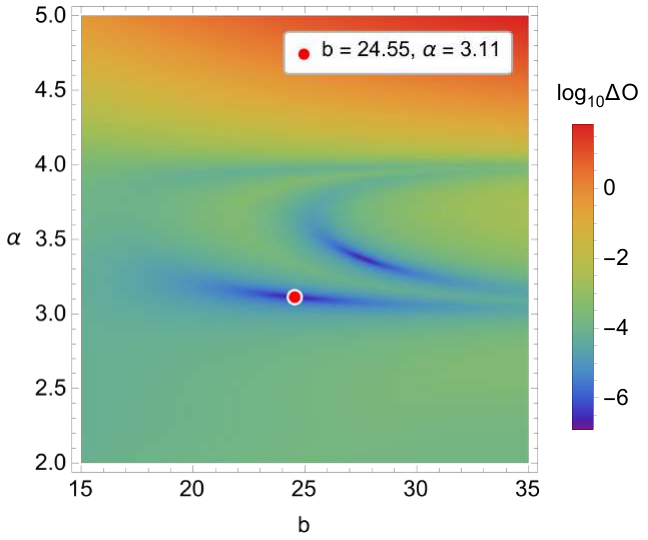}}\hfill
			\subfloat[]{\includegraphics[width=0.33\textwidth]{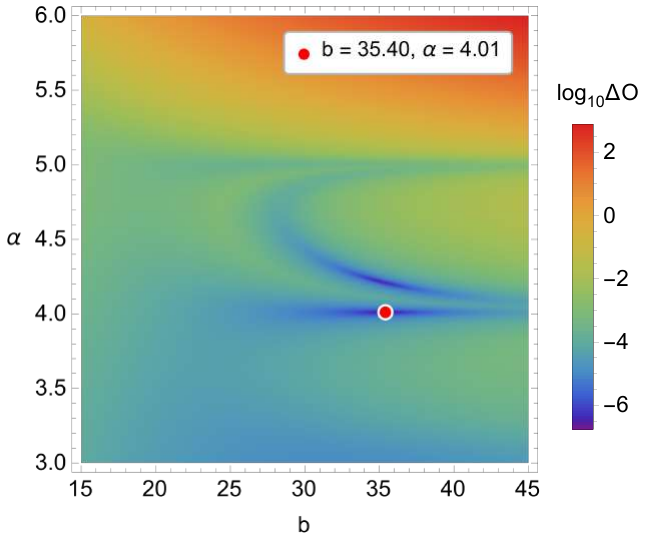}}\hfill
			\subfloat[]{\includegraphics[width=0.33\textwidth]{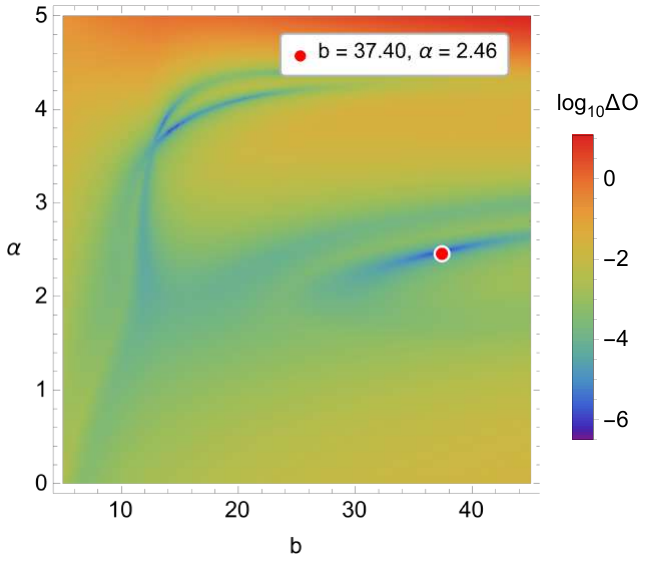}}\hfill
			\subfloat[]{\includegraphics[width=0.33\textwidth]{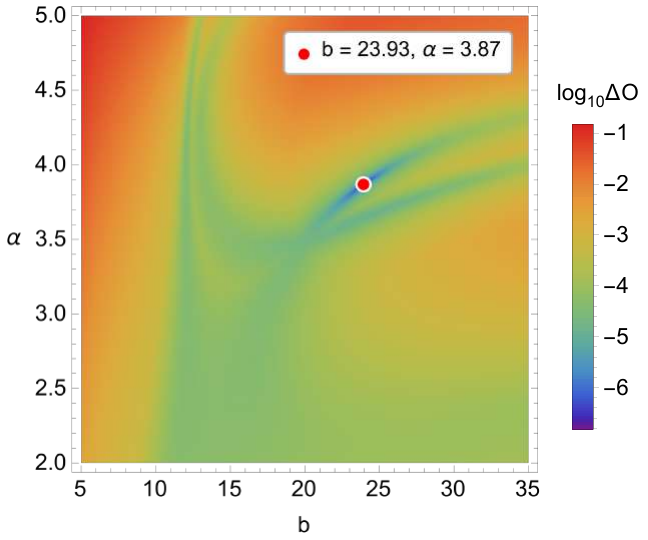}}\hfill
			\subfloat[]{\includegraphics[width=0.33\textwidth]{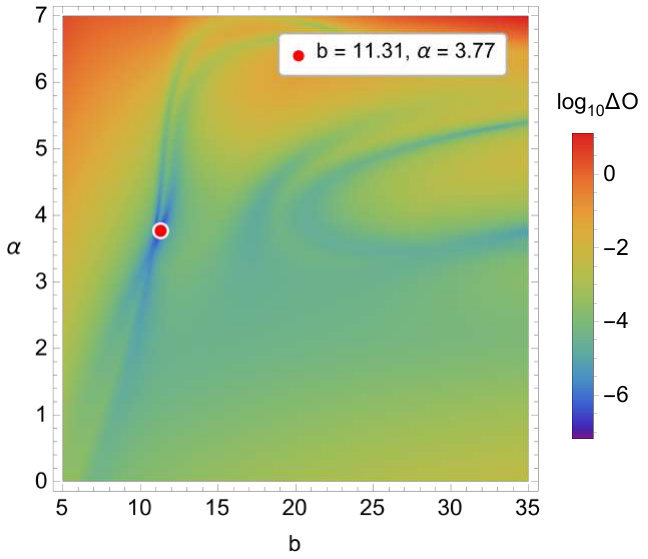}}\
		\centering
		\caption{Density plots of $\log_{10} \Delta \mathcal{O}$ [Eq.~(\ref{func_serone})] in the $(b,\alpha)$ parameter space. The rows illustrate different critical exponents across successive loop orders: the first row shows $\eta$ at six- [(a)], seven- [(b)], and eight-loop [(c)] orders; the second row shows $\nu$ at five- [(d)], six- [(e)], and seven-loop [(f)] orders; and the third row shows the correction-to-scaling exponent $\omega$ at five- [(g)], six- [(h)], and seven-loop [(i)] orders. The color scale indicates the degree of stability, with darker regions corresponding to smaller values of $\Delta \mathcal{O}$. Red dots mark the optimal points $(b_{\mathrm{opt}}, \alpha_{\mathrm{opt}})$ determined from the combined PMS and PFAC criteria.}
				\label{two_param_density}
	\end{figure}

To test the flexibility of this two-parameter scheme, we varied the perturbation windows, $\Delta b$ and $\Delta \alpha$, based on the specific nature and sensitivity of each resummation parameter. To visualize these stability properties, we map the functional $\Delta \mathcal{O}$ [Eq.~(\ref{func_serone})] in the $(b, \alpha)$ parameter space for several loop orders; the resulting profiles are displayed in Fig.~\ref{two_param_density}. As seen from these plots, the functional $\Delta \mathcal{O}$ exhibits a wide, extended plateau along the $b$-axis. Consequently, our numerical minimization consistently drifts toward the lower bound of the search grid, making $\Delta b = 5.0$ the optimal baseline choice across all orders to properly capture this landscape feature without artificially inflating the error.

In contrast, the resummation procedure proves highly sensitive to the parameter $\alpha$, as evidenced by a much narrower and sharper stability valley. To account for this, the window $\Delta \alpha$ was adjusted dynamically across different exponents and loop orders to accurately trace the region of minimal mutual sensitivity. The resulting parameters and their corresponding critical exponent estimates are summarized in Table~\ref{tab:appendix_results}.

\begin{table}[h!]
	\centering
	\caption{Comparison of critical exponent estimates obtained using fixed stability windows ($\Delta b = 5.0$, $\Delta \alpha = 0.5$) 
	versus the optimized window scheme, where $\Delta \alpha$ is varied dynamically at a fixed $\Delta b = 5.0$.}
	\begin{tabular}{c|c|c|c|c}
		\hline
		\hline
		 &  & Fixed window & \multicolumn{2}{c}{Optimized window} \\
		Exponent & Order & Estimate ($\Delta \alpha = 0.5$) & $\Delta \alpha$ & Estimate \\
		\hline
		$\eta$ & $\epsilon^{6} $ & $0.0356(10)$  & 1.25 & $0.03555(74)$ \\
		       & $\epsilon^{7}$  & $0.03596(76)$ & 1.5  & $0.03596(55)$ \\
		       & $\epsilon^{8}$  & $0.03623(87)$ & 1.5  & $0.03623(53)$ \\
		\hline
		$\nu$  & $\epsilon^{5}$  & $0.6279(83)$  & 0.5  & $0.6279(83)$ \\ 
		       & $\epsilon^{6}$  & $0.6266(43)$  & 0.75 & $0.6266(42)$ \\
		       & $\epsilon^{7}$  & $0.6295(36)$  & 0.75 & $0.6295(35)$ \\
		\hline
		$\omega$ & $\epsilon^{5}$ & $0.789(19)$  & 0.75 & $0.789(19)$ \\
		       & $\epsilon^{6}$  & $0.817(14)$   & 0.5  & $0.817(14)$ \\
		       & $\epsilon^{7}$  & $0.821(21)$   & 1.5  & $0.821(12)$ \\ 
		\hline
		\hline
	\end{tabular}
	\label{tab:appendix_results}
\end{table}
	
	A closer inspection of Table~\ref{tab:appendix_results} reveals varied trends in the error behavior. In the fixed-window scheme (column 3), one observes an increase in the uncertainty for certain consecutive orders, such as between the seven-loop and eight-loop orders for $\eta$, or between the six-loop and seven-loop orders for $\omega$. While this might initially seem counterintuitive, it reflects the conservative nature of the fixed-window error estimate, which transparently captures the shifting structure and sign fluctuations of the higher-order perturbation coefficients.
	
	Conversely, allowing the algorithm to optimize the window size (column 5) frequently yields narrower error bars. However, this local reduction in uncertainty does not always translate into better convergence properties. For instance, for $\eta$ at the $\epsilon^6$ order, the optimized window produces an overly optimistic error bound. A similar behavior is apparent for $\nu$ between the $\epsilon^6$ and $\epsilon^7$ orders, as well as for $\omega$ when comparing the $\epsilon^5$ result with successive orders. This indicates that while the adaptive selection of $\Delta \alpha$ minimizes immediate parameter sensitivity, it can lead to an underestimation of the systematic error.
	
	Consequently, varying the stability windows within the two-parameter scheme yields inconclusive results. This lack of systematic improvement underscores the need to expand the parameter space. To achieve better convergence and reliable error margins, we introduce a third parameter via the homographic transformation detailed in Section~\ref{homographic_section}.

	\newpage

	\newpage
	\section{Critical exponents within the three-parameter resummation scheme}
	\label{figures_app}

\begin{figure}[h!] 
			\subfloat[]{\includegraphics[width=0.34\textwidth]{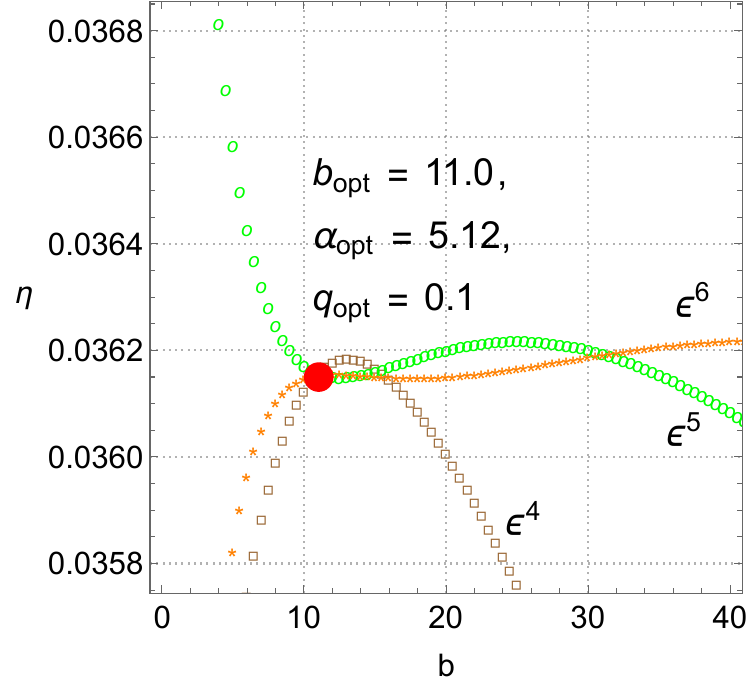}}\hfill
			\subfloat[]{\includegraphics[width=0.34\textwidth]{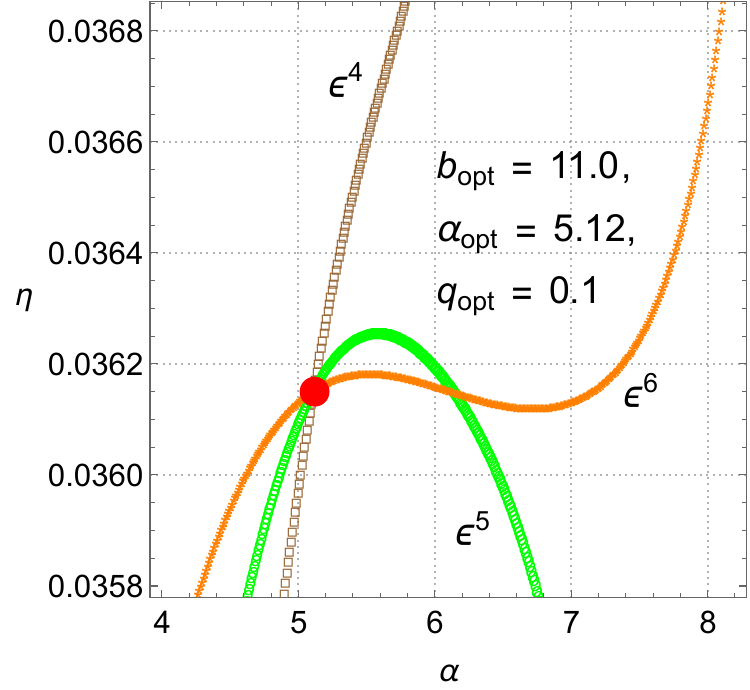}}\hfill
			\subfloat[]{\includegraphics[width=0.32\textwidth]{Figures/eaeta6.pdf}}\
		\centering
		\caption{Dependence of the critical exponent $\eta$ on the resummation parameters $b$ (a) and $\alpha$ (b) around the optimal point $(b, \alpha, q) = (11.0, 5.12, 0.1)$ at six-loop order. Panel (c) displays the normalized error $E/\overline{E}$ distribution across the parameter space. The global minimum (GM, red cross) and weighted mean (WM, dashed line) are shown with uncertainties of $3\overline{E}$ and $\overline{E} + 2\sigma_w$, respectively, accounting for both local stability and the statistical spread of the candidate set within the $E \leq 3\overline{E}$ region [see Eq.~(\ref{weight})]. Our final estimates, $\eta = 0.03615(54)$ obtained via {\it Method I} and $\eta = 0.03615(68)$ via {\it Method II}, are consistent with the value reported in \cite{kompaniets2017}.}
		\label{eta6__n1}
	\end{figure}

\begin{figure}[h!] 
		\subfloat[]{\includegraphics[width=0.34\textwidth]{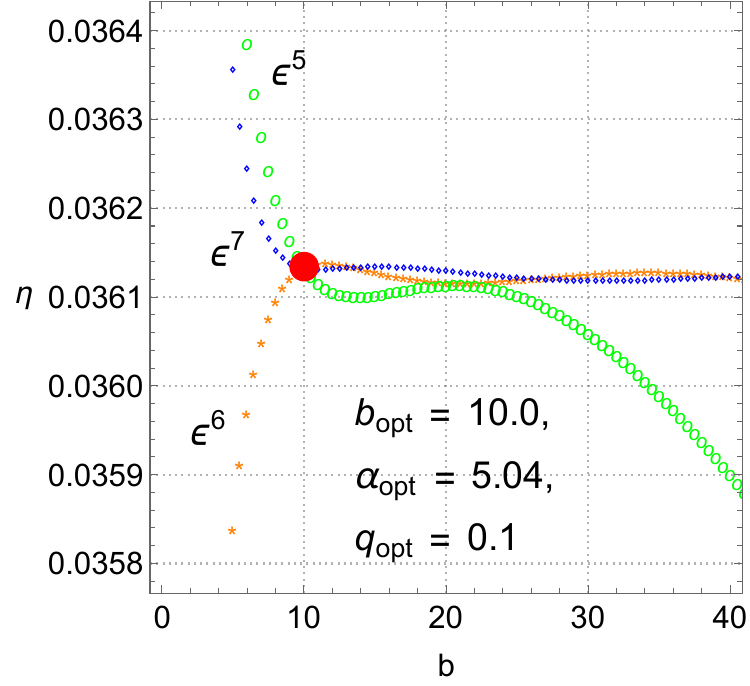}}\hfill
		\subfloat[]{\includegraphics[width=0.34\textwidth]{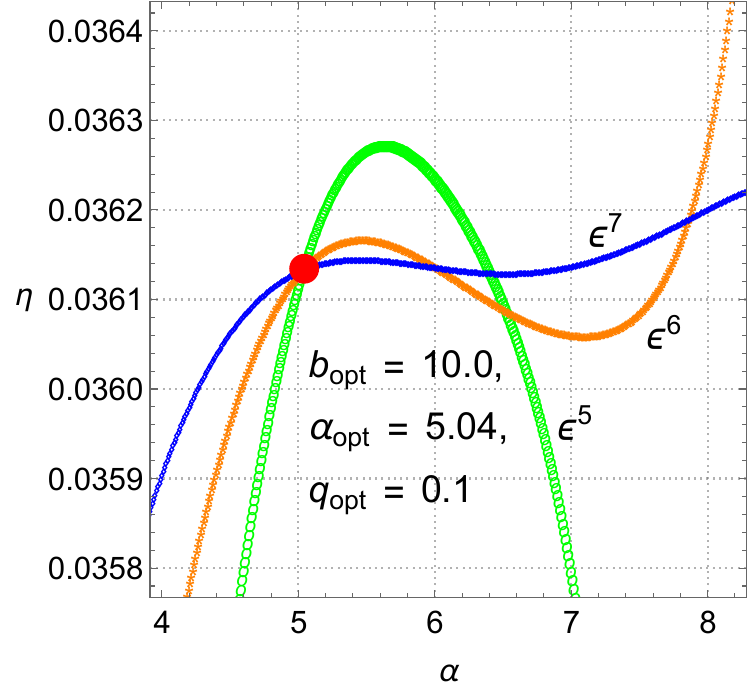}}\hfill
		\subfloat[]{\includegraphics[width=0.32\textwidth]{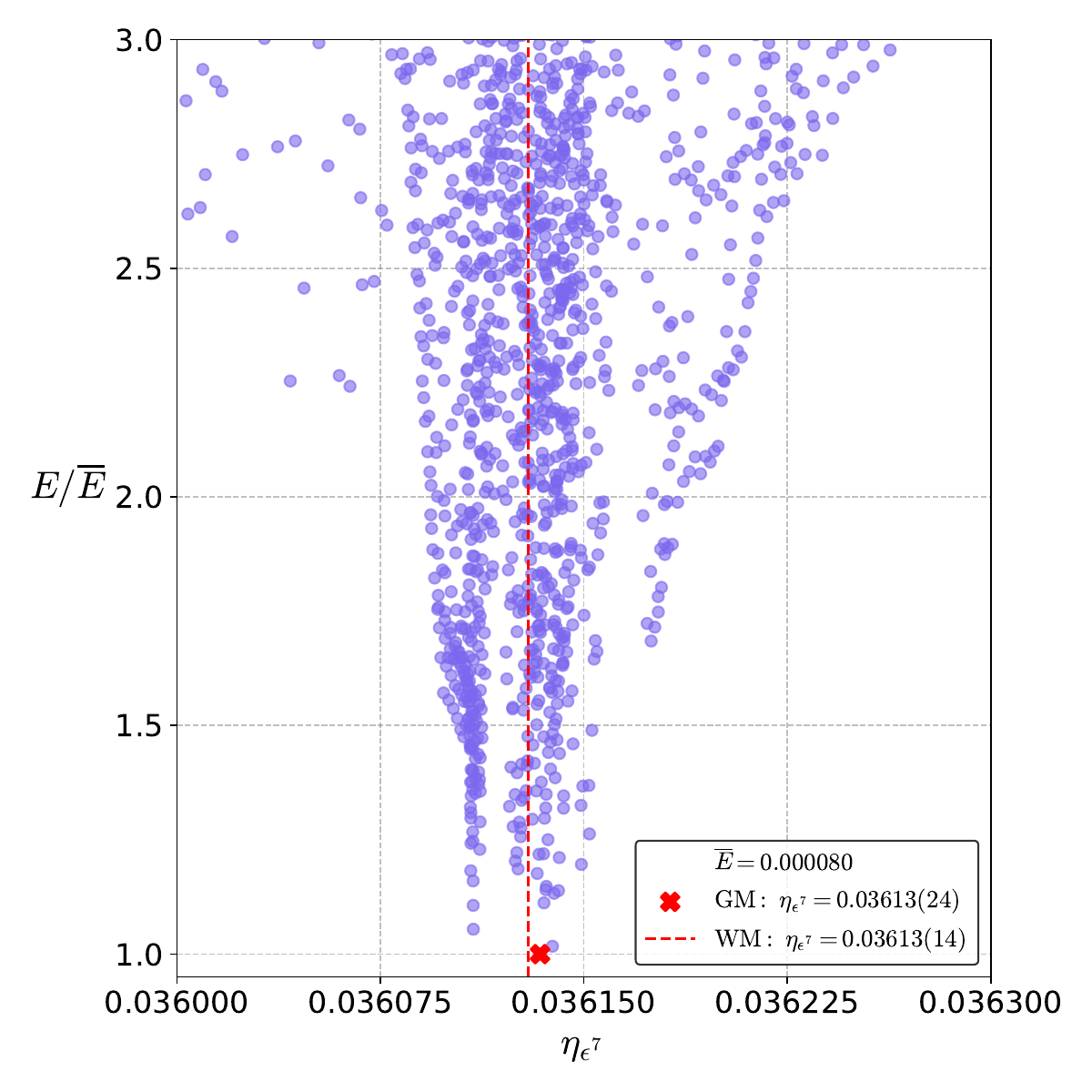}}\hfill
		\subfloat[]{\includegraphics[width=0.23\textwidth]{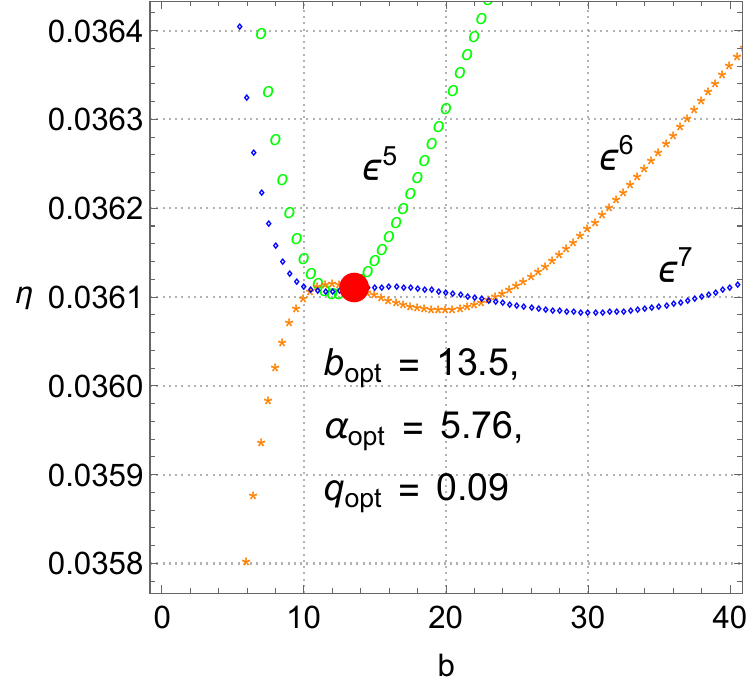}}\hfill
		\subfloat[]{\includegraphics[width=0.23\textwidth]{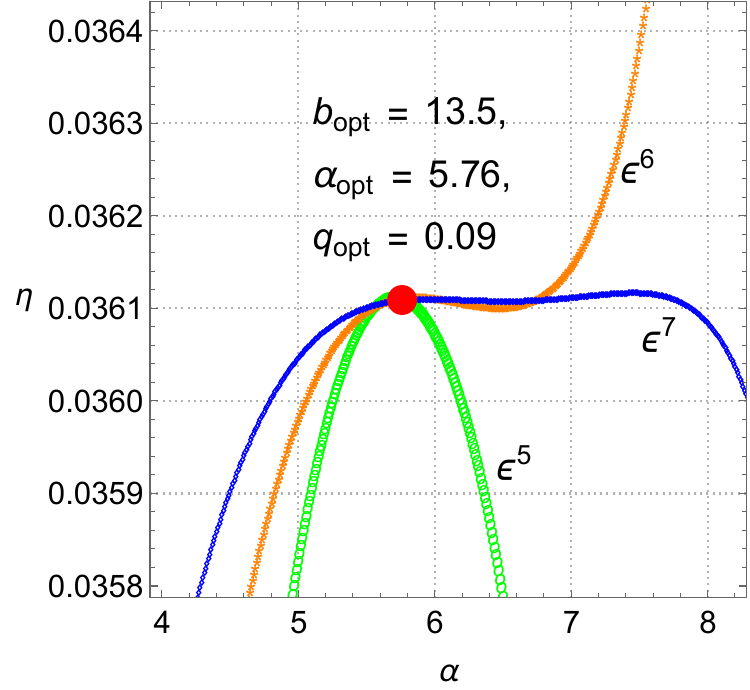}}\hfill
		\subfloat[]{\includegraphics[width=0.26\textwidth]{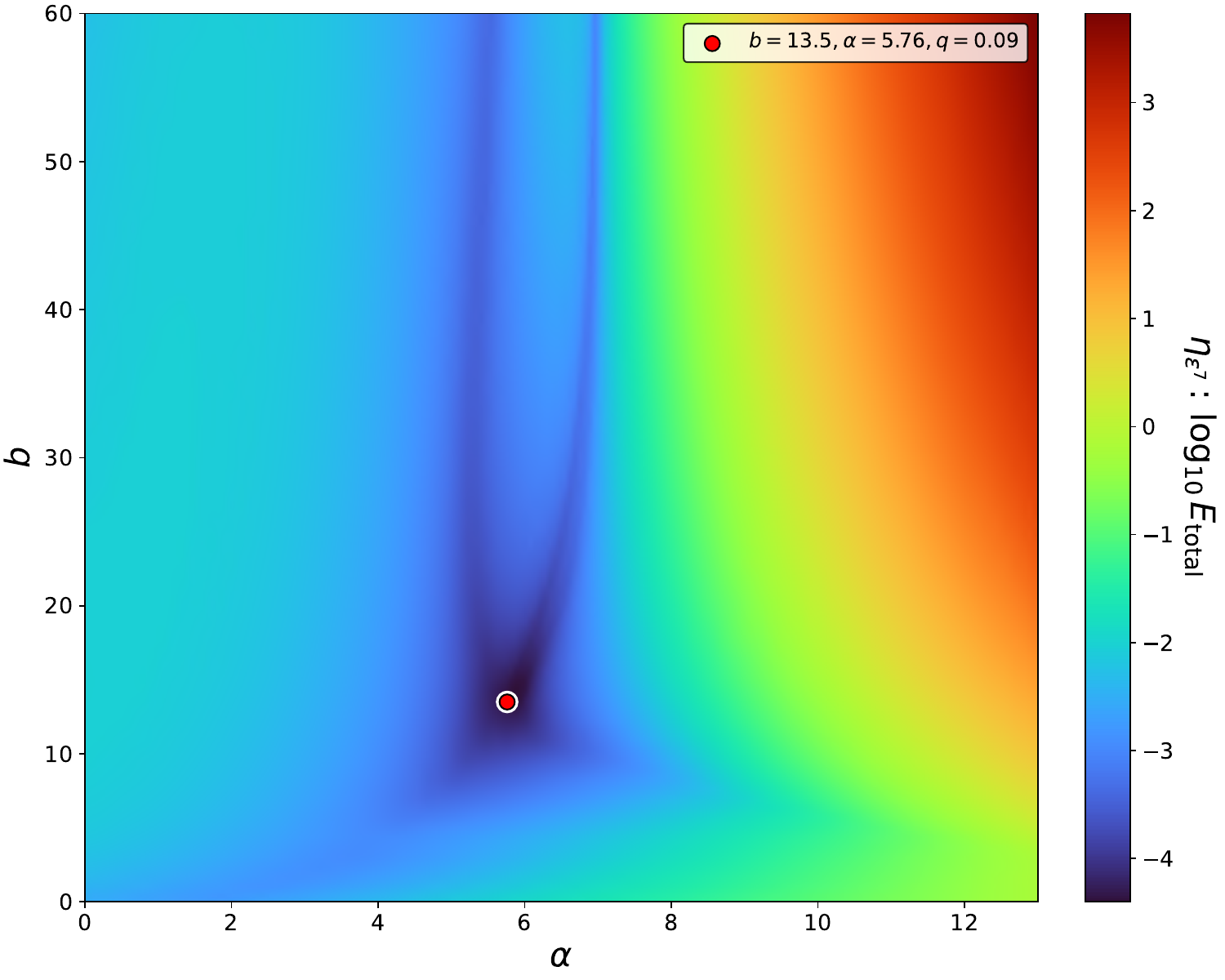}}\hfill
		\subfloat[]{\includegraphics[width=0.26\textwidth]{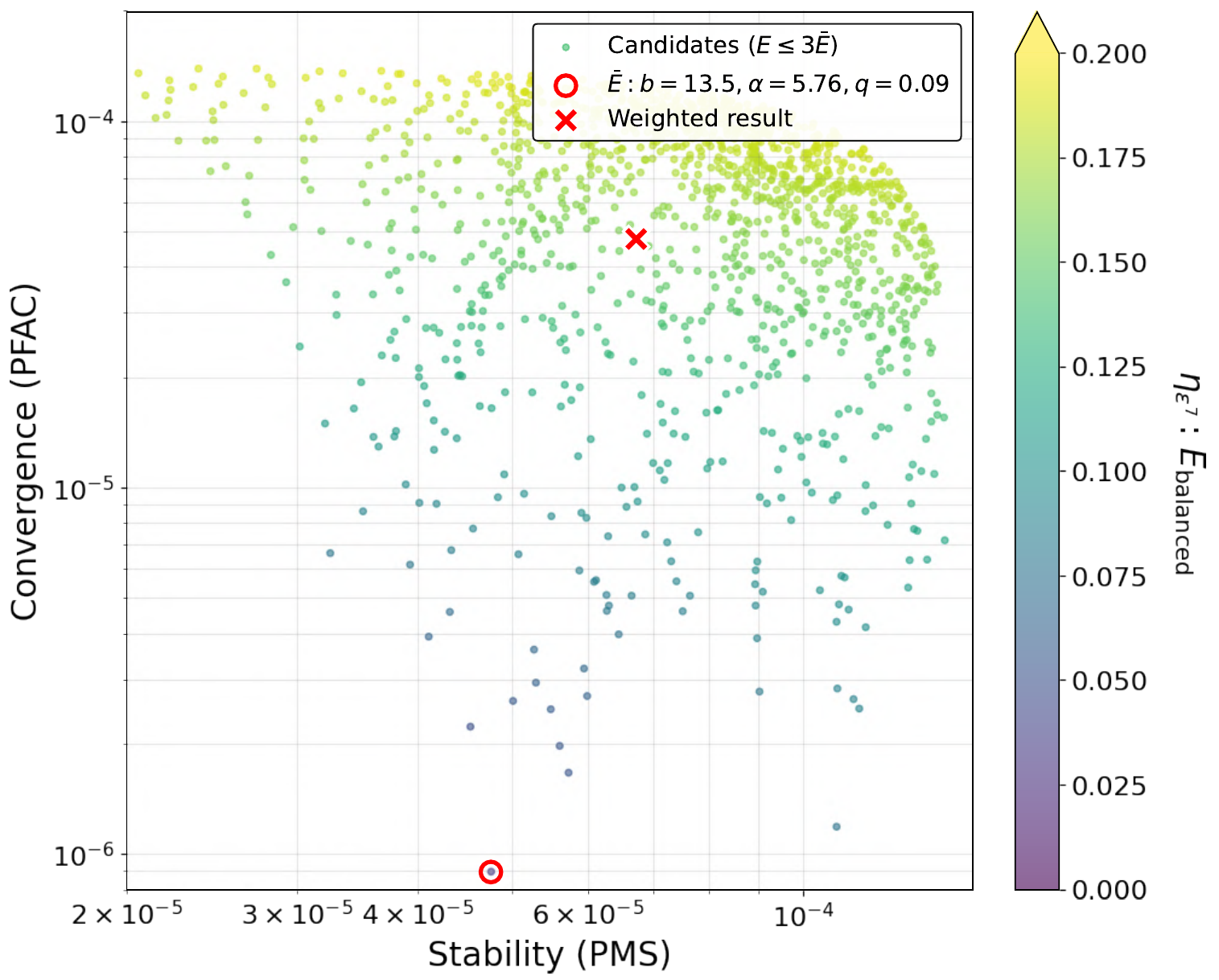}}
		\centering
		\caption{Dependence of the critical exponent $\eta$ on the resummation parameters $b$ [(a), (d)] and $\alpha$ [(b), (e)] at seven-loop order. The top row corresponds to the optimal point $(b, \alpha, q) = (10.0, 5.04, 0.10)$ ({\it Method I}, Section~\ref{error_estimations}), and the bottom row corresponds to the optimal point $(b, \alpha, q) = (13.5, 5.76, 0.09)$ ({\it Method II}, Section~\ref{rms_error_estimations}). Panel (c) displays the normalized error $E/\overline{E}$ distribution across the parameter space. The global minimum (GM, red cross) and weighted mean (WM, dashed line) are shown with uncertainties of $3\overline{E}$ and $\overline{E} + 2\sigma_w$, respectively, accounting for both local stability and the statistical spread of the candidate set within the $E \leq 3\overline{E}$ region [see Eq.~(\ref{weight})]. Panels (f) and (g) show the error landscape in the $(b, \alpha)$ plane at the optimal $q$ value and the distribution of calculated points in the plane of stability ($E_{\mathrm{pms}}$) versus convergence ($E_{\mathrm{pfac}}$), respectively. Our final estimates are $\eta = 0.03613(24)$ via {\it Method I} and $\eta = 0.03611(14)$ via {\it Method II}.}
		\label{eta7__n1}
	\end{figure}

\begin{figure}[h!] 
		\subfloat[]{\includegraphics[width=0.34\textwidth]{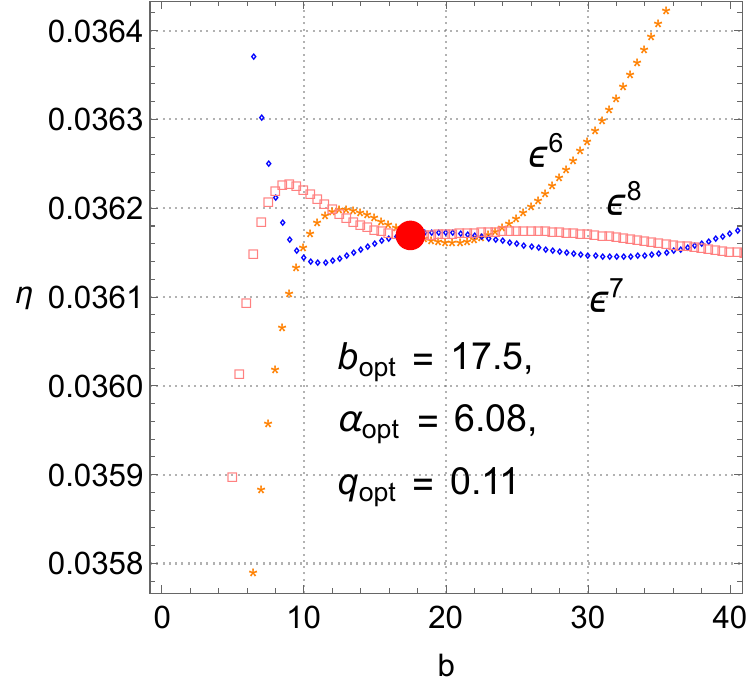}}\hfill
		\subfloat[]{\includegraphics[width=0.34\textwidth]{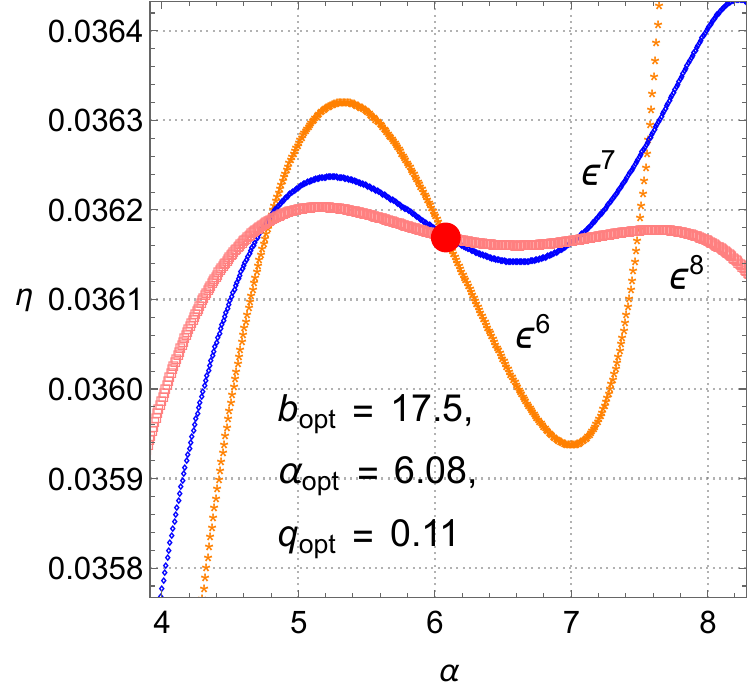}}\hfill
		\subfloat[]{\includegraphics[width=0.32\textwidth]{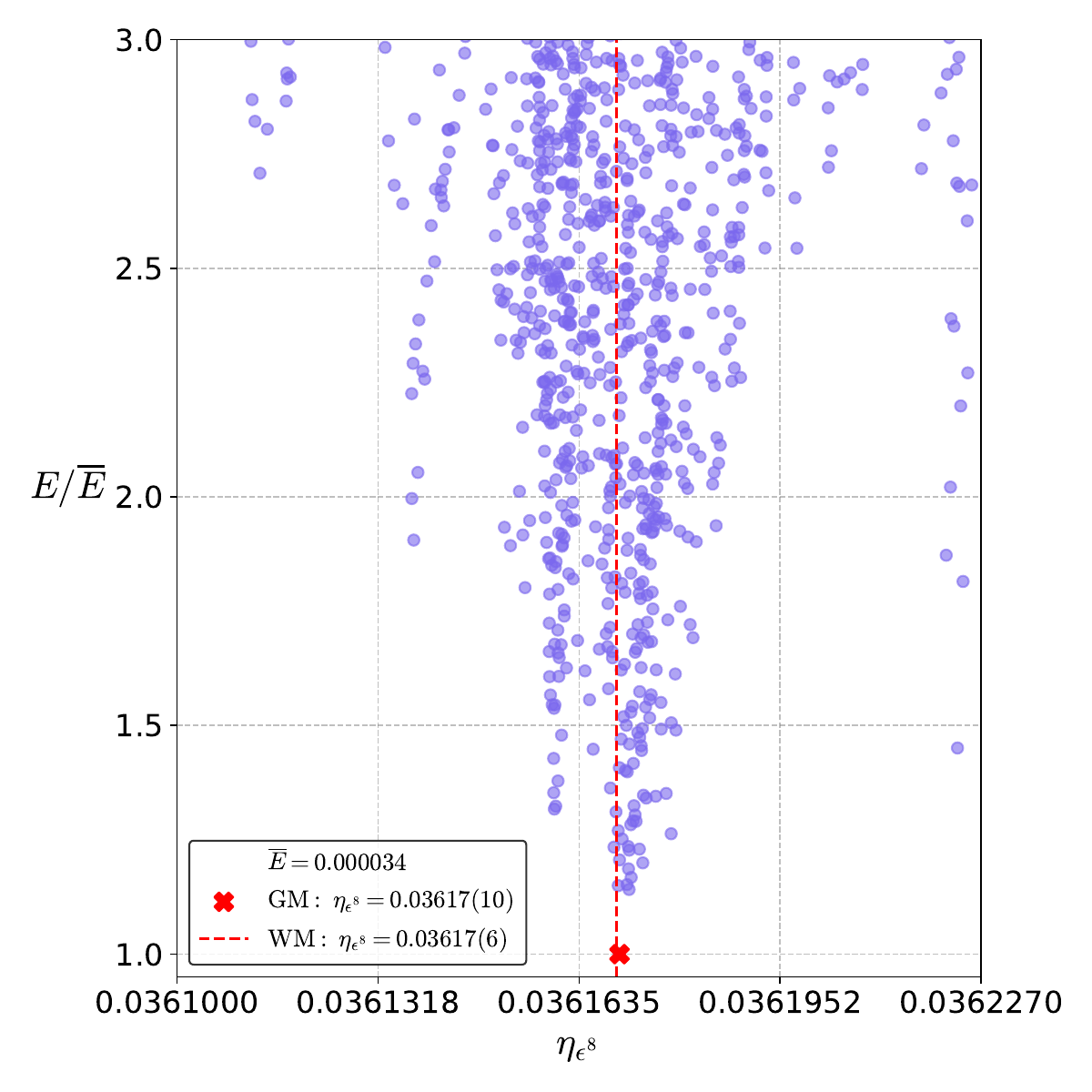}}\hfill
		\subfloat[]{\includegraphics[width=0.23\textwidth]{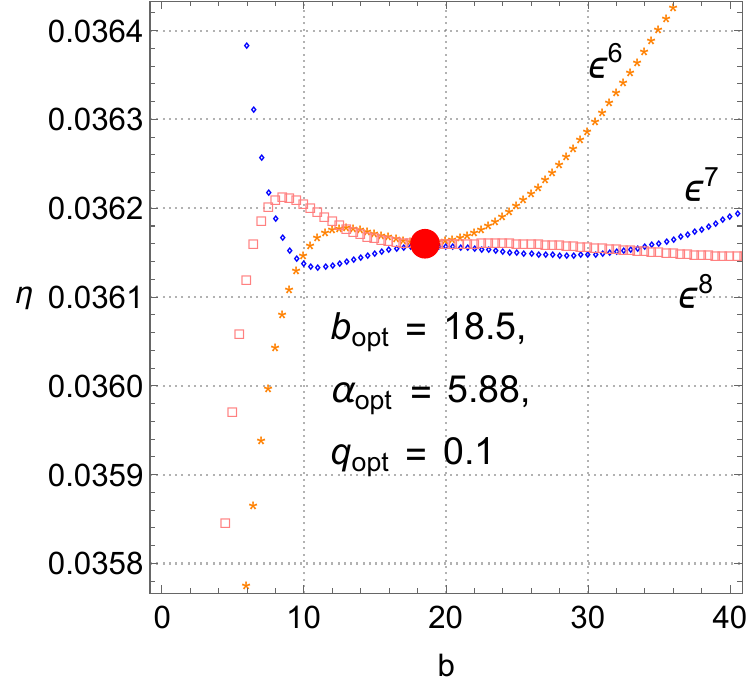}}\hfill
		\subfloat[]{\includegraphics[width=0.23\textwidth]{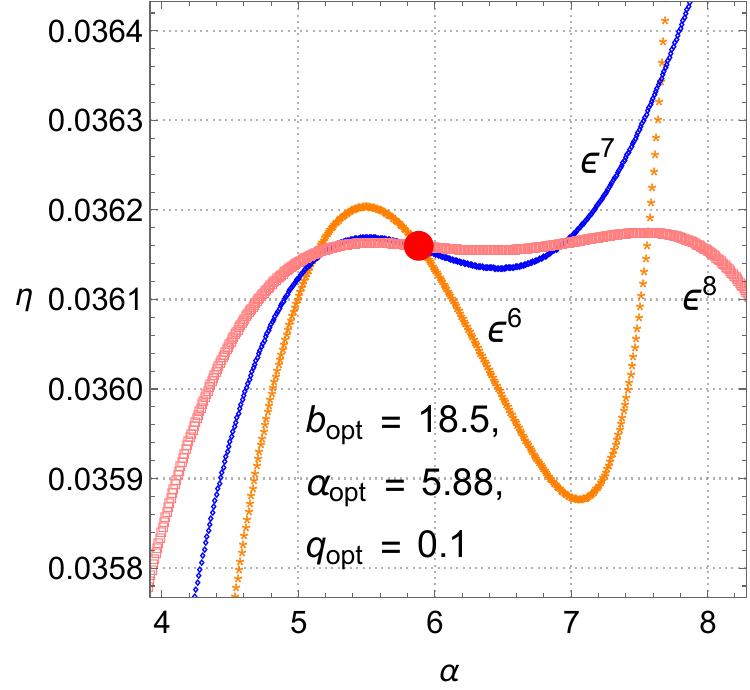}}\hfill
		\subfloat[]{\includegraphics[width=0.26\textwidth]{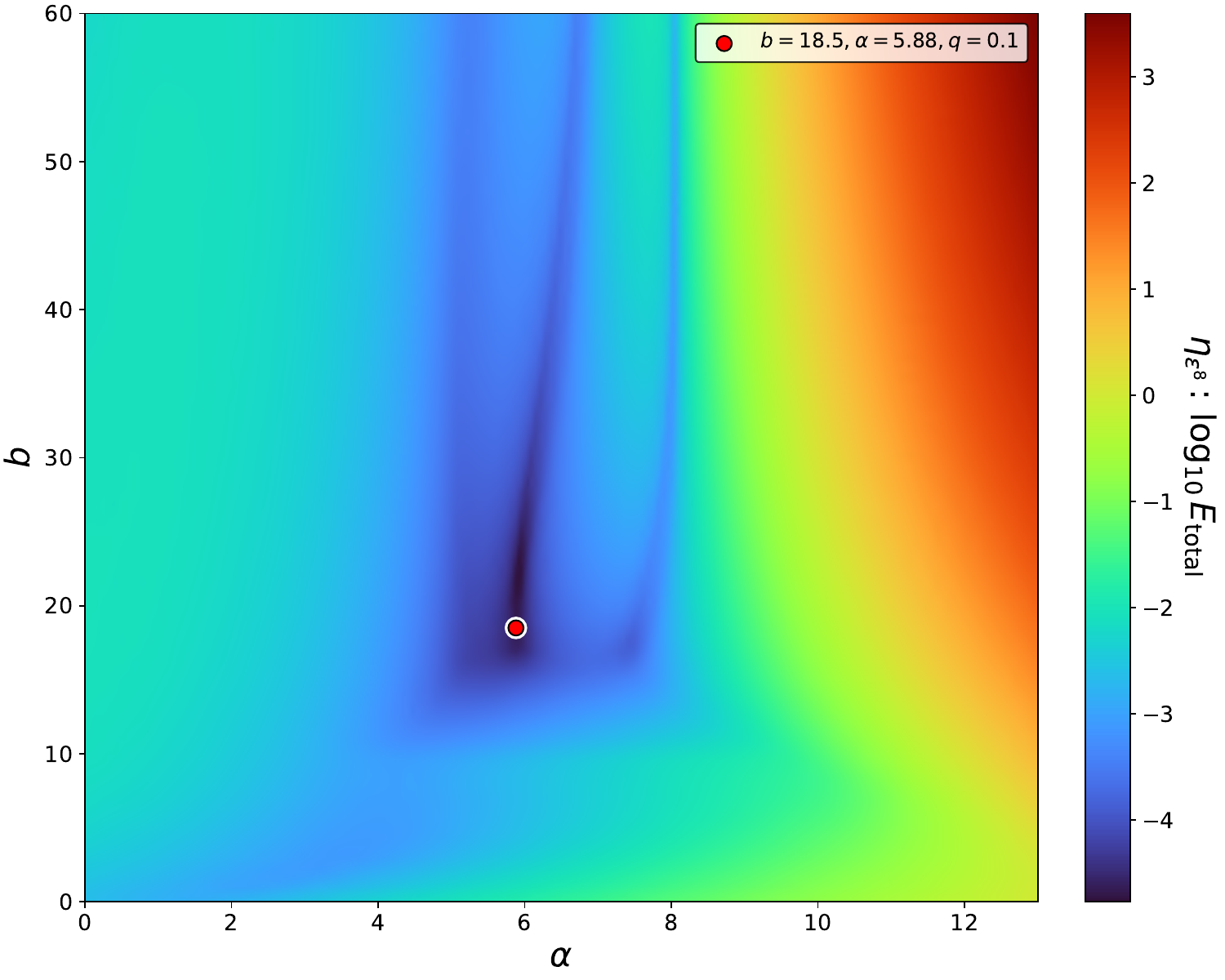}}\hfill
		\subfloat[]{\includegraphics[width=0.26\textwidth]{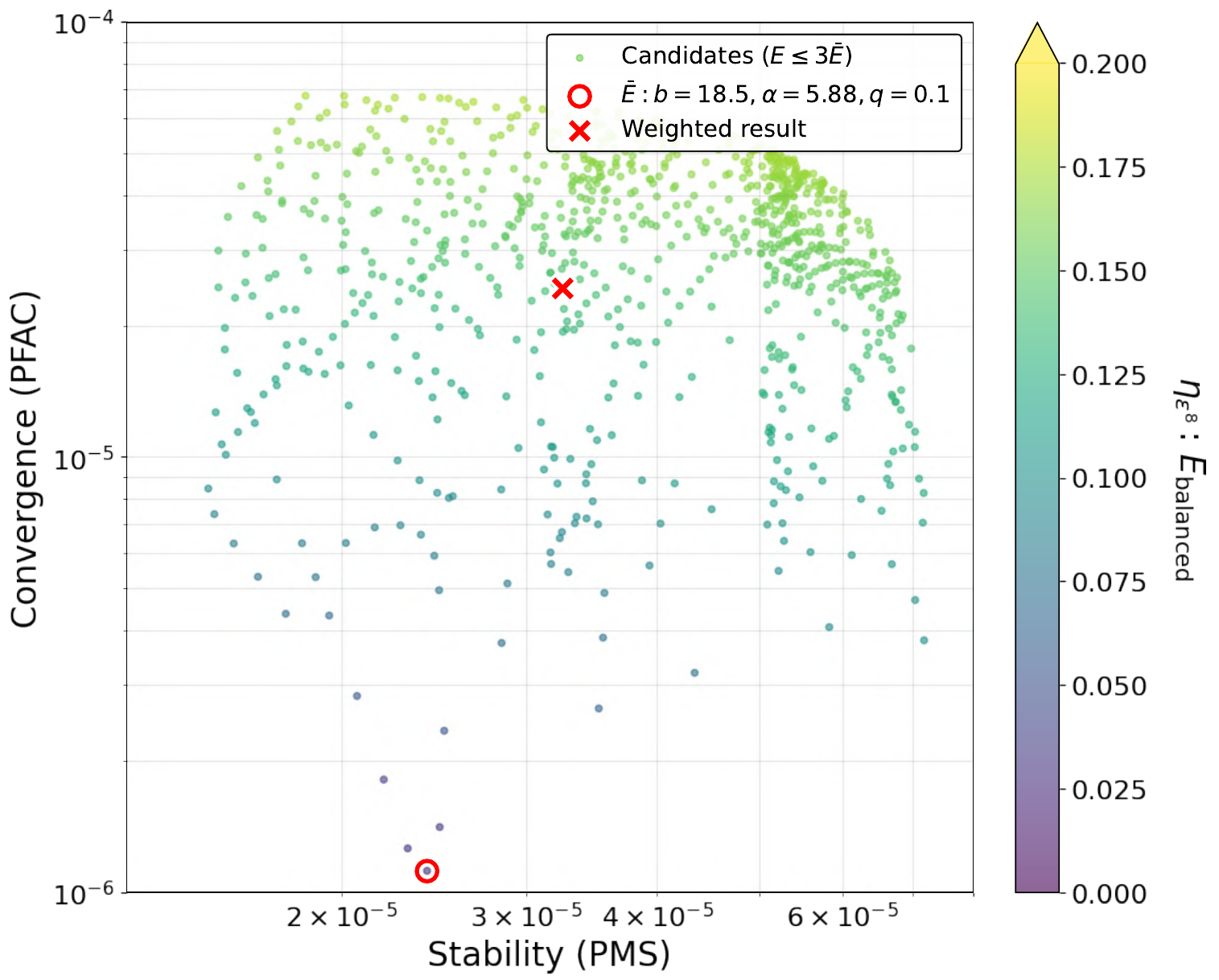}}
		\centering
		\caption{Dependence of the critical exponent $\eta$ on the resummation parameters $b$ [(a), (d)] and $\alpha$ [(b), (e)] at eight-loop order. The top row corresponds to the optimal point $(b, \alpha, q) = (17.5, 6.08, 0.11)$ ({\it Method I}, Section~\ref{error_estimations}), and the bottom row corresponds to the optimal point $(b, \alpha, q) = (18.5, 5.88, 0.1)$ ({\it Method II}, Section~\ref{rms_error_estimations}). Panel (c) displays the normalized error $E/\overline{E}$ distribution across the parameter space. The global minimum (GM, red cross) and weighted mean (WM, dashed line) are shown with uncertainties of $3\overline{E}$ and $\overline{E} + 2\sigma_w$, respectively, accounting for both local stability and the statistical spread of the candidate set within the $E \leq 3\overline{E}$ region [see Eq.~(\ref{weight})]. Panels (f) and (g) show the error landscape in the $(b, \alpha)$ plane at the optimal $q$ value and the distribution of calculated points in the plane of stability ($E_{\mathrm{pms}}$) versus convergence ($E_{\mathrm{pfac}}$), respectively. Our final estimates are $\eta = 0.03617(10)$ via {\it Method I} and $\eta =  0.03616(7)$ via {\it Method II}.}
		\label{eta8__n1}
	\end{figure}

\begin{figure}[h!] 
			\subfloat[]{\includegraphics[width=0.34\textwidth]{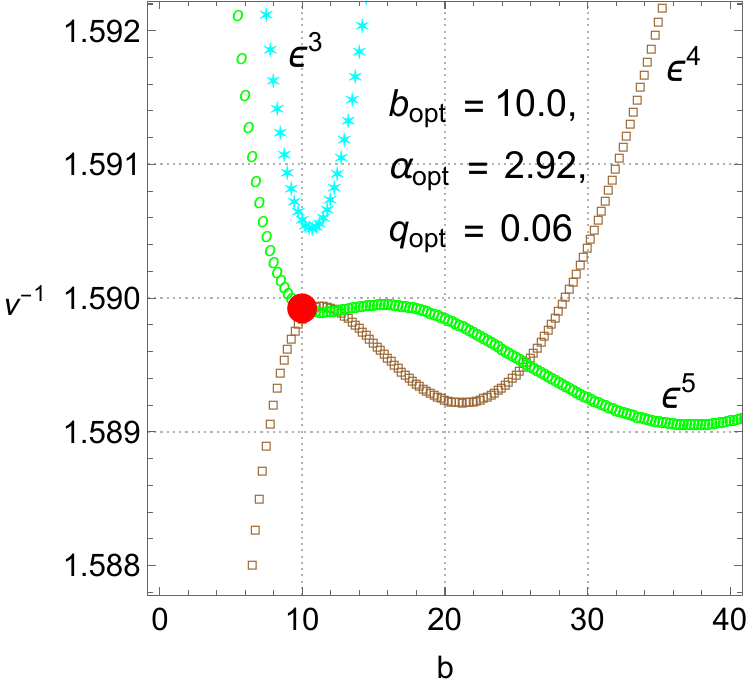}}\hfill
			\subfloat[]{\includegraphics[width=0.34\textwidth]{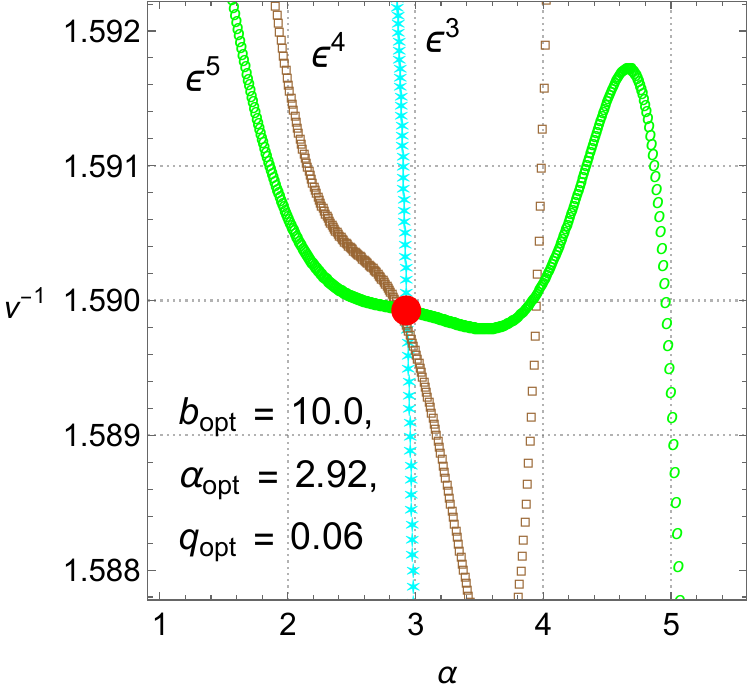}}\hfill
			\subfloat[]{\includegraphics[width=0.32\textwidth]{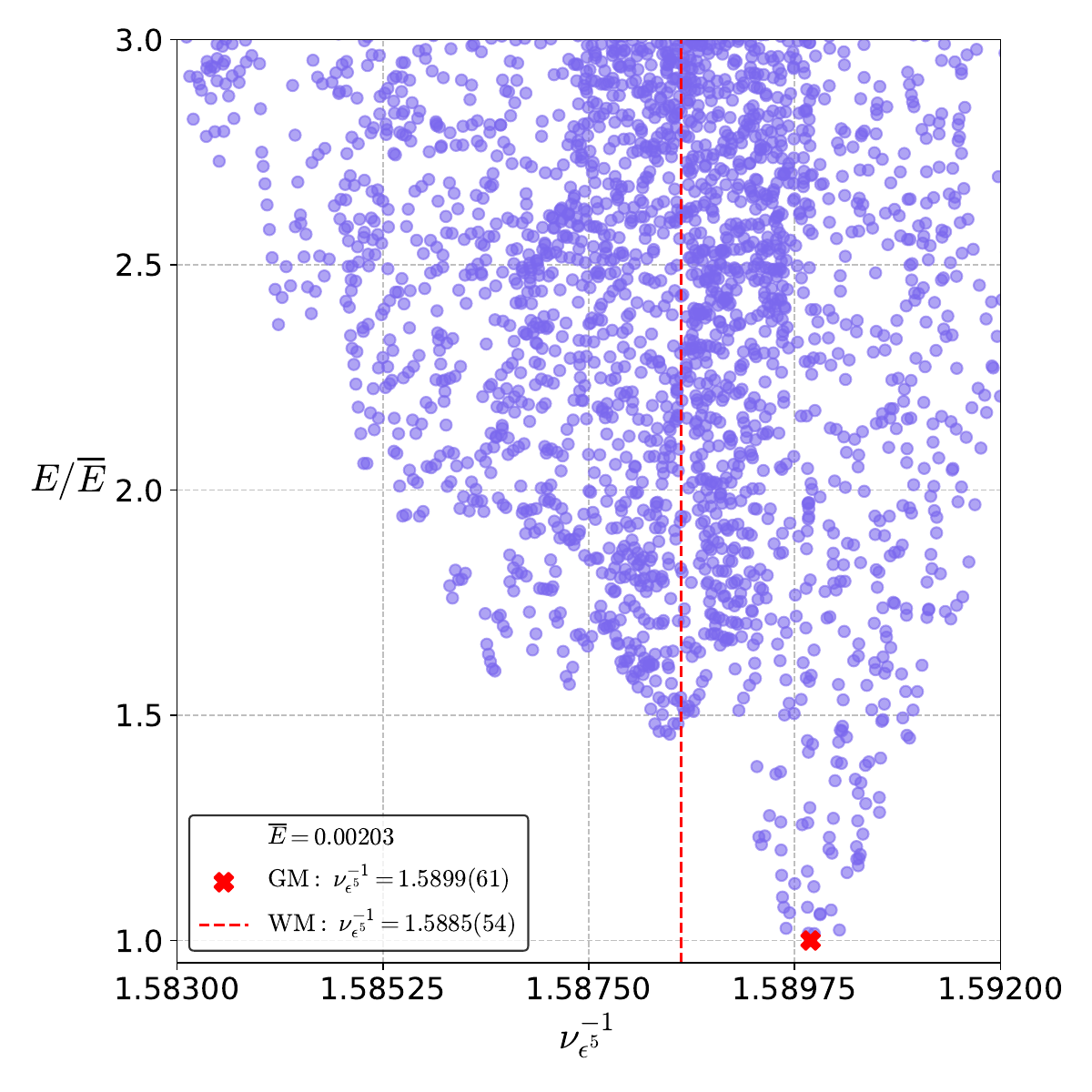}}\hfill
			\subfloat[]{\includegraphics[width=0.23\textwidth]{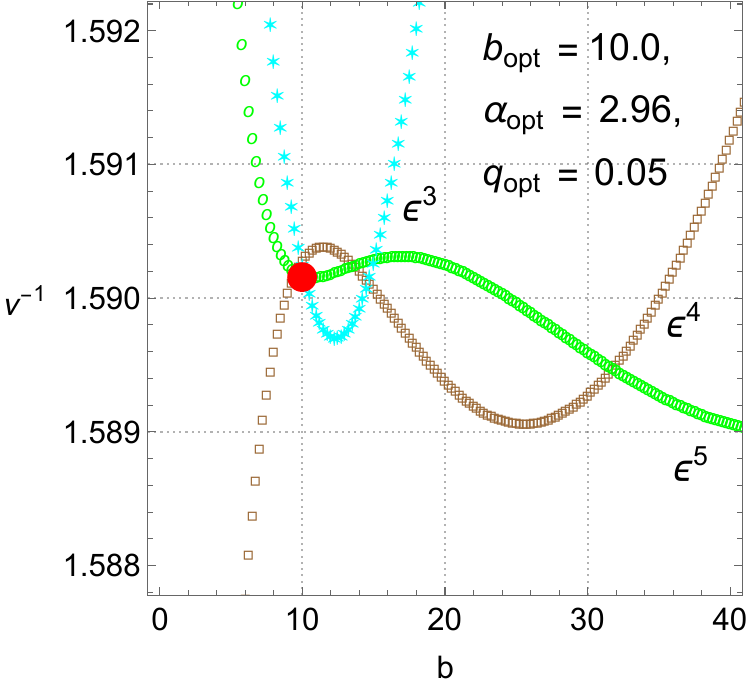}}\hfill 
			\subfloat[]{\includegraphics[width=0.23\textwidth]{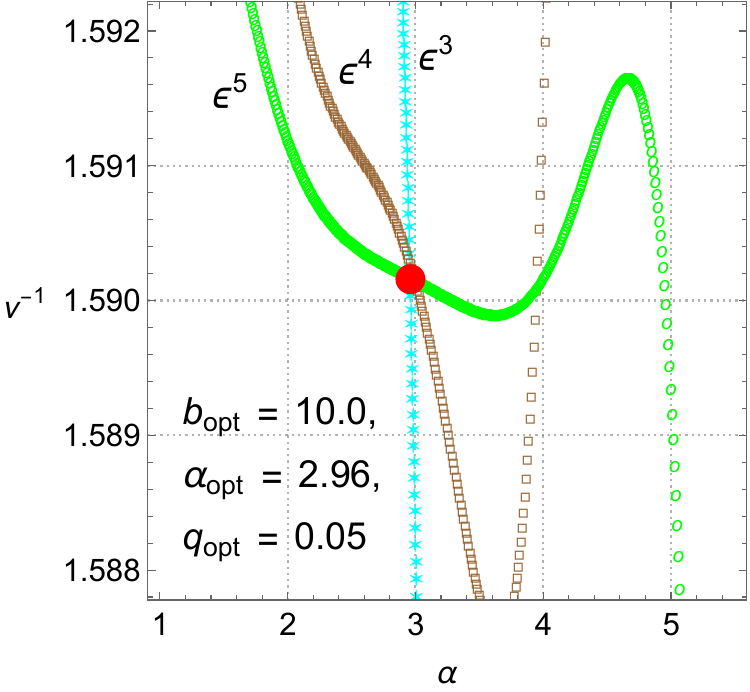}}\hfill
			\subfloat[]{\includegraphics[width=0.26\textwidth]{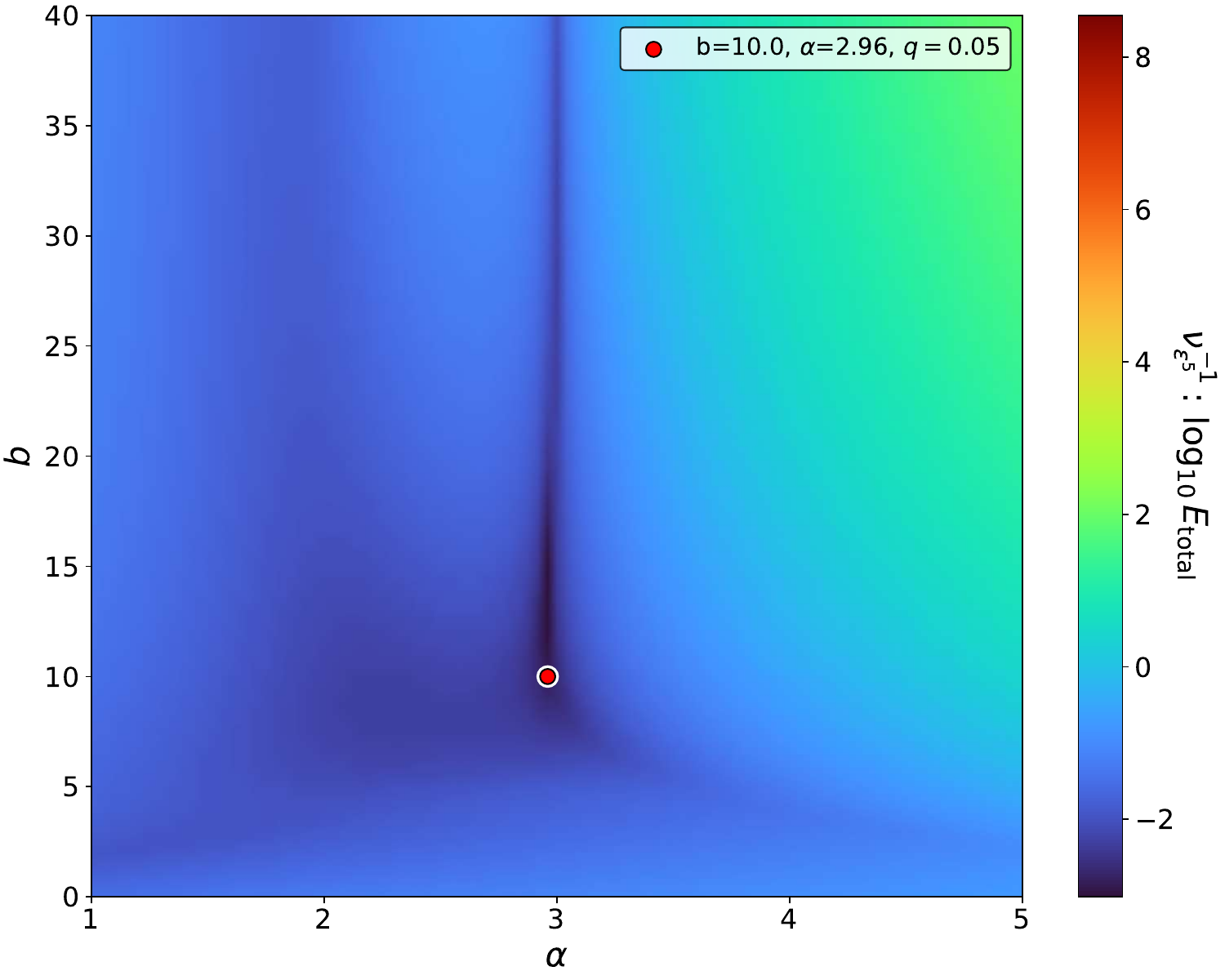}}\hfill
			\subfloat[]{\includegraphics[width=0.26\textwidth]{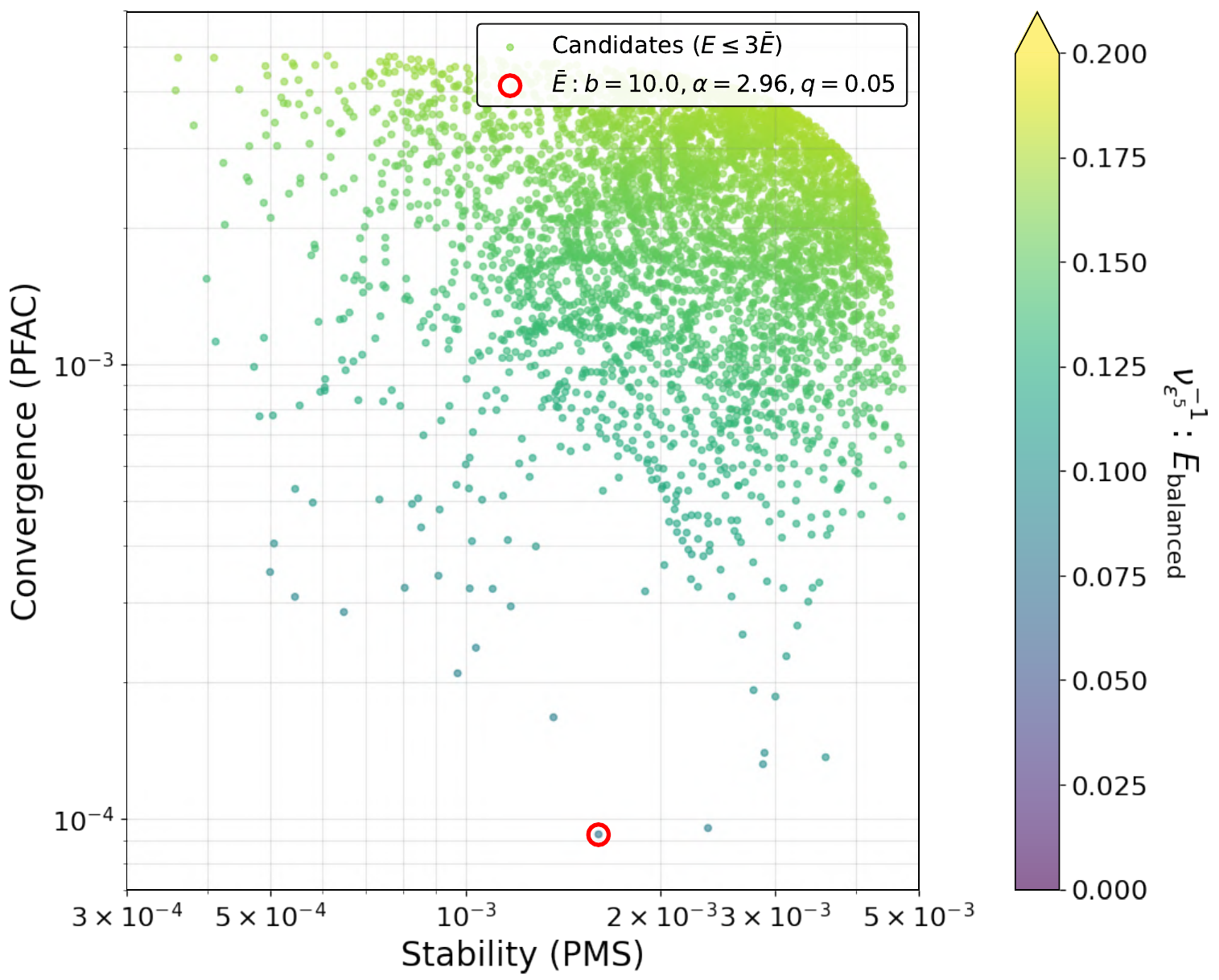}}
		\centering
		\caption{Dependence of the critical exponent $\nu^{-1}$ on the resummation parameters $b$ [(a), (d)] and $\alpha$ [(b), (e)] at five-loop order. The top row corresponds to the optimal point $(b, \alpha, q) = (10.0, 2.92, 0.06)$ ({\it Method I}, Section~\ref{error_estimations}), and the bottom row corresponds to the optimal point $(b, \alpha, q) = (10.0, 2.96, 0.05)$ ({\it Method II}, Section~\ref{rms_error_estimations}). Panel (c) displays the normalized error $E/\overline{E}$ distribution across the parameter space. The global minimum (GM, red cross) and weighted mean (WM, dashed line) are shown with uncertainties of $3\overline{E}$ and $\overline{E} + 2\sigma_w$, respectively, accounting for both local stability and the statistical spread of the candidate set within the $E \leq 3\overline{E}$ region [see Eq.~(\ref{weight})]. Panels (f) and (g) show the error landscape in the $(b, \alpha)$ plane at the optimal $q$ value and the distribution of calculated points in the plane of stability ($E_{\mathrm{pms}}$) versus convergence ($E_{\mathrm{pfac}}$), respectively. Our final estimates are $\nu = 0.6290(24)$ via {\it Method I} and $\nu = 0.6289(19)$ via {\it Method II}.}
		\label{nu5__n1}
	\end{figure}

\begin{figure}[h!] 
			\subfloat[]{\includegraphics[width=0.34\textwidth]{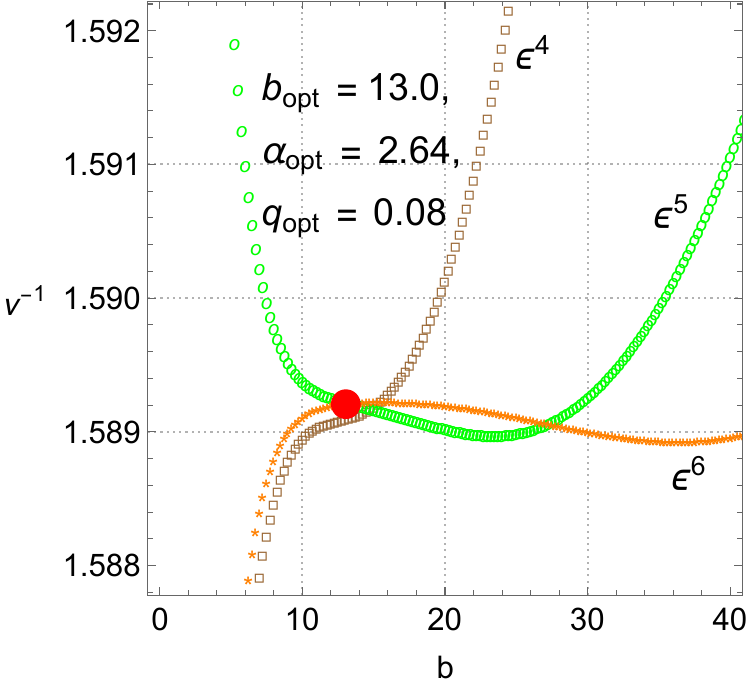}}\hfill
			\subfloat[]{\includegraphics[width=0.34\textwidth]{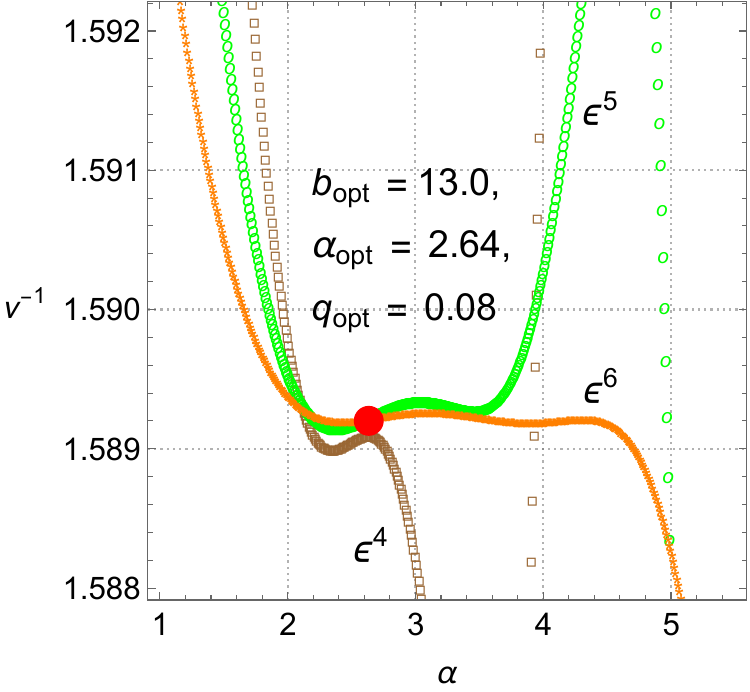}}\hfill
			\subfloat[]{\includegraphics[width=0.32\textwidth]{Figures/eanu6.pdf}}\hfill
			\subfloat[]{\includegraphics[width=0.23\textwidth]{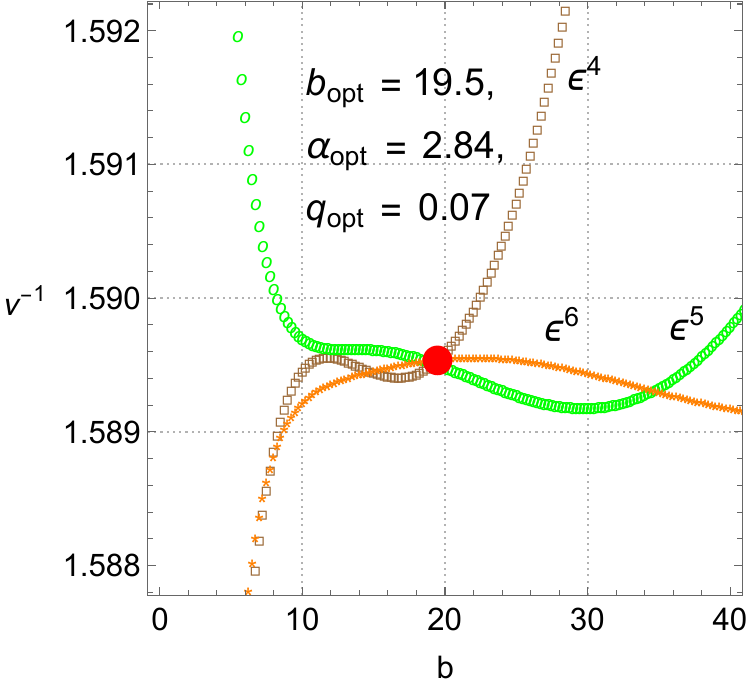}}\hfill 
			\subfloat[]{\includegraphics[width=0.23\textwidth]{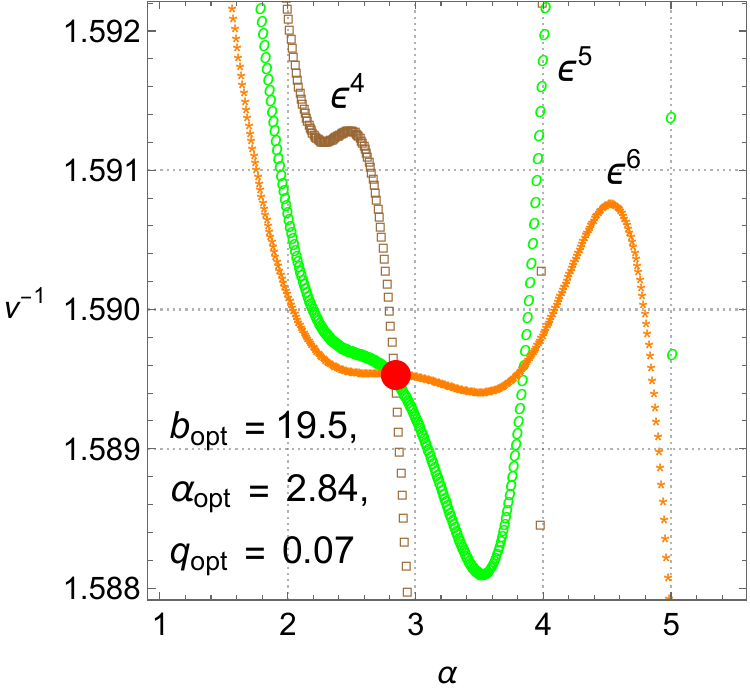}}\hfill
			\subfloat[]{\includegraphics[width=0.26\textwidth]{Figures/elanu6.pdf}}\hfill
			\subfloat[]{\includegraphics[width=0.26\textwidth]{Figures/epanu6.pdf}}
		\centering
		\caption{Dependence of the critical exponent $\nu^{-1}$ on the resummation parameters $b$ [(a), (d)] and $\alpha$ [(b), (e)] at six-loop order. The top row corresponds to the optimal point $(b, \alpha, q) = (13.0, 2.64, 0.08)$ ({\it Method I}, Section~\ref{error_estimations}), and the bottom row corresponds to the optimal point $(b, \alpha, q) = (19.5, 2.84, 0.07)$ ({\it Method II}, Section~\ref{rms_error_estimations}). Panel (c) displays the normalized error $E/\overline{E}$ distribution across the parameter space. The global minimum (GM, red cross) and weighted mean (WM, dashed line) are shown with uncertainties of $3\overline{E}$ and $\overline{E} + 2\sigma_w$, respectively, accounting for both local stability and the statistical spread of the candidate set within the $E \leq 3\overline{E}$ region [see Eq.~(\ref{weight})]. Panels (f) and (g) show the error landscape in the $(b, \alpha)$ plane at the optimal $q$ value and the distribution of calculated points in the plane of stability ($E_{\mathrm{pms}}$) versus convergence ($E_{\mathrm{pfac}}$), respectively. Our final estimates are $\nu = 0.62925(76)$ via {\it Method I} and $\nu = 0.62912(67)$ via {\it Method II}.}
		\label{nu6__n1}
	\end{figure}

\begin{figure}[h!] 
			\subfloat[]{\includegraphics[width=0.34\textwidth]{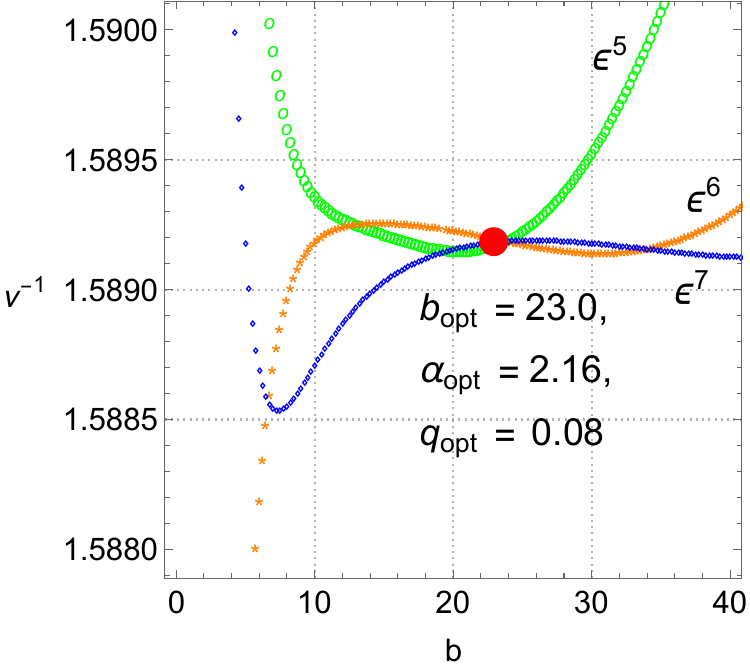}}\hfill
			\subfloat[]{\includegraphics[width=0.34\textwidth]{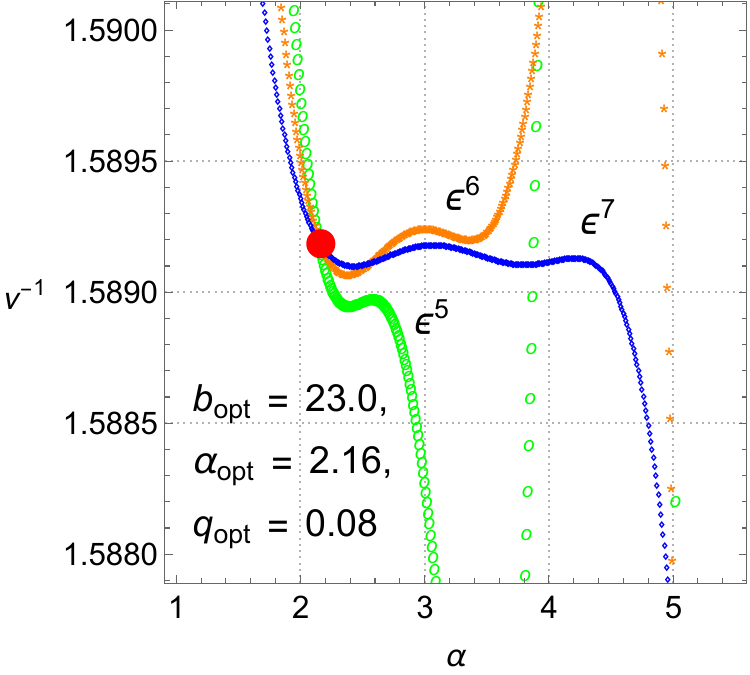}}\hfill
			\subfloat[]{\includegraphics[width=0.32\textwidth]{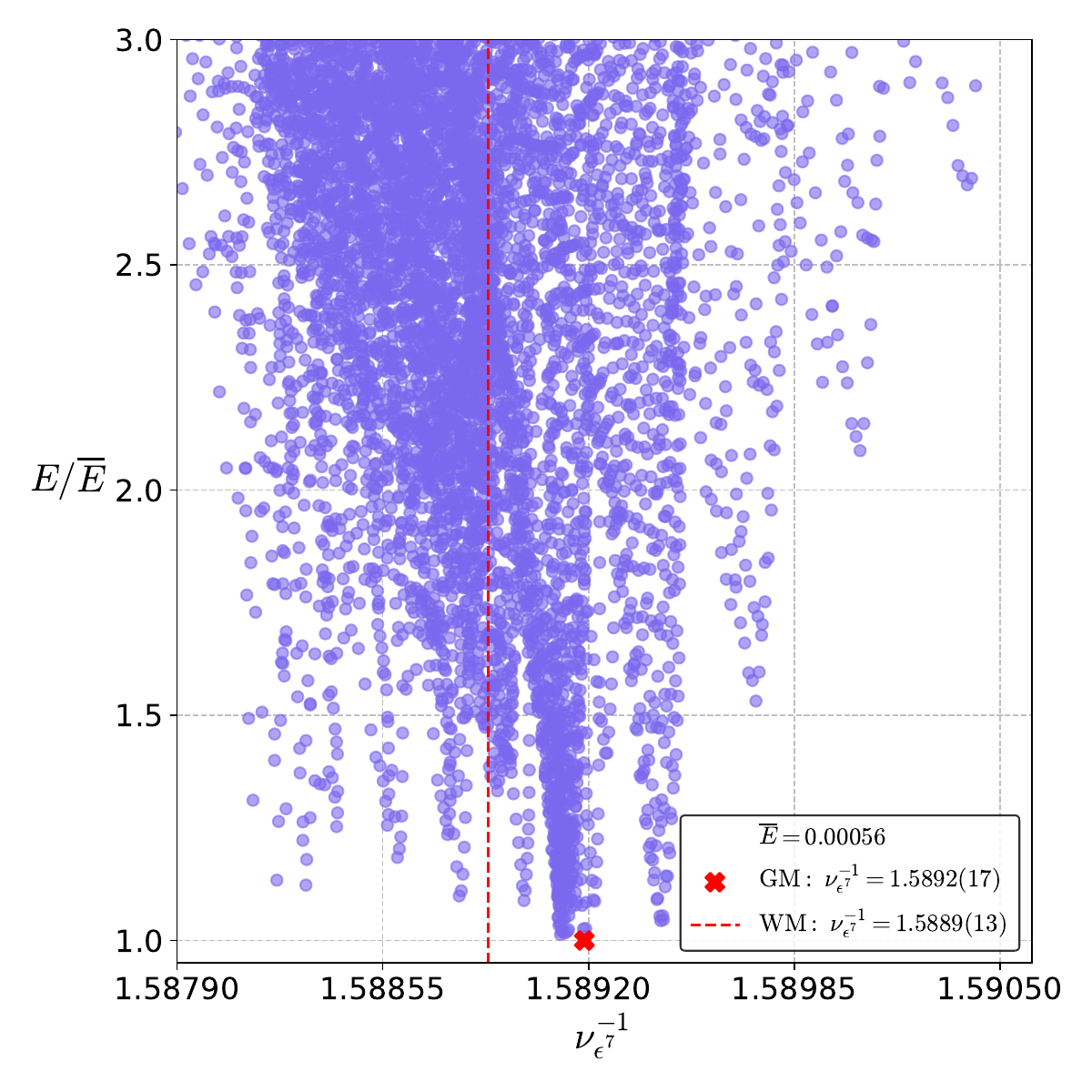}}\hfill
			\subfloat[]{\includegraphics[width=0.23\textwidth]{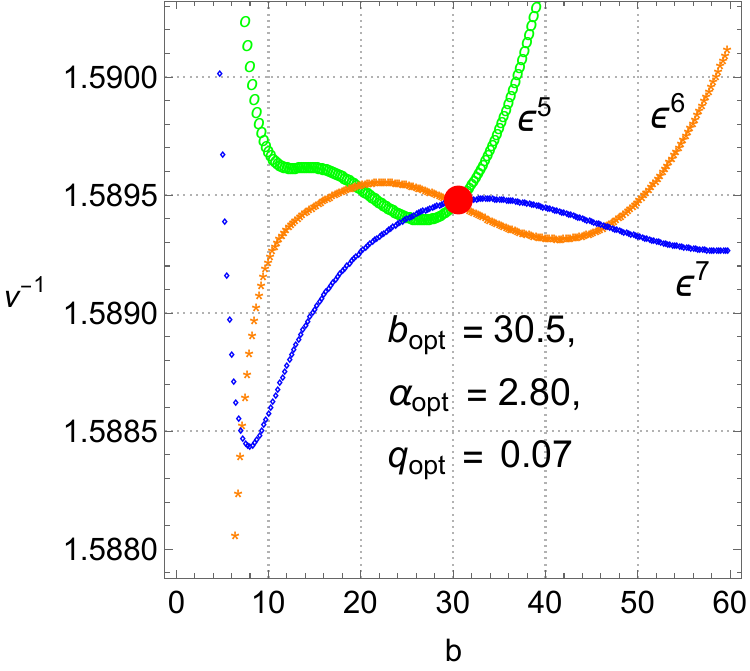}}\hfill
			\subfloat[]{\includegraphics[width=0.23\textwidth]{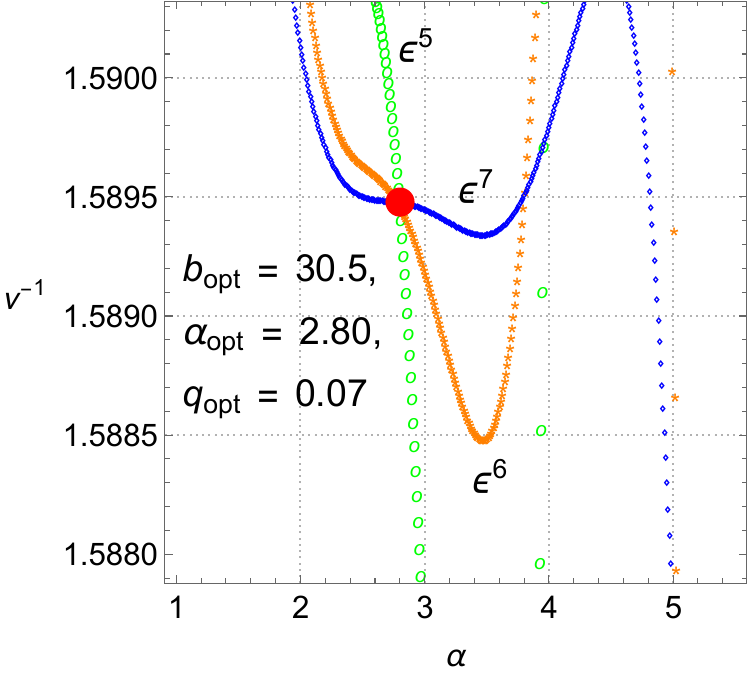}}\hfill
			\subfloat[]{\includegraphics[width=0.26\textwidth]{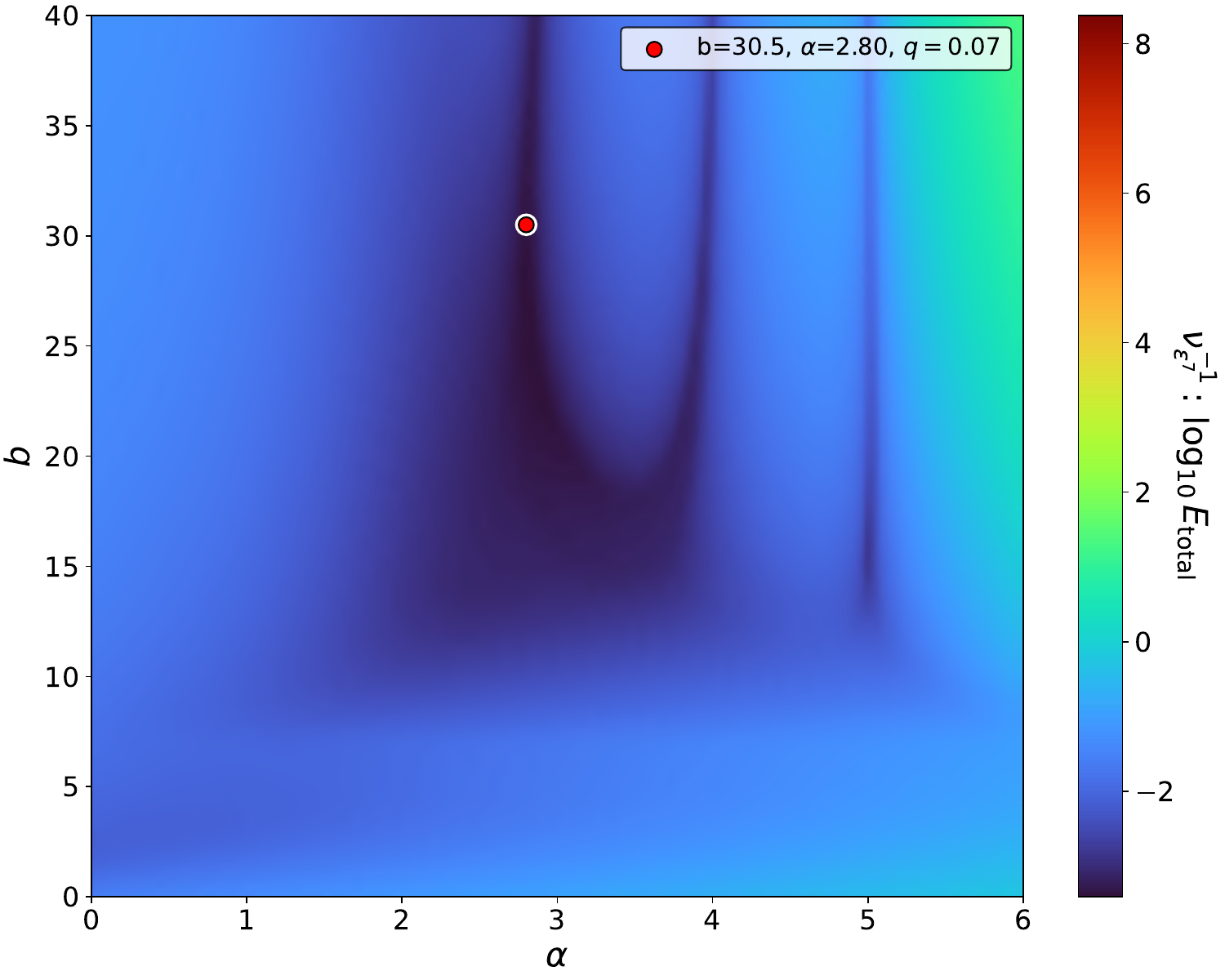}}\hfill
			\subfloat[]{\includegraphics[width=0.26\textwidth]{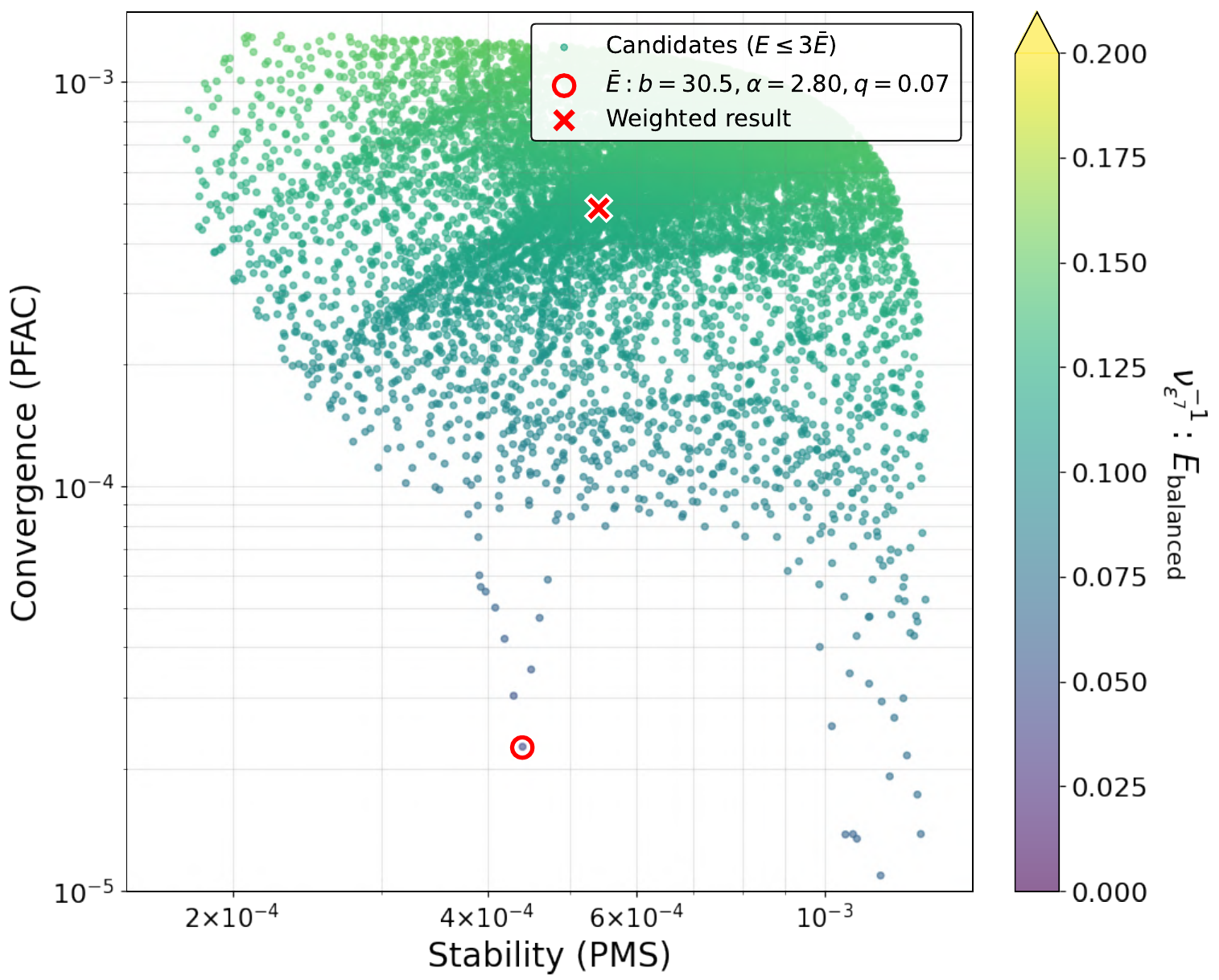}}\
		\centering
		\caption{Dependence of the critical exponent $\nu^{-1}$ on the resummation parameters $b$ [(a), (d)] and $\alpha$ [(b), (e)] at seven-loop order. The top row corresponds to the optimal point $(b, \alpha, q) = (23.0, 2.16, 0.08)$ ({\it Method I}, Section~\ref{error_estimations}), and the bottom row corresponds to the optimal point $(b, \alpha, q) = (30.5, 2.8, 0.07)$ ({\it Method II}, Section~\ref{rms_error_estimations}). Panel (c) displays the normalized error $E/\overline{E}$ distribution across the parameter space. The global minimum (GM, red cross) and weighted mean (WM, dashed line) are shown with uncertainties of $3\overline{E}$ and $\overline{E} + 2\sigma_w$, respectively, accounting for both local stability and the statistical spread of the candidate set within the $E \leq 3\overline{E}$ region [see Eq.~(\ref{weight})]. Panels (f) and (g) show the error landscape in the $(b, \alpha)$ plane at the optimal $q$ value and the distribution of calculated points in the plane of stability ($E_{\mathrm{pms}}$) versus convergence ($E_{\mathrm{pfac}}$), respectively. Our final estimates are $\nu = 0.62925(67)$ via {\it Method I} and $\nu = 0.62914(54)$  via {\it Method II}.}
		\label{nu7__n1}
	\end{figure}

\begin{figure}[h!] 
			\subfloat[]{\includegraphics[width=0.34\textwidth]{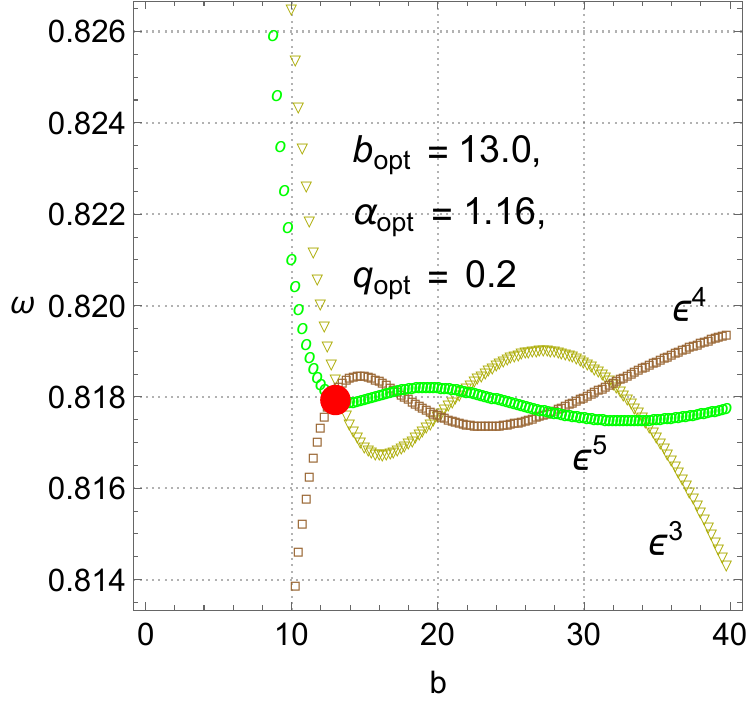}}\hfill
			\subfloat[]{\includegraphics[width=0.34\textwidth]{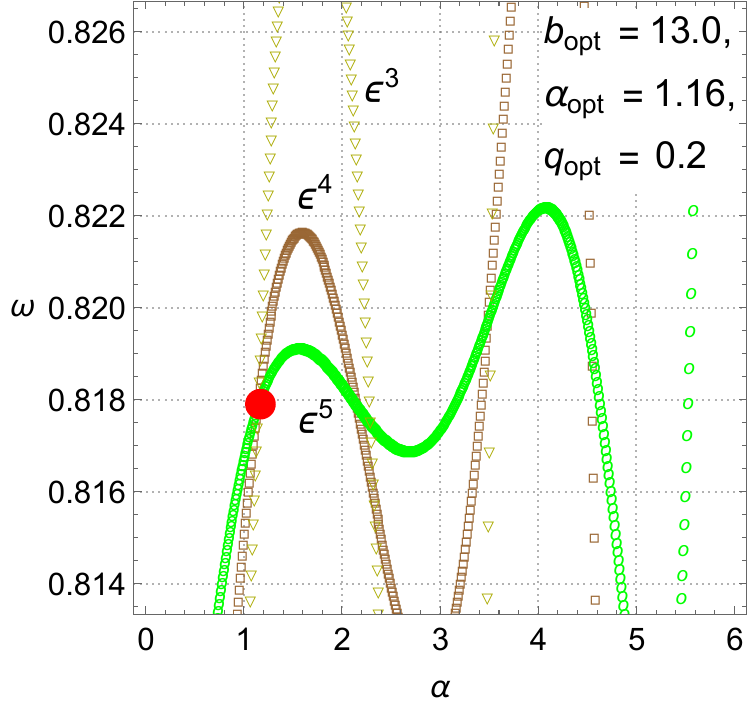}}\hfill
			\subfloat[]{\includegraphics[width=0.32\textwidth]{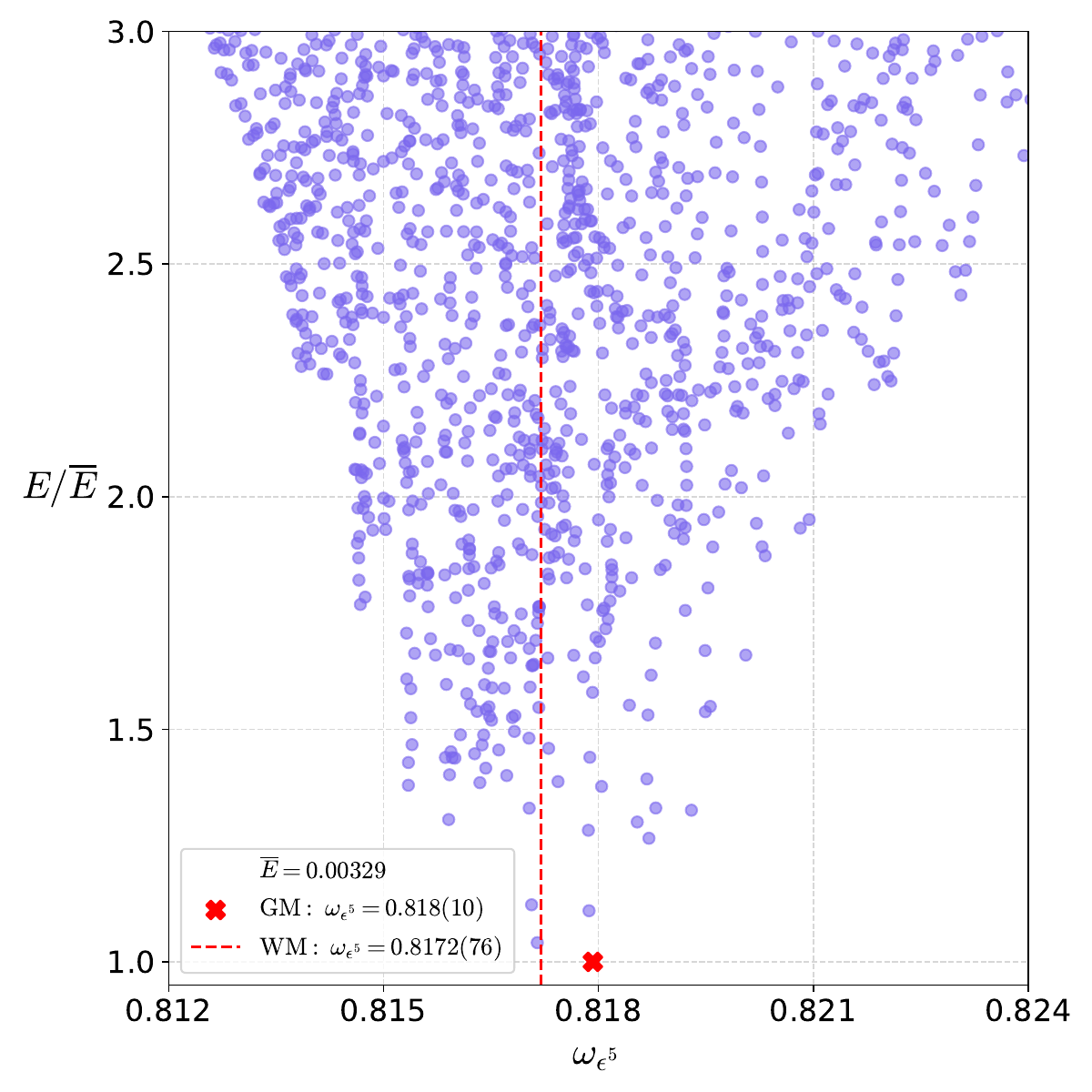}}\hfill
			\subfloat[]{\includegraphics[width=0.23\textwidth]{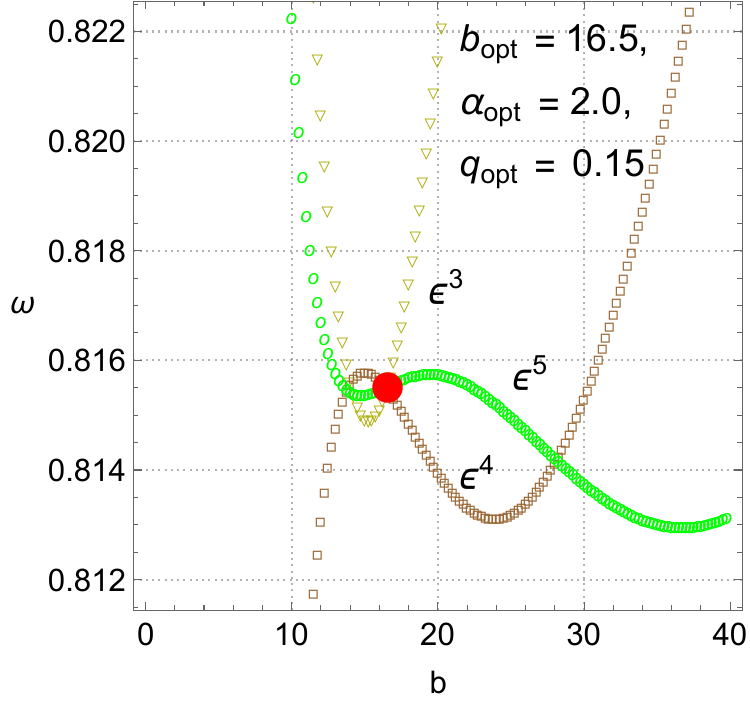}}\hfill
			\subfloat[]{\includegraphics[width=0.23\textwidth]{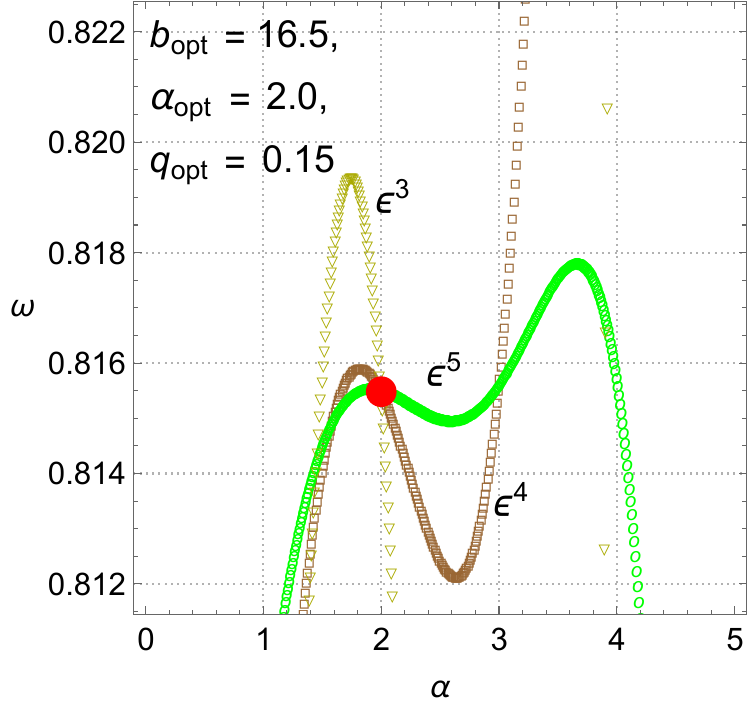}}\hfill
			\subfloat[]{\includegraphics[width=0.26\textwidth]{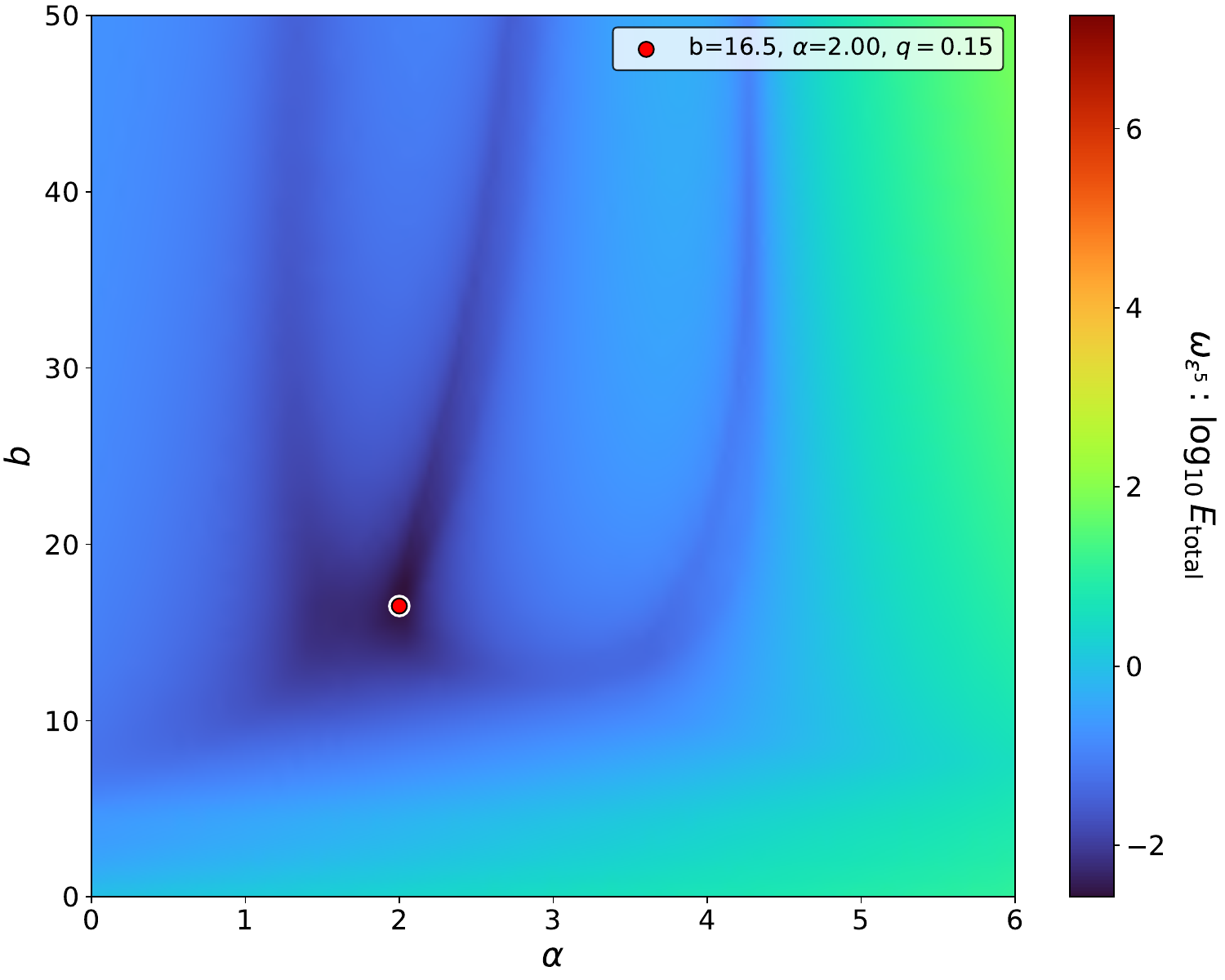}}\hfill
			\subfloat[]{\includegraphics[width=0.26\textwidth]{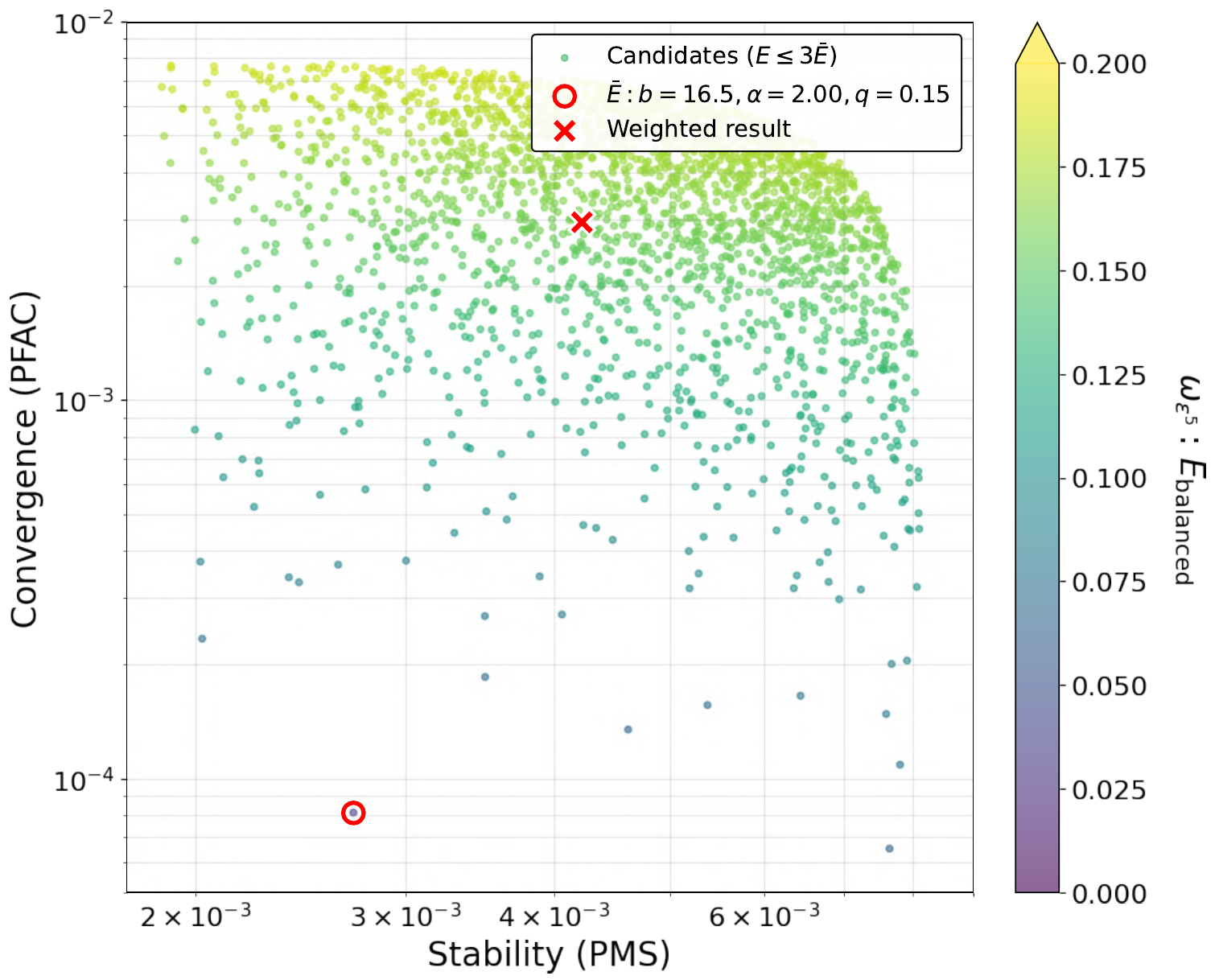}}\
		\centering
		\caption{Dependence of the correction-to-scaling exponent $\omega$ on the resummation parameters $b$ [(a), (d)] and $\alpha$ [(b), (e)] at five-loop order. The top row corresponds to the optimal point $(b, \alpha, q) = (13.0, 1.16, 0.2)$ ({\it Method I}, Section~\ref{error_estimations}), and the bottom row corresponds to the optimal point $(b, \alpha, q) = (16.5, 2.0, 0.15)$ ({\it Method II}, Section~\ref{rms_error_estimations}). Panel (c) displays the normalized error $E/\overline{E}$ distribution across the parameter space. The global minimum (GM, red cross) and weighted mean (WM, dashed line) are shown with uncertainties of $3\overline{E}$ and $\overline{E} + 2\sigma_w$, respectively, accounting for both local stability and the statistical spread of the candidate set within the $E \leq 3\overline{E}$ region [see Eq.~(\ref{weight})]. Panels (f) and (g) show the error landscape in the $(b, \alpha)$ plane at the optimal $q$ value and the distribution of calculated points in the plane of stability ($E_{\mathrm{pms}}$) versus convergence ($E_{\mathrm{pfac}}$), respectively. Our final estimates are $\omega = 0.818(10)$ via {\it Method I} and  $\omega = 0.816(8)$  via {\it Method II}.}
		\label{omega5__n1}
	\end{figure}

\begin{figure}[h!] 
			\subfloat[]{\includegraphics[width=0.34\textwidth]{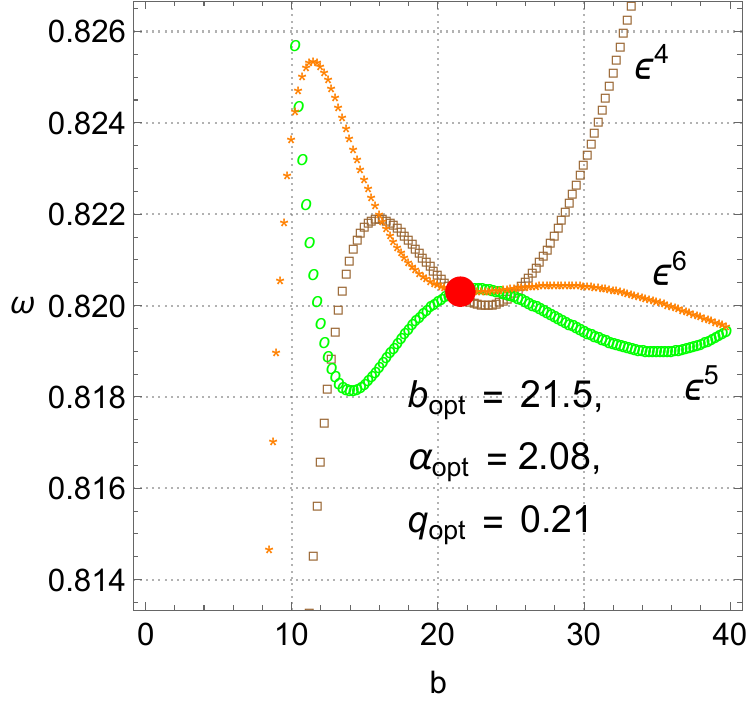}}\hfill
			\subfloat[]{\includegraphics[width=0.34\textwidth]{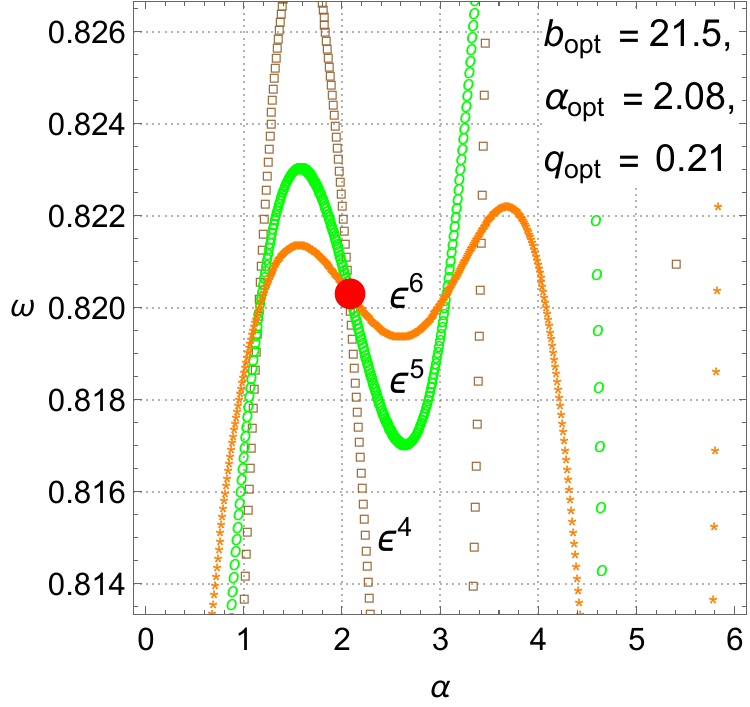}}\hfill
			\subfloat[]{\includegraphics[width=0.32\textwidth]{Figures/eaomega6.pdf}}\hfill
		\subfloat[]	{\includegraphics[width=0.23\textwidth]{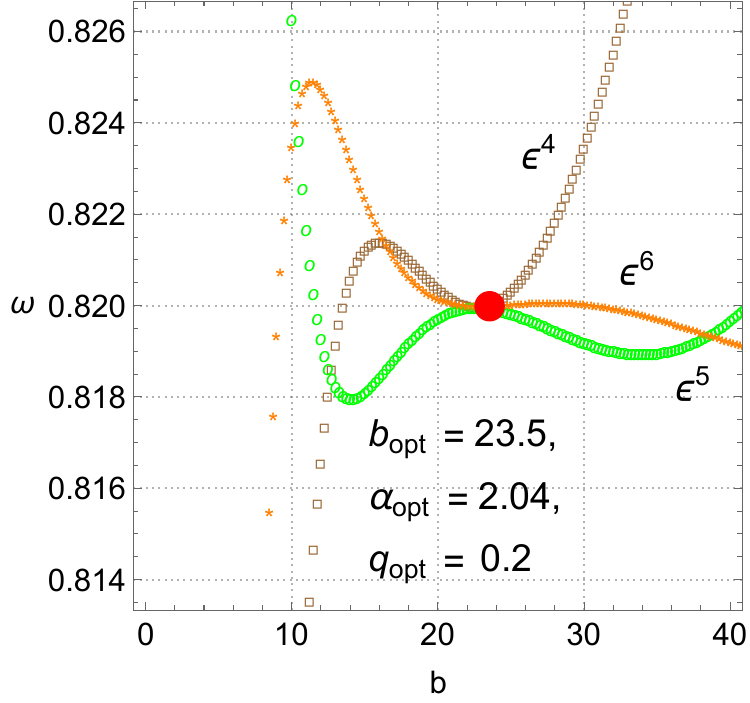}}\hfill
			\subfloat[]{\includegraphics[width=0.23\textwidth]{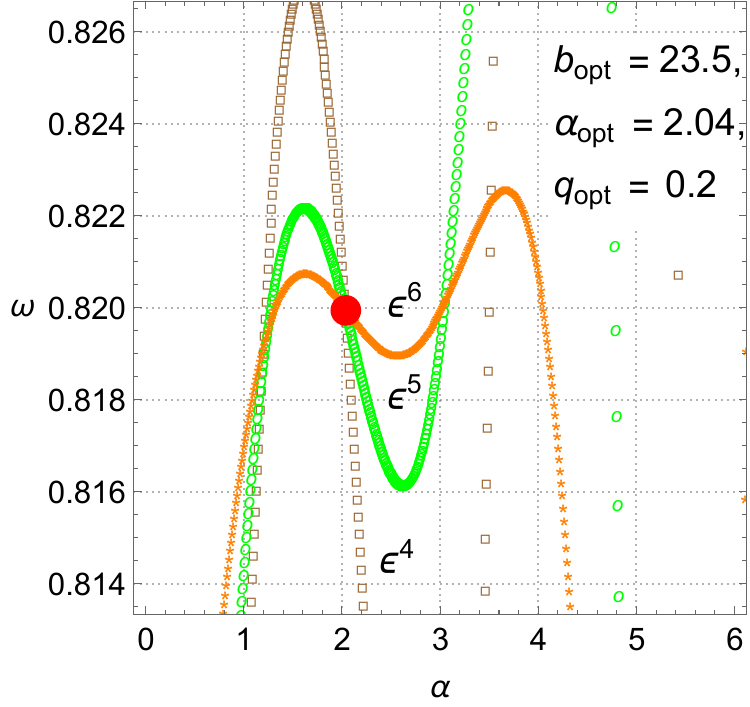}}\hfill
			\subfloat[]{\includegraphics[width=0.26\textwidth]{Figures/elaomega6.pdf}}\hfill
			\subfloat[]{\includegraphics[width=0.26\textwidth]{Figures/epaomega6.pdf}}\
		\centering
		\caption{Dependence of the correction-to-scaling exponent $\omega$ on the resummation parameters $b$ [(a), (d)] and $\alpha$ [(b), (e)] at six-loop order. The top row corresponds to the optimal point $(b, \alpha, q) = (21.5, 2.08, 0.21)$ ({\it Method I}, Section~\ref{error_estimations}), and the bottom row corresponds to the optimal point $(b, \alpha, q) = (23.5, 2.04, 0.2)$ ({\it Method II}, Section~\ref{rms_error_estimations}). Panel (c) displays the normalized error $E/\overline{E}$ distribution across the parameter space. The global minimum (GM, red cross) and weighted mean (WM, dashed line) are shown with uncertainties of $3\overline{E}$ and $\overline{E} + 2\sigma_w$, respectively, accounting for both local stability and the statistical spread of the candidate set within the $E \leq 3\overline{E}$ region [see Eq.~(\ref{weight})]. Panels (f) and (g) show the error landscape in the $(b, \alpha)$ plane at the optimal $q$ value and the distribution of calculated points in the plane of stability ($E_{\mathrm{pms}}$) versus convergence ($E_{\mathrm{pfac}}$), respectively. Our final estimates are $\omega = 0.8203(76)$  via {\it Method I} and $\omega = 0.8200(52)$ via {\it Method II}.}
		\label{omega6__n1}
	\end{figure}

\begin{figure}[h!] 
		\subfloat[]{\includegraphics[width=0.34\textwidth]{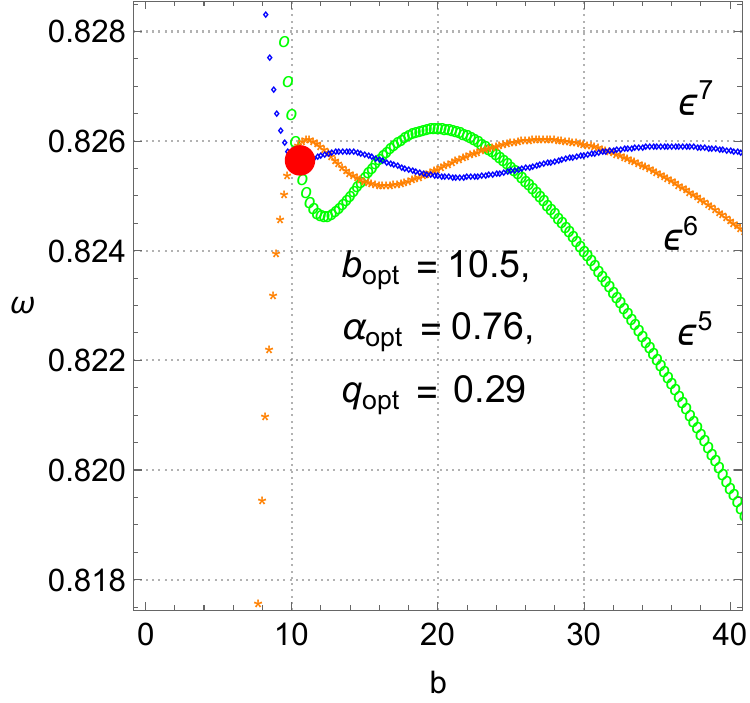}}\hfill
		\subfloat[]{\includegraphics[width=0.34\textwidth]{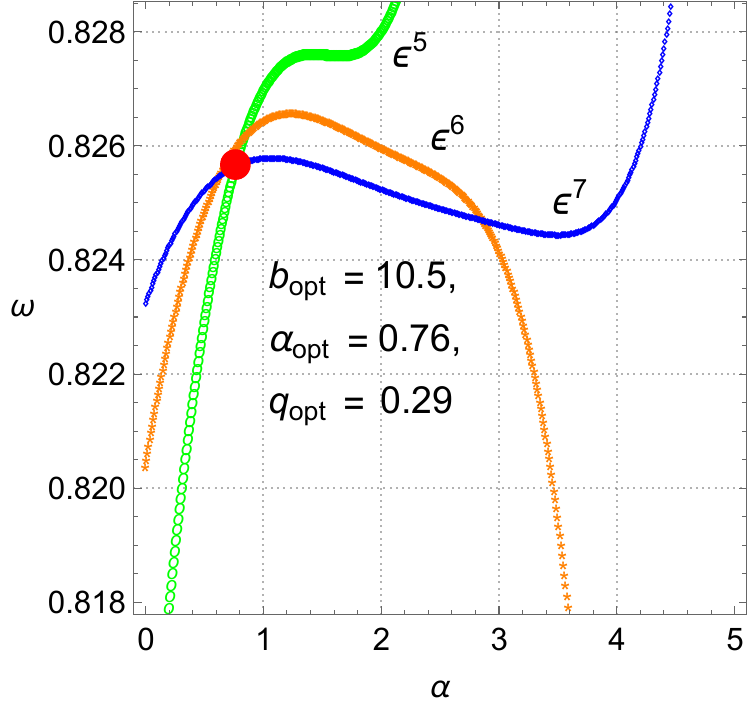}}\hfill
		\subfloat[]{\includegraphics[width=0.32\textwidth]{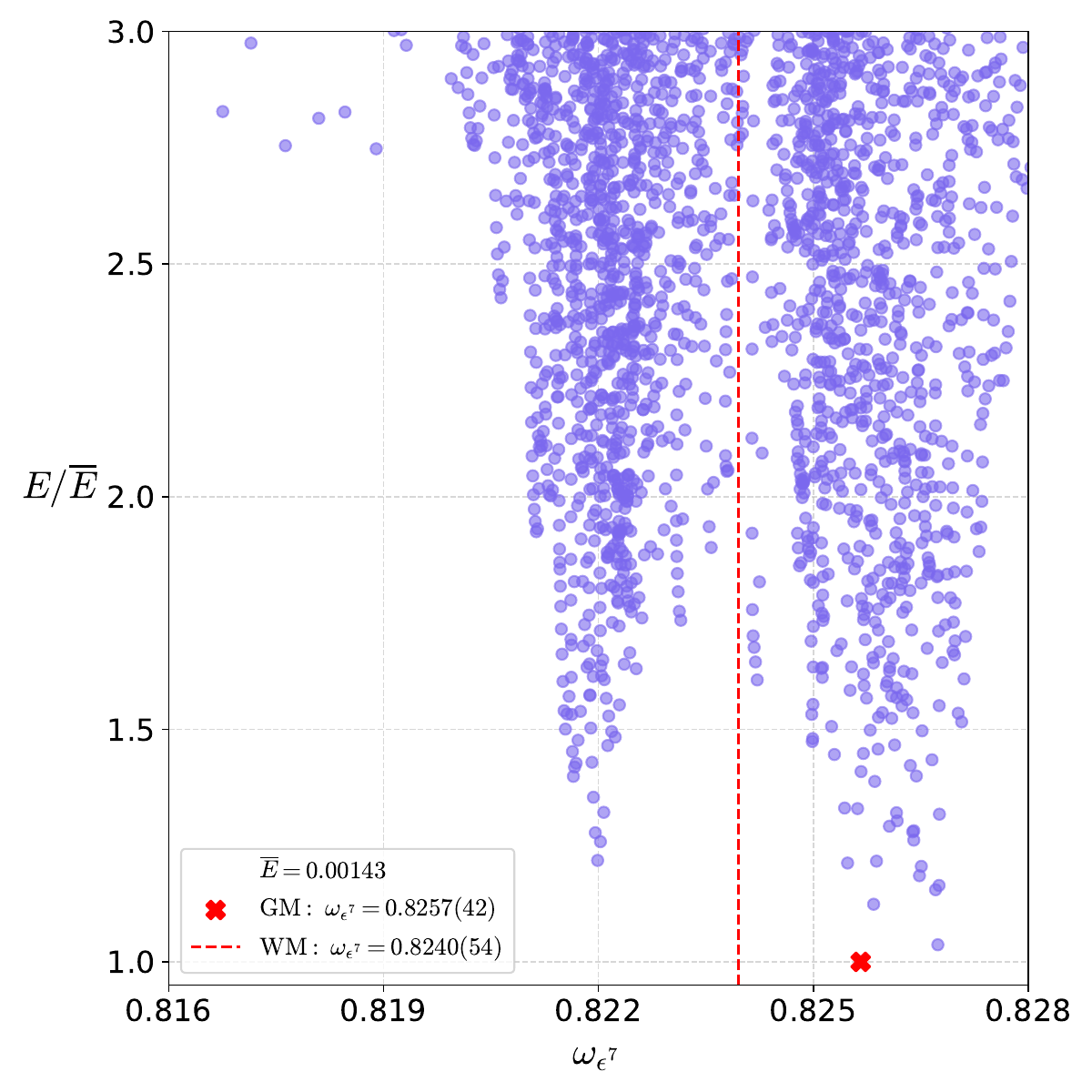}}\hfill
		\subfloat[]{\includegraphics[width=0.34\textwidth]{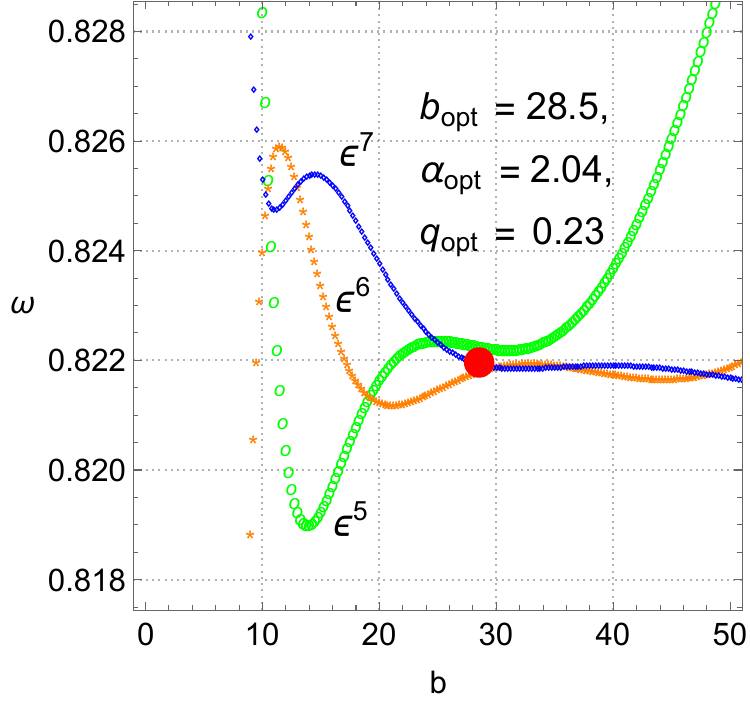}}\hfill
		\subfloat[]{\includegraphics[width=0.34\textwidth]{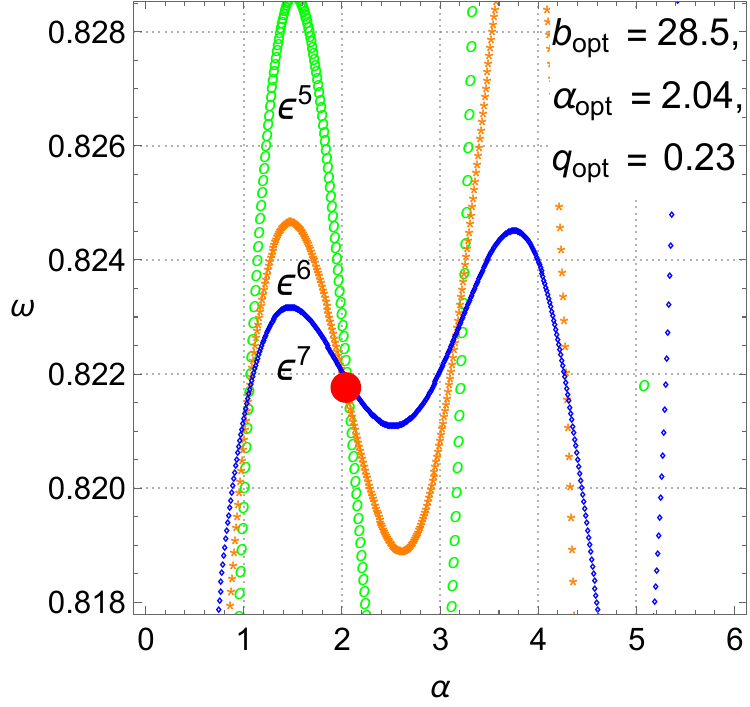}}\hfill
		\subfloat[]{\includegraphics[width=0.3\textwidth]{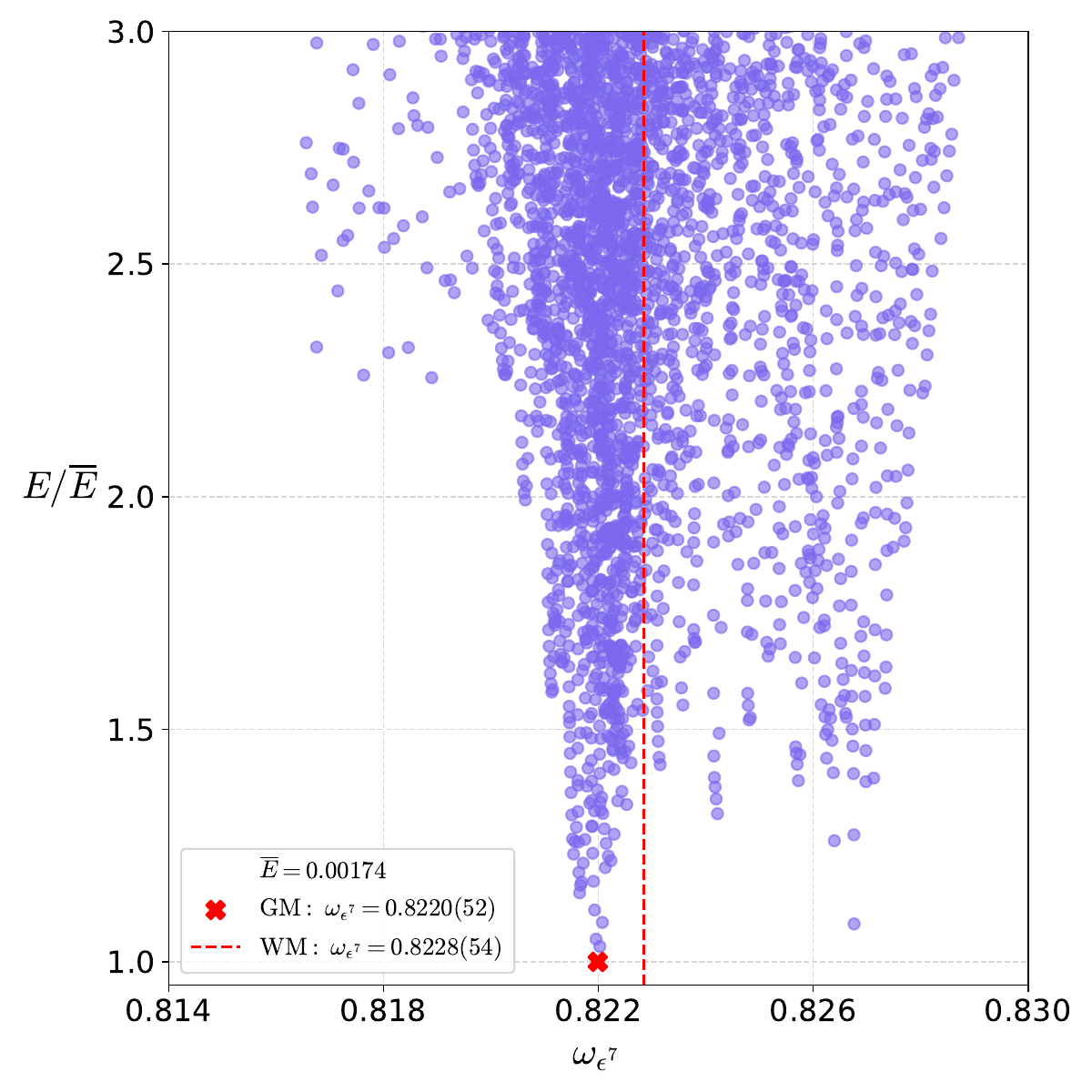}}\hfill
		\subfloat[]{\includegraphics[width=0.23\textwidth]{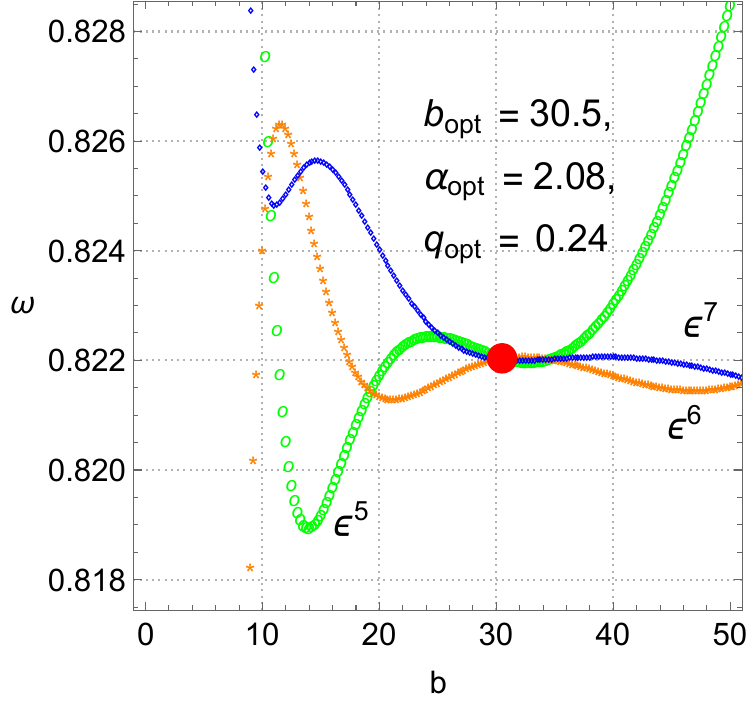}}\hfill
		\subfloat[]{\includegraphics[width=0.23\textwidth]{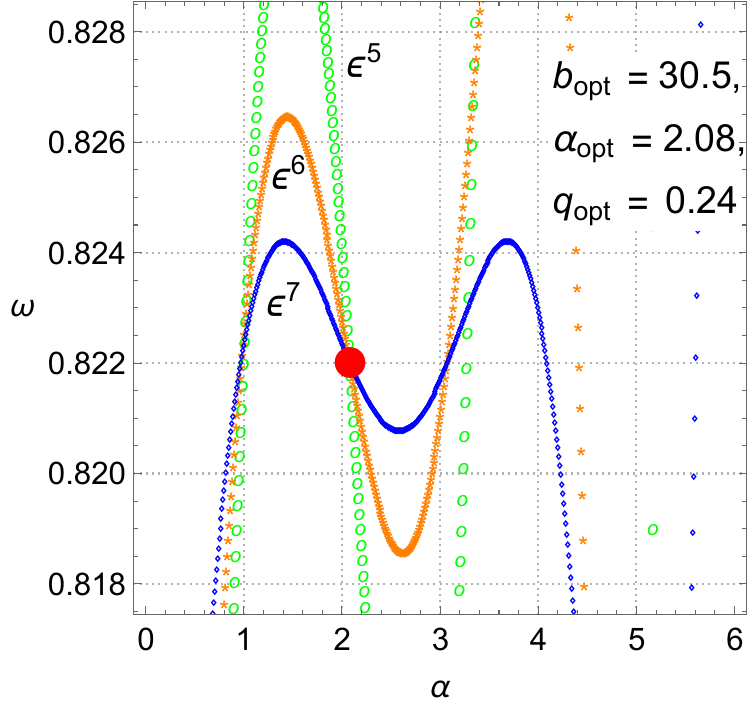}}\hfill
		\subfloat[]{\includegraphics[width=0.26\textwidth]{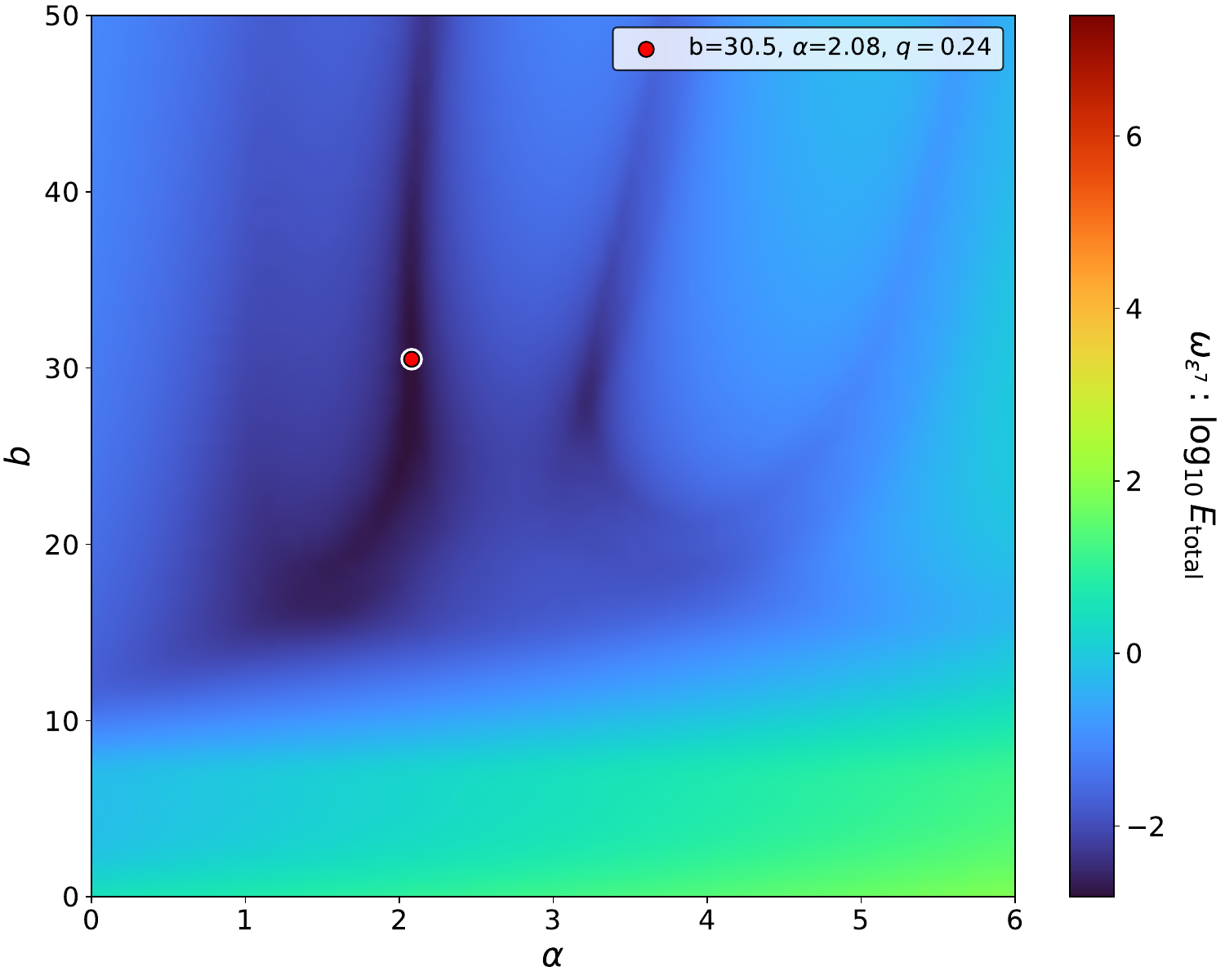}}\hfill
		\subfloat[]{\includegraphics[width=0.26\textwidth]{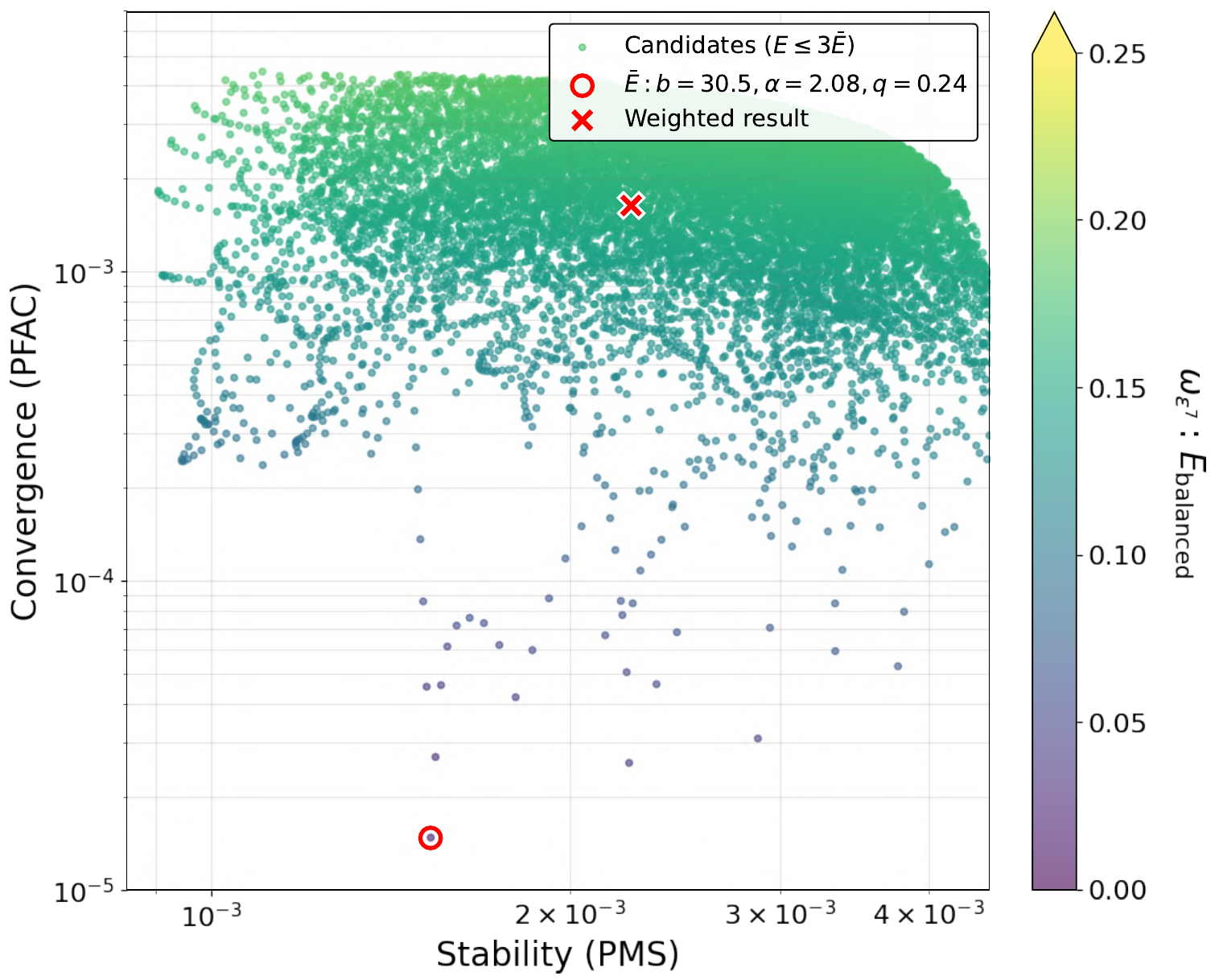}}\
		\centering
		\caption{Dependence of the correction-to-scaling exponent $\omega$ on the resummation parameters $b$ [(a), (d)] and $\alpha$ [(b), (e)] at seven-loop order. The top row corresponds to the smaller-$b$ initialization around the optimal point $(b, \alpha, q) = (10.5, 0.76, 0.29)$ ({\it Method I}, Section~\ref{error_estimations}). The middle row corresponds to the larger-$b$ initialization around the optimal point $(b, \alpha, q) = (28.5, 2.04, 0.23)$ ({\it Method I}, Section~\ref{error_estimations}). The bottom row corresponds to the optimal point $(b, \alpha, q) = (30.5, 2.08, 0.24)$ ({\it Method II}, Section~\ref{rms_error_estimations}). Panels (c), (f) display the normalized error $E/\overline{E}$ distribution across the parameter space. The global minimum (GM, red cross) and weighted mean (WM, dashed line) are shown with uncertainties of $3\overline{E}$ and $\overline{E} + 2\sigma_w$, respectively, accounting for both local stability and the statistical spread of the candidate set within the $E \leq 3\overline{E}$ region [see Eq.~(\ref{weight})]. Panels (i) and (j) show the error landscape in the $(b, \alpha)$ plane at the optimal $q$ value and the distribution of calculated points in the plane of stability ($E_{\mathrm{pms}}$) versus convergence ($E_{\mathrm{pfac}}$), respectively. Our final estimates are $\omega = 0.8257(42)$  via {\it Method I} and $\omega = 0.8220(46)$ via {\it Method II}.}
		\label{omega7__n1}
	\end{figure}

	In this Appendix, we provide the supporting data and complementary illustrative material regarding the numerical analysis of the resummed $\epsilon$-expansion series. This material serves as the basis for the numerical estimates discussed in Sections~\ref{error_estimations}, \ref{rms_error_estimations}, and \ref{results} -- which are summarized in Tables~\ref{tab:comparison__two_methods} and \ref{tab:final_results} -- as well as the convergence trends shown in Fig.~\ref{final_trend}. To ensure clarity and avoid repetition across the various loop approximations for the critical exponents ($\eta$, $\nu$, and $\omega$), we present the core analysis through an extensive set of figures. Specifically, Figs.~\ref{eta6__n1}--\ref{eta8__n1} show the results for $\eta$ at six-, seven-, and eight-loop orders, Figs.~\ref{nu5__n1}--\ref{nu7__n1} show those for $\nu^{-1}$ at five-, six-, and seven-loop orders, and Figs.~\ref{omega5__n1}--\ref{omega7__n1} show those for $\omega$ at five-, six-, and seven-loop orders, respectively. Detailed explanations of the parameter spaces, stability landscapes, and specific resummation configurations are provided directly within the respective figure captions. 
	
	
\end{document}